\documentclass[11pt]{amsart}
\usepackage{amsmath}
\usepackage{amstext}
\usepackage{amssymb}
\usepackage{latexsym}
\numberwithin{equation}{section}

\advance\textheight3truecm
\advance\hoffset-2truecm
\advance\voffset-1.7truecm
\theoremstyle{remark}
\newtheorem{remark*}{Remark}

\let\Bbb\mathbb \let\Cal\mathcal \let\frak\mathfrak
\let\boldkey\mathbf \let\bold\mathbf
\newcommand\sptilde{^\sim}

\newcommand\spr[2]{\langle #1,#2\rangle}
\newcommand\into{\int_\Omg}
\newcommand\jedna{{\boldkey1}}
\newcommand\TT{{\bold T}}
\newcommand\Obs{Obs}
\newcommand\AAA{{\Cal A}}
\newcommand\h{^{(h)}}
\newcommand\Omg{{\bigam}}   %the phase space - used to be {M},
\newcommand\OMEGA{\Omega}
\newcommand\Oph{\operatorname{Op}\h}
\newcommand\KK{{\Cal K}}
\newcommand\XXX{{\Cal X}}     \newcommand\xX{\XXX}
\newcommand\XX{{\bfrakQ}}
\newcommand\mfr[2]{\raise 0.2\baselineskip\hbox{$\scriptstyle #1$}\!/\hskip
                 -0.1\baselineskip{\scriptstyle #2}}

\newcommand\FF{\Cal F}
\newcommand\PP{{\Cal P}}
\newcommand\curv{\operatorname{curv}}
\newcommand\DD{{\Cal D}}
\newcommand\MD{{\Omg/\DD}}
\newcommand\mhD{$-\tfrac12$-$\DD$}
\newcommand\BB{\Cal B}
\newcommand\LL{\Cal L}
\newcommand\EE{{\Cal E}}
\newcommand\FnPC{\FF^n\PP^\CC}
\newcommand\hatFnPC{\hat\FF^n\PP^\CC}
\newcommand\tildeFnPC{\tilde\FF^n\PP^\CC}
\newcommand\mhP{$-\tfrac12$-$\PP$}
\newcommand\tBP{\tilde\BB^\PP}
\newcommand\ZZ{\bold Z}
\newcommand\GG{{\PP'}}
\newcommand\hatDD{{\hat\DD}}
\newcommand\hatEE{\hat\EE}
\newcommand\FwM{\FF^\omega \Omg}
\newcommand\tFwM{\tilde\FF^\omega \Omg}
\newcommand\Lag{\LL^\omega \Omg}
\newcommand\tildeLag{\tilde\LL^\omega \Omg}
\newcommand\rank{\operatorname{rank}}
\newcommand\ad{\operatorname{ad}}
\newcommand\opP{\overline\partial_\PP}
\newcommand\Ker{\operatorname{Ker}}
\newcommand\Ran{\operatorname{Ran}}

\newcommand\HH{\mathfrak H}
\newcommand\RR{\mathbb R}

\newtheorem{defi}{Definition}[section]
\newtheorem{lem}[defi]{Lemma}
\newtheorem{prop}[defi]{Proposition}
\newtheorem{theorem}[defi]{Theorem}
\newtheorem{cor}[defi]{Corollary}

\newcommand{\bedefin}{\begin{defi}}

\newcommand{\betheo}{\begin{theorem}}
\newcommand{\enth}{\end{theorem}}
\newcommand{\entheo}{\end{theorem}}

\newcommand{\becor}{\begin{cor}}
\newcommand{\encor}{\end{cor}}

\newcommand{\belem}{\begin{lem}}
\newcommand{\enlem}{\end{lem}}

\newcommand{\beprop}{\begin{prop}}
\newcommand{\enprop}{\end{prop}}

\newcommand{\be}{\begin{equation}}
\newcommand{\en}{\end{equation}}

\newcommand{\bea}{\begin{eqnarray}}
\newcommand{\ena}{\end{eqnarray}}

\newcommand{\beano}{\begin{eqnarray*}}
\newcommand{\enano}{\end{eqnarray*}}

\newcommand{\bee}{\begin{enumerate}}
\newcommand{\ene}{\end{enumerate}}

\newcommand{\bei}{\begin{itemize}}
\newcommand{\eni}{\end{itemize}}

\newcommand{\betab}{\begin{tabular}}
\newcommand{\entab}{\end{tabular}}

\newcommand{\bd}{\begin{displaymath}}

\newcommand{\htil}{\widetilde{\mathfrak H}}

\newcommand{\s }{\mathbf}

\newcommand{\prhs}{{\mathbb C\mathbb P}( \HH)}

\newcommand{\CC}{\C}

\newcommand{\bpsi}{\mbox{\boldmath $\psi$}}

\newcommand{\bomega}{\mbox{\boldmath $\omega$}}

\newcommand{\bfeta}{\mbox{\boldmath $\eta$}}

\newcommand{\bigam}{\mbox{\boldmath $\Gamma$}}
\newcommand{\bxi}{\mbox{\boldmath $\xi$}}

\newcommand{\bfrakQ}{\mbox{\boldmath $\mathfrak Q$}}
\newcommand{\bfrakX}{\mbox{\boldmath $\mathfrak X$}}

\newcommand{\bfrakL}{\mbox{\boldmath $\mathfrak L$}}

\newcommand{\bA}{\mathbf A}

\newcommand{\bB}{\mathbf B}

\newcommand{\bk}{\mathbf k}
\newcommand{\bp}{\mathbf p}
\newcommand{\bq}{\mathbf q}

\newcommand{\bx}{\mathbf x}

\newcommand{\by}{\mathbf y}

\newcommand{\C}{\mathbb C}

\newcommand{\I}{\mathbb I}

\newcommand{\R}{\mathbb R}

\newcommand{\BMP}{\mathbf{Mp}}
\newcommand{\bmp}{\mathbf{mp}}
\newcommand{\CinfRQ}{C^\infty(\bfrakQ)_{\RR}}

\begin{document}
\title[Quantization methods]
{Quantization methods: a~guide\\ for physicists and analysts}
%{Quantization methods for physicists\\ and analysts -- a rapid survey}
\author[S.T.~Ali, M.~Engli\v s]{S.~Twareque Ali$^\dag$ and
   Miroslav Engli\v s$^{\ddagger}$ }
\address{$^\dag$Department of Mathematics and Statistics, Concordia University,
Montr\'eal, Qu\'ebec, CANADA H4B~1R6}
 \email{stali{@}mathstat.concordia.ca}
\address{$^{\ddagger}$M\'U AV \v CR, \v Zitn\'a 25, 11567 Praha 1,
 Czech Republic}
\email{englis{@}math.cas.cz}
\begin{thanks}{The work of the first author (STA) was partially supported by
a grant from the Natural Sciences and Engineering Research Council (NSERC) of
Canada. The~second author (ME) acknowledges support from GA~AV~\v CR grants
A1019005 and A1019304.}\end{thanks}

\maketitle

\tableofcontents

\section{Introduction} \label{sec1}

Quantization is generally understood as the transition from classical to
quantum mechanics. Starting with a classical system, one often
wishes to formulate a quantum theory, which in an appropriate
limit, would reduce back to the classical system of departure. In a
more general setting, quantization is also understood as a
correspondence between a classical and a quantum theory. In~this
context, one also talks about {\sl dequantization\/,} which is a
procedure by which one starts with a quantum theory and arrives
back at its classical counterpart. It is well-known however, that
not every quantum system has a meaningful classical counterpart
and moreover, different quantum systems may reduce to the same
classical theory. Over the years, the processes of quantization
and dequantization have evolved into mathematical theories in
their own right, impinging on areas of group representation theory
and symplectic geometry. Indeed, the programme of geometric quantization
is in many ways an offshoot of group representation theory on
coadjoint orbits, while other techniques borrow heavily from the
theory of representations of diffeomorphism groups.

   In these pages we attempt to present an overview of some of the better known
quantization techniques found in the current literature and used both by
physicists and mathematicians. The treatment will be more descriptive than
rigorous, for we aim to reach both physicists and mathematicians, including
non-specialists in the field. It is our hope that an overview such as this will
put into perspective the relative successes as well as shortcomings of the various
techniques that have been developed and,  besides delineating  their usefulness
in understanding the nature of the quantum regime, will also demonstrate
the mathematical richness of the attendant structures. However, as will become clear,
no one method solves the problem of quantization completely and we shall
try to point out both the successes and relative shortcomings of each method.

\subsection{The problem}

The original concept of quantization (nowadays usually referred to as
{\sl canonical quantization}), going back to Weyl, von Neumann, and Dirac
 \cite{bib:Dirac} \cite{bib:vNeu} \cite{bib:WeyGrleh}, consists in assigning
(or~rather, trying to assign) to the observables of  classical mechanics, which
are real-valued functions $f(\bp,\bq)$ of $(\bp,\bq)=(p_1,\dots,p_n,q_1,\dots,
q_n)\in\RR^n\times\RR^n$ (the phase space),  self-adjoint operators $Q_f$ on
the Hilbert space $L^2(\RR^n)$ in such a way that
\begin{enumerate}
\item[(q1)] the correspondence $f\mapsto Q_f$ is linear;
\item[(q2)] $Q_\jedna=I$, where $\jedna$ is the constant function, equal to one
everywhere, and $I$ the identity operator;
\item[(q3)] for any function $\phi:\RR\to\RR$ for which $Q_{\phi\circ f}$ and
$\phi(Q_f)$ are well-defined,
$Q_{\phi\circ f}=\phi(Q_f)$; and
\item[(q4)] the operators $Q_{p_j}$ and $Q_{q_j}$ corresponding to the
coordinate functions $p_j,q_j$ ($j=1,\dots,n$) are given by
\begin{equation}  Q_{q_j} \psi = q_j \psi,
\qquad Q_{p_j} \psi =-\frac{ih}{2\pi} \,\frac{\partial \psi} {\partial q_j}
\qquad \text{for } \psi \in L^2(\RR^n,d\bq).
\label{tag:Schro}  \end{equation}
\end{enumerate}
The condition (q3) is usually known as the {\sl von Neumann rule.\/}
The domain of definition of the mapping $Q:f\mapsto Q_f$ is called the space of
{\sl quantizable observables,\/} and one would of course like to make it as
large as possible --- ideally, it should include at least the
infinitely differentiable functions $C^\infty(\RR^n)$, or
some other convenient function space. The parameter $h$, on which the
quantization map $Q$ also depends, is usually a small positive number,
identified with the {\sl Planck constant.}
(One~also often uses the shorthand notation $\hbar$ for the ratio $h/2\pi$).

An~important theorem of Stone and von Neumann  \cite{bib:vNeu}
states that up to unitary
equivalence, the operators (\ref{tag:Schro}) are the unique operators acting on
a Hilbert space $\HH$, which satisfy $(a)$ the irreducibility condition,
\begin{equation}  \begin{aligned}
& \text{there are no subspaces $\HH_0\subset\HH$, other than $\{0\}$ and $\HH$
itself, that are stable} \\
\noalign{\vskip-2pt}
& \text{under the action  of all the operators $Q_{p_j}$ and $Q_{q_j}$,
$j=1,\dots,n$,} \end{aligned}  \label{tag:IRE}  \end{equation}
and $(b)$ the commutation relations
\begin{equation}  [Q_{p_j},Q_{p_k}] = [Q_{q_j},Q_{q_k}] =0, \qquad
[Q_{q_k}, Q_{p_j}] = \frac{ih}{2\pi} \delta_{jk} I . \label{tag:CCR}
\end{equation}

The physical interpretation is as follows\footnote{It~is
precisely because of this interpretation that one actually has to
insist on the operators $Q_f$ being self-adjoint (not just
symmetric or~``formally self-adjoint'').
See~Gieres~\cite{bib:Gieres} for a thorough discussion of this
issue.}. The classical system, of $n$ linear degrees of freedom,
moves on the phase space $\RR^n \times \RR^n$, with $q_j, p_j$ being
the canonical position and momentum observables, respectively. Any classical
state is given as a probability distribution (measure) on phase space.
The states of the quantum system correspond to
one-dimensional subspaces $\CC u$ ($\|u\|=1$) of $L^2(\RR^n)$, and
the result of measuring an observable $f$ in the state $u$ leads to the
probability distribution $\spr{\Pi(Q_f) u}u$, where
$\Pi(Q_f)$ is the spectral measure of~$Q_f$. In~particular, if
$Q_f$ has pure point spectrum consisting of eigenvalues
$\lambda_j$ with unit eigenvectors $u_j$, the possible outcomes of
measuring $f$ will be $\lambda_j$ with probability $|\spr
u{u_j}|^2$; if $u=u_j$ for some $j$, the measurement will be
deterministic and will always return~$\lambda_j$. Noncommutativity
of operators corresponds to the impossibility of measuring
simultaneously the corresponding observables. In~particular, the
canonical commutation relations (\ref{tag:CCR}) above express the
celebrated Heisenberg uncertainty principle.

Evidently, for $f=f(\bq)$ a polynomial in the position variables
$q_1,\dots,q_n$, the linearity (q1) and the von Neumann rule (q3) dictate that
$Q_{f(\bq)}=f(Q_\bq)$ in the sense of spectral theory (functional calculus for
commuting self-adjoint operators); similarly for polynomials $f(\bp)$ in~$\bp$.
The canonical commutation relations then imply that for any functions $f,g$
which are at most linear in either $\bp$ or $\bq$,
\begin{equation}  [Q_f,Q_g] = \frac{ih}{2\pi} Q_{\{f,g\}} \, ,  \label{tag:PB}
\end{equation}
where
\begin{equation}  \{f,g\} = \sum_{j=1}^n \bigg( \frac{\partial f}{\partial q_j}
\frac{\partial g}{\partial p_j} -\frac{\partial f}{\partial p_j}
\frac{\partial g}{\partial q_j} \bigg)   \label{tag:POAB}   \end{equation}
is the {\sl Poisson bracket\/} of $f$ and $g$. It~turns out that another
desideratum on the quantization operator~$Q$, motivated by physical
considerations~(\cite{bib:Dirac},~pp.~87-92), is that
\begin{enumerate}
\item[(q5)] the correspondence (\ref{tag:PB}), between the classical Poisson
bracket and the quantum commutator bracket,  holds for all quantizable
observables $f$ and $g$.
\end{enumerate}
Thus we are lead to the following problem: find a vector space $\Obs$ (as large
as possible) of real-valued functions $f(\bp,\bq)$ on $\RR^{2n}$, containing
the coordinate functions $p_j$ and $q_j$ ($j=1,\dots,n$), and a mapping
$Q:f\mapsto Q_f$ from $\Obs$ into self-adjoint operators on $L^2(\RR^n)$ such
that (q1)--(q5) are satisfied.

(Note that the axiom (q2)~is, in~fact, a~consequence of either~(q3) (taking
$\phi=\jedna$) or~(q5) (taking $f=p_1$, $g=q_1$); we~have stated it separately
for reasons of exposition.)

\subsection{Stumbling blocks}

Unfortunately, it turns out that the axioms (q1)--(q5) are not quite
consistent. First of all, using (q1)--(q4) it is possible to express $Q_f$ for
$f(\bp,\bq)=p_1^2 q_1^2=(p_1 q_1)^2$ in two ways with two different results
(see~\cite{bib:Foll}, p.~17; or Arens and Babbitt~\cite{bib:ArensB}). Namely,
let us temporarily write just $p,q$ instead of $p_1,q_1$ and $P,Q$ instead of
$Q_{p_1}$ and $Q_{q_1}$, respectively. Then by the von Neumann rule (q3) for
the squaring function $\phi(t)=t^2$ and~(q1),
$$ pq = \frac{(p+q)^2-p^2-q^2}2 \quad \implies \quad
Q_{pq} = \frac{(P+Q)^2-P^2-Q^2}2=\frac{PQ+QP}2;  $$
and similarly
$$ p^2 q^2 = \frac{(p^2+q^2)^2-p^4-q^4}2 \quad \implies \quad
Q_{p^2 q^2} = \frac{P^2 Q^2+Q^2 P^2}2 .  $$
However, a small computation using only the canonical commutation
relations~(\ref{tag:CCR}) (which are a consequence of either (q4) or~(q5))
shows that
$$ \frac{P^2 Q^2+Q^2 P^2}2 \neq \bigg(\frac{PQ+QP}2\bigg)^2.   $$
Thus neither (q4) nor (q5) can be satisfied if (q1) and (q3) are and $p_1^2,
q_1^2,p_1^4,q_1^4,p_1 q_1$ and $p_1^2 q_1^2\in\Obs$.

Secondly, it is a result of Groenewold \cite{bib:Groe}, later elaborated
further by van Hove \cite{bib:vHove}, that (q5) fails whenever (q1) and (q4)
are satisfied and $\Obs$ contains all polynomials in $\bp,\bq$ of degree not
exceeding four. To~see this, assume, for simplicity, that $n=1$ (the argument
for general~$n$ is the same), and let us keep the notations $p,q,P,Q$ of the
preceding paragraph and for the sake of brevity also set $c=-\dfrac{ih}{2\pi}$.
Note first of all that for any self-adjoint operator~$X$,
\begin{equation}  [X,P]=[X,Q]=0 \quad\implies\quad X=dI\text{ for some }
d\in\CC. \label{tag:XPXQ} \end{equation}
(Indeed, any spectral projection $E$ of $X$ must then commute with $P,Q$, hence
the range of $E$ is a subspace invariant under both $P$ and~$Q$;
by~irreducibility, this forces $E=0$ or~$I$.) Set~now $X=Q_{pq}$; then, since
$$ \{pq,p\}=p, \qquad \{pq,q\}=-q,  $$
we must have by~(q5)
$$ [X,P]=-cP, \qquad [X,Q]=cQ.  $$
As~also
$$ [ \tfrac{PQ+QP}2,P]=-cP, \qquad [\tfrac{PQ+QP}2,Q]=cQ,  $$
it~follows from (\ref{tag:XPXQ}) that
$$ Q_{pq} \equiv X = \frac{PQ+QP}2 + d I \qquad\text{for some }d\in\CC.  $$
Next set $X=Q_{q^m}$ ($m=1,2,\dots$); then from
$$ \{q^m,q\}=0, \qquad \{q^m,p\}=m q^{m-1}  $$
we~similarly obtain
$$ X=Q^m + d_m I \qquad\text{for some }d_m\in\CC.  $$
Furthermore, since
$$ \{pq,q^m\}= - m q^m,  $$
it~follows that
$$ cmX=[\tfrac{PQ+QP}2+dI,Q^m+d_m I] = [\tfrac{PQ+QP}2,Q^m] =cm Q^m.  $$
Thus (using also a similar argument for $X=Q_{p^m}$)
$$ Q_{q^m} = Q^m, \qquad Q_{p^m}=P^m, \qquad \forall m=1,2,\dots.  $$
Now from
$$ \{ p^2,q^3 \} = - 6q^2 p  $$
we~obtain that
$$ 6c Q_{q^2 p} = [P^2,Q^3] = 3c P Q^2+3c Q^2 P ,  $$
so
$$ Q_{q^2 p} = \frac{PQ^2+Q^2P}2  $$
and similarly for $Q_{p^2 q}$. Thus finally, we have on the one hand
$$ \{p^3,q^3\}=- 9p^2 q^2 \quad\implies\quad Q_{p^2 q^2} = \frac1{9c} [P^3,Q^3]
= Q^2 P^2 +2c QP + \frac23 c^2,  $$
while on the other hand
$$ \{p^2 q,p q^2\} = - 3 p^2 q^2 \quad\implies\quad Q_{p^2 q^2}= \frac1{3c}
\Big[\frac{P^2Q+QP^2}2,\frac{PQ^2+Q^2P}2\Big] = Q^2 P^2+2c QP+ \frac13c^2,  $$
yielding a contradiction.

Thirdly, it can be shown that one arrives (by arguments of a similar nature as
above) at a contradiction even if one insists on the axioms (q3), (q4)
and~(q5), but discards~(q1) (linearity); see~\cite{bib:ETuy}. (Note that
by (q3) with  $\phi(t)=c t$, we still have at least homogeneity, i.e. $Q_{cf}=c
Q_f$ for any constant~$c$.)

In conclusion, we  see that not only the axioms (q1)--(q5) taken together,
but even \underbar{any} \underbar{three} of the axioms (q1), (q3), (q4)
and~(q5) are inconsistent.

\begin{remark*} The idea of discarding the linearity axiom (q1) may seem a
little wild at first sight, but there seems to be no physical motivation for
assuming linearity, though it is definitely convenient from the computational
point of view (cf.~Tuynman \cite{bib:TuyWis},~\S5.1). In~fact, nonlinear
assignments $f\mapsto Q_f$ do actually occur already in some existing
approaches to geometric quantization, namely when one defines the quantum
observables $Q_f$ using the Blattner-Kostant-Sternberg kernels;
cf.~(\ref{tag:BAOZ}) in \S\ref{sec27} below.   \qed  \end{remark*}

\begin{remark*} The inconsistencies among the axioms above actually go even
further. Namely, an analysis of the argument in \cite{bib:ETuy} shows that,
in~fact, it only requires (q3) and (q5) alone to produce a contradiction. The
combination (q1)+(q3) is satisfied e.g.~by the map assigning to $f$ the
operator of multiplication by $f$, however this is uninteresting from the point
of view of physics (noncommutativity is lost). Similarly, (q1)+(q4) can be
satisfied but the outcome is of no physical relevance. The combination
(q1)+(q5) is satisfied by the prequantization of van~Hove (to~be discussed in
detail in \S\ref{sec21} below). In~conclusion, it thus transpires that with the
exception of (q1)+(q5), and possibly also of (q4)+(q3) and (q4)+(q5), even
any \underbar{two} of the axioms (q1), (q3), (q4) and~(q5) are either
inconsistent or lead to something trivial.   \qed  \end{remark*}

\begin{remark*} From a purely mathematical viewpoint, it can, in fact, be shown
that already (q3) and the canonical commutation relations (\ref{tag:CCR}) by
themselves lead to a contradiction if one allows the space $\Obs$ to contain
sufficiently ``wild'' functions (i.e.~not $C^\infty$ --- for instance, the
Pe\'ano curve function $f$ mapping $\RR$ continuously onto~$\RR^{2n}$).
See again~\cite{bib:ETuy}.   \qed  \end{remark*}

\subsection{Getting out of the quagmire}

There are two traditional approaches on how to handle this disappointing
situation. The first is to keep the four axioms (q1), (q2), (q4) and (q5)
(possibly giving up only the von Neumann rule~(q3)) but restrict the space
$\Obs$ of quantizable observables. For instance, we have seen above that it may
not contain simultaneously $p_j^2,q_j^2$ and $p_j^2 q_j^2$, for any~$j$;
however, taking $\Obs$ to be the set of all functions at most linear
in~$\bp$,~i.e.
$$ f(\bp,\bq) = f_0(\bq) + \sum_j f_j(\bq) p_j, \qquad
 f,f_j\in C^\infty(\RR^n), $$
and setting
$$ Q_f = f_0(\widehat{\bq} ) + \frac12 \sum_j [ f_j(\widehat{\bq}) Q_{p_j} +
 Q_{p_j} f_j(\widehat{\bq}) ],  $$
where we have written $\widehat{\bq}$ for the vector operator $Q_\bq$,
it is not difficult to see that all of (q1),(q2),(q4) and (q5) are satisfied.
Similarly one can use functions at most linear in~$\bq$, or, more generally, in
$a\bp+b\bq$ for some fixed constants $a$ and $b$.

The second approach is to keep (q1),(q2) and (q4), but require (q5) to hold
only asymptotically as the Planck constant $h$ tends to zero. The simplest way
to achieve this is as follows. By~the remarks above, we know that the operator
$Q_f$ corresponding to $f(\bp,\bq)=e^{i\bfeta\cdot \bq}$ ($\bfeta\in\RR^n$) is
$Q_f=e^{i\bfeta\cdot\widehat{\bq}}$, and similarly for~$\bp$.  Now an
``arbitrary'' function $f(\bp,\bq)$ can be expanded into exponentials via the
Fourier transform,
$$ f(\bp,\bq) = \iint \hat f(\bxi,\bfeta) \,
e^{2\pi i (\bxi\cdot \bp + \bfeta \cdot\bq)} \, d\bxi \, d\bfeta .  $$
Let us now postulate that
$$ Q_f = \iint \hat f(\bxi,\bfeta) \,
e^{2\pi i(\bxi\cdot\widehat{\bp} + \bfeta \cdot\widehat{\bq})} \, d\bxi \,
d\bfeta =: W_f,  $$
where again, $\widehat{\bp} = Q_\bp$.
After a simple manipulation, the operator $W_f$ can be rewritten as the
oscillatory integral
\begin{equation}  W_f g(\bx) = h^{-n} \iint f \Big(  \bp ,\frac{\bx+\by}2 \Big)
\, e^{2\pi i(\bx-\by)\cdot \bp /h}  g(\by) \, d\by \, d \bp  .
\label{tag:WEY}  \end{equation}
This is the celebrated Weyl calculus of pseudodifferential operators (see
H\"ormander \cite{bib:HormW}, Shubin \cite{bib:Shu}, Taylor \cite{bib:Tayl},
for instance). The last formula allows us to define $W_f$ as an operator from
the Schwartz space $\Cal S(\RR^n)$ into the space $\Cal S'(\RR^n)$ of tempered
distributions; conversely, it follows from the Schwartz kernel theorem that any
continuous operator from $\Cal S$ into $\Cal S'$ is of the form $W_f$ for some
$f\in\Cal S'(\RR^{2n})$. In~particular, if $f,g\in \Cal S'(\RR^{2n})$ are such
that $W_f$ and $W_g$ map $\Cal S(\RR^n)$ into itself (this is the case, for
instance, if $f,g\in\Cal S(\RR^{2n})$), then so does their composition
$W_f W_g$.  Thus, $W_f W_g=W_{f\,\sharp\, g}$ for some $f\,\sharp\, g\in\Cal
S'(\RR^{2n})$ and we  call $f\,\sharp\, g$ the {\sl twisted\/} (or {\sl
Moyal\/}) product of $f$ and~$g$. Now it turns out that under appropriate
hypotheses on $f$ and $g$ (for instance, if $f,g\in\Cal S(\RR^{2n})$, but much
weaker assumptions will~do), one has the asymptotic expansion
\begin{equation}  f\,\sharp\, g = \sum_{j=0}^\infty h^j \rho_j(f,g)
\quad\text{as }h\to0,
\text{ where }\rho_0(f,g)=fg,\ \rho_1(f,g)= \tfrac i{4\pi}\{f,g\}.
\label{tag:MOY} \end{equation}
Hence, in particular,
\begin{equation}  f\,\sharp\, g - g\,\sharp\, f = \frac{ih}{2\pi} \{f,g\} +
O(h^2) \qquad\text{as }h\to0.  \label{tag:AS}  \end{equation}
This is the asymptotic version of~(q5). (Incidentally, for $\phi$ a polynomial,
one also gets an asymptotic version of the von Neumann rule~(q3).) The validity
of (q1), (q2), and (q4) follows immediately from the construction. See
Chapter~2 in \cite{bib:Foll} for the details.

\begin{remark*} An elegant general calculus for non-commuting tuples of
operators (of which (\ref{tag:Schro}) are an example), building essentially on
(q1), (q2) and a version of~(q3), was developed by Nelson~\cite{bib:Nels}.
Generalizations of the Weyl calculus were studied by
Anderson~\cite{bib:Anders}. \qed  \end{remark*}

\medskip

The basic problem of quantization is to extend these two approaches from
$\RR^{2n}$ to any symplectic manifold. The first of the above approaches leads
to {\sl geometric quantization,\/} and the second to {\sl deformation
quantization.\/} We shall discuss the former in Section~\ref{sec2} and  the
latter in Sections~\ref{sec3} and~\ref{sec4}, and then mention some other
approaches in Sections~\ref{sec5}--\ref{sec6}. Prior to that, we~review in
Section~\ref{sec-canquant} two other approaches, the {\sl Segal quantization\/}
and the {\sl Borel quantization\/}, which are straightforward generalizations
of the canonical scheme. They take a slightly different route by working only
with the configuration space~$\bfrakQ$ (the~phase space $\Omg$ is basically
forgotten completely, and its symplectic structure $\omega$ is used solely for
the purpose of defining the Poisson bracket), and quantizing only functions on
$\bfrakQ$ and vector fields on it instead of functions on~$\Omg$. This is the
Segal quantization; the Borel quantization enhances it further by allowing for
internal degrees of freedom (such as spin) with the aid of tools from
representation theory --- systems of imprimitivity and projection-valued
measures. As mentioned earlier and as will emerge from the discussion,
no one method completely solves the
the problem of quantization, nor does it adequately answer all the questions raised.
Consequently, we refrain from promoting one over the other,
inviting the reader to formulate their own preference.

\section{Canonical quantization and its generalizations}\label{sec-canquant}

We discuss in some detail in this section the original idea of quantization,
introduced in the early days of quantum mechanics  -- rather simple minded and
ad hoc, but extremely effective -- and some later refinements of it. Some
useful references are \cite{bib:-doebtol}, \cite{bib:ET},
\cite{bib:-golmensh1}, \cite{bib:-golmensh2}, \cite{bib:Groe}, \cite{bib:-GS},
\cite{bib:vHove}, \cite{bib:-nattermann}, \cite{bib:Segal} and
\cite{bib:-simsudmu1}.

\subsection{The early notion of quantization}
\label{sec-earlynotion}

The originators of quantum theory used the following simple technique
for quantizing a classical system: As before, let
 $q_{i}, p_{i}, \quad i = 1,2,
\ldots , n$, be the canonical position and momentum coordinates, respectively,
of a free classical system with $n$ degrees of freedom. Then their quantized
counterparts, $\widehat{q}_{i}, \widehat{p}_{i}$, are to be realized as
operators on the Hilbert space $\HH = L^{2}({\mathbb R}^{n}, d\bx)$, by the
prescription (see (\ref{tag:Schro})):
\be
  (\widehat{q}_{i}\psi )(\bx ) = x_{i}\psi (\bx )  \qquad
    (\widehat{p}_{i}\psi )(\bx ) = -i\hbar
    \frac {\partial}{\partial x_{i}}\psi (\bx),
\label{canquant1}
\en
on an appropriately chosen dense set of vectors $\psi$ in $\HH$. This simple
procedure is known as {\em canonical quantization\/}.   Then, as mentioned
earlier, the Stone-von Neumann {\em uniqueness theorem} \cite{bib:vNeu} states
that, up to unitary equivalence,
this is the only
representation which realizes the canonical commutation relations (CCR):
\be
 [\widehat{q}_{i}, \widehat{p}_{j}] = i\hbar I\; \delta_{ij},
      \qquad i,j = 1,2, \ldots, n,
\label{CCR4}
\en
irreducibly on a separable Hilbert space. Let us examine this question
of irreducibility a little more closely.

The operators $\widehat{q}_{i}, \widehat{p}_{j}$ and $I$ are the
generators of a representation of the Weyl-Heisenberg group on
$L^{2}({\mathbb R}^{n}, d\bx)$. This group (for a system with $n$ degrees of
freedom), which we denote by $G_{WH}(n)$, is topologically isomorphic to
${\mathbb R}^{2n+1}$ and consists of elements $(\theta , \bfeta )$, with
$\theta \in \mathbb R$ and $\bfeta \in {\mathbb R}^{2n}$, obeying the product
rule
\be
   (\theta , \bfeta )(\theta' , \bfeta' ) = (\theta + \theta'
   + \xi (\bfeta , \bfeta' ), \bfeta + \bfeta' ),
\label{WeylHeisenprod}
\en
where, the  {\em multiplier} $\xi$ is given by
\be
 \xi(\bfeta , \bfeta' ) = \frac 12\; \bfeta^{\dagger}\bomega\bfeta'
      = \frac 12\; (\bp\cdot\bq' - \bq\cdot\bp' )\; , \qquad
      \bomega = \begin{pmatrix} \mathbf 0 & -\mathbb I_n\\
      \mathbb I_n & \mathbf 0\end{pmatrix}\; ,
\label{multiplier}
\en
${\mathbb I}_n$ being the $n\times n$ identity matrix. This group is unimodular
and  nilpotent, with Haar measure $d\theta \; d{\bfeta}, \;\; d{\bfeta}$ being
the Lebesgue measure of ${\mathbb R}^{2n}$. Each unitary irreducible
representation (UIR) of $G_{WH}(n)$ is characterized by a non zero real number,
which we write as $\displaystyle{\frac 1\hbar}$, and eventually identify
$h = 2\pi\hbar$ with Planck's constant (of course, for a specific value of it).
Each UIR is carried by the Hilbert space $\HH =L^{2}({\mathbb R}^{n}, d\bx )$
via the following unitary operators:
\bea
  (U^{\hbar}(\theta , \bfeta )\psi )(\bx )&  = &
   (\exp \left[\frac i\hbar\;\left\{\theta + {\bfeta}^{\dagger}
  \bomega\widehat{\bfeta}\right\}\right]\psi )(\bx )
   \nonumber\\
     & = & \exp \left[\frac i\hbar\;\left\{\theta + \bp\cdot\bx -\frac 12
      \bp\cdot\bq \right\}\right]\psi (\bx -\bq ), \qquad
  \psi \in \HH .
\label{WeylHeisenrep}
\ena
This shows that the $2n$ quantized (unbounded)
operators, $\widehat{\eta}_i = \widehat{q}_i\;\;
i=1,2, \ldots , n$ and $\widehat{\eta}_i = \widehat{p}_{i-n},\;\;
i=n+1, n+2, \ldots , 2n$, which are the components of $\widehat{\bfeta}$,
along with the identity operator $I$ on $\HH$, are
the infinitesimal generators spanning the representation of the Lie algebra
${\mathfrak g}_{WH}(n)$ of the Weyl-Heisenberg group $G_{WH}(n)$. Since the
representation (\ref{WeylHeisenrep}) is irreducible, so also is the
representation (\ref{canquant1}) of the Lie algebra. This is the precise
mathematical sense in which we say that the algebra of Poisson brackets
$\{q_i, p_{j}\} = \delta_{ij}$ is irreducibly realized by the representation
(\ref{CCR4}) of the CCR.

One could justifiably ask at this point, how many other elements
could be added to the set ${\mathfrak g}_{WH}(n)$ and the
resulting enlarged algebra still be represented irreducibly on the
same Hilbert space $\HH$. In other words, does there exist a larger
algebra, containing ${\mathfrak g}_{WH}(n)$, which is also irreducibly
represented on $\HH = L^{2}({\mathbb R}^{n}, d\bx )$? To analyze this
point further, let us look at functions
$u$ on ${\mathbb R}^{2n}$ which are real-valued homogeneous polynomials in the
variables $q_i$ and $p_j$ of degree two. Any such polynomial can be written~as:
\be
   u(\bfeta ) = \frac 12 \sum_{i,j = 1}^{2n}\eta_i U_{ij}\eta_j
        = \frac 12 \bfeta^{T}U\bfeta ,
\label{homogpolyn1}
\en
where the $U_{ij}$ are the elements of a $2n\times 2n$ real, symmetric matrix
$U$. Set
\be
     U = JX(u)\; , \qquad J = \bomega^{-1}\; ,
\label{homogpolyn2}
\en
with $X(u) = -JU$, a $2n\times 2n$ real matrix satisfying
\be
    X(u) = JX(u)^{T}J.
\label{homogpolyn3}
\en
It follows, therefore, that every such homogeneous real-valued
polynomial $u$ is characterized
by a $2n\times 2n$ real matrix  $X(u)$ satisfying (\ref{homogpolyn3}),
and conversely, every such matrix represents a homogeneous
real-valued
polynomial of degree two via
\be
    u(\bfeta )= \frac 12 \bfeta^{T}JX(u)\bfeta .
\label{homogpolyn4}
\en
Computing the Poisson bracket of two such polynomials $u$ and $v$, we easily
see that
\be
  \{u ,v\} = \frac 12 \bfeta^{T}J[X(u), X(v)]\bfeta,
\quad \mbox{\rm where,} \quad [X(u), X(v)] = X(u)X(v)- X(v)X(u).
\label{homogpolyn5}
\en
In other words, the set of homogeneous, real-valued, quadratic polynomials
constitutes a closed algebra under the Poisson bracket operation, which we
denote by ${\mathfrak P}_2$, and the corresponding set of matrices $X(u)$ is
closed under the bracket relation,
\be
  [X(u), X(v)] = X(\{u, v\}),
\label{homogpolyn6}
\en
constituting thereby a matrix realization of the same algebra,
${\mathfrak P}_2$. In~fact, it is not hard to see that this is a maximal
subalgebra of the Poisson algebra $(C^\infty(\RR^{2n}),\{\cdot,\cdot\})$
of all smooth functions on $\RR^{2n}$ with respect to the Poisson bracket
(i.e.,~any other subalgebra which contains ${\mathfrak P}_2$ must necessarily
be the entire Poisson algebra). Moreover, we also see that
\be
  \{\eta_i, u\} = (X(u)\bfeta)_i, \qquad i=1,2, \ldots , 2n,
\label{semidirprod1}
\en
or compactly,
\be
   \{\bfeta, u\} = X(u)\bfeta,
\label{semidirprod2}
\en
which can be thought of as giving the action of the Poisson algebra of
quadratic polynomials on~${\mathbb R}^{2n}$.

Consider now the {\em symplectic group}
$\mbox{\rm Sp}(2n, \mathbb R )$, of $2n\times 2n$ real matrices $S$, satisfying
$SJS^T = J$ and $\text{det} S = 1$. Let
$S = \displaystyle{e^{\varepsilon X}}$ be an element of this group, close to
the identity, where $\varepsilon > 0$ and $X$ is a $2n\times 2n$ real matrix.
The fact that $S$ can be written this way is guaranteed by the exponential
mapping theorem for Lie groups. The defining condition $SJS^{T} = J$,
for an element of $\mbox{\rm Sp}(2n, \mathbb R )$, then implies,
 $$ ({\mathbb I}_{2n} + \varepsilon X)J({\mathbb I}_{2n} + \varepsilon X)^T
         + {\mathcal O}(\varepsilon^{3}) = J. $$
Simplifying and dividing by $\varepsilon$,
$$ XJ + JX^{T} + \varepsilon XJX^{T} + {\mathcal O}(\varepsilon^{2}) = 0.$$
Hence, letting $\varepsilon \rightarrow 0$, we find that
\be
   XJ + JX^{T} = 0 \quad \Rightarrow \quad X = JX^{T}J.
\label{symplgroup1}
\en
Thus, $JX$ is a symmetric matrix and $X$ a matrix of the type
(\ref{homogpolyn3}) with an associated second degree, homogeneous, real-valued
polynomial:
\be
   X = X(u), \qquad u(\bfeta) = \frac 12 \bfeta^{T}JX\bfeta .
\label{symplgroup2}
\en
On the other hand, the matrices $X$ in
$S = \displaystyle{e^{\varepsilon X}}$ constitute the Lie algebra
$\mbox{\rm sp}(2n, \mathbb R )$ of the Lie group
$\mbox{\rm Sp}(2n, \mathbb R )$, and thus we have established
an algebraic
isomorphism ${\mathfrak P}_{2}\simeq  \mbox{\rm sp}(2n, \mathbb R )$.
Moreover, the relations (\ref{homogpolyn5}) and (\ref{semidirprod2})
together then constitute the Lie algebra of the {\em metaplectic
group}~\footnote{Due to some existing terminological confusion in the
literature, this is a \underbar{different} metaplectic group from the one we
will encounter in~\S\ref{sec25} below.}, which is the semi-direct product
$\BMP(2n, \mathbb R ) =
G_{WH}(n)\rtimes\mbox{\rm Sp}(2n, \mathbb R )$. The Lie algebra,
$\bmp(2n, \mathbb R )$, of this group consists, therefore, of
all real-valued, first order and second order homogeneous polynomials
in the variables $q_i, p_{i} \;\; i=1,2, \ldots , n$. The group
$\BMP(2n, \mathbb R )$ has elements
$(\theta , \bfeta , S)$ and the multiplication rule is:
\be
(\theta , \bfeta , S)(\theta', \bfeta', S') = (\theta + \theta'
  + \xi (\bfeta , S\bfeta'), \bfeta + S\bfeta', SS'),
\label{metaplectgroup1}
\en
with the same multiplier $\xi$ as in (\ref{multiplier}).

The metaplectic group has a UIR on the same space $\HH$, extending
the representation of $U^{\hbar}$ of $G_{WH}(n)$ given in
(\ref{WeylHeisenrep}). We denote this representation again by
$U^{\hbar}$ and see that since $(\theta , \bfeta , S) =
(\theta , \bfeta , {\mathbb I}_{2n})(0 , \mathbf 0 , S)$,
\be
   U^{\hbar} (\theta , \bfeta , S) =
      U^{\hbar}(\theta , \bfeta)U^{\hbar}(S),
\label{metaplectrep1}
\en
where for $S = \displaystyle{e^{\varepsilon X(u)}}$, the unitary operator
$U^{\hbar}(S)$ can be shown \cite{bib:-simsudmu1} to be
\be
  U^{\hbar}(S) = \exp\left[-\frac {i\varepsilon}\hbar \widehat{X}(u)
     \right], \qquad \widehat{X}(u) = -\frac 12 \widehat{\bfeta}^{T}
        JX(u)\widehat{\bfeta}.
\label{metaplectrep2}
\en
Furthermore, using the unitarity of $U^{\hbar}(S)$, it is easily shown that
\be
  [\widehat{X}(u), \widehat{X}(v)] = i\hbar\; \widehat{X}(\{u,v\}),
\label{metaplectrep3}
\en
that is, the quantization of $\bfeta$ now extends to second degree,
homogeneous polynomials in the manner $u \rightarrow  \widehat{X}(u) :=
\widehat{u}$. The  self-adjoint operators $\widehat{\bfeta}$ and
$\widehat{X}(u)$ of the representation of the Lie algebra
$\bmp(2n, \mathbb R )$, on the Hilbert space $\HH$, satisfy the full
set of commutation relations,
\bea
 [\widehat{\bfeta}, \widehat{X}(u)] & = & i\hbar X(u)\widehat{\bfeta},
           \nonumber\\[0pt]
  [\widehat{X}(u), \widehat{X}(v)]  &  = & i\hbar\;\widehat{X}(\{u,v\}).
\label{metaplectrep4}
\ena

    In the light of the Groenewold-van Hove results, mentioned earlier,
this is the best one can do. In other words, it is
not possible to find an algebra larger than
$\bmp(2n, \mathbb R )$, which could also be irreducibly
represented on $L^{2}({\mathbb R}^{n}, d\bx )$. On the other hand,
van Hove also showed that if one relaxes the irreducibility condition, then
on $L^{2}({\mathbb R}^{2n}, d\bfeta )$,
it is possible to represent the full Poisson algebra
of~${\mathbb R}^{2n}$. This is the so-called
{\em prequantization} result, to which we shall return later.

Given the present scheme of canonical quantization, a number of questions
naturally arise.

\begin{itemize}

\item  Let $\bfrakQ$ be the position space manifold of the classical
system and $q$ any point in it. Geometrically, the phase space of the system
is the cotangent bundle $\bigam = T^{*}\bfrakQ$.
 If $\bfrakQ$ is linear, i.e., $\bfrakQ
\simeq {\mathbb R}^n$, then the replacement $q_i \rightarrow x_i,
\;\;
p_{j} \rightarrow -i\hbar\;{\displaystyle\frac {\partial}{\partial x_j}}$
works fine. But what if $\bfrakQ$ is not a linear space?

\item How do we quantize observables which involve higher powers
of $q_i, p_{j}$, such as for example $f(q,p) = (q_i)^{n}(p_{j})^{m}$,
when $m+n \geq 3$?

\item How should we  quantize more general phase spaces, which are symplectic
manifolds but not necessarily cotangent bundles?

\end{itemize}

In the rest of this  Section we review two procedures which have been proposed
to extend  canonical quantization to provide, among others, the answer to the
first of these questions.

\subsection{Segal and Borel quantization}
\label{sec-segborelquant}

A method for quantizing on an arbitrary configuration space manifold $\bfrakQ$
was proposed by Segal~\cite{bib:Segal}, as a generalization of canonical
quantization and very much within the same spirit. A group theoretical method
was suggested by Mackey~\cite{bib:Mack}, within the context of the theory of
{\it induced representations} of finite dimensional groups. A~much more general
method, combining the Segal and Mackey approaches, was later developed by
Doebner, Tolar, Pasemann, Mueller, Angermann and Nattermann \cite{bib:-doebtol,
bib:DoebNa,bib:-nattermann}. It~cannot be applied to an arbitrary symplectic
manifold, but only to cotangent bundles; the~reason is that it distinguishes
between the position variables $q\in\XX$ (the configuration space) and the
momentum variables $X\in T \XX$ in an essential way. Functions $f(q)$ of the
spatial variables are quantized by the multiplication operators
$(\widehat f \phi)(q) =f(q) \phi(q)$ on $L^2(\XX,\mu)$ with some measure~$\mu$,
while vector fields $X$ are quantized~by
$$ \widehat X \phi = -\tfrac{ih}{2\pi}
  (X\phi + \operatorname{div}_\mu X\cdot\phi)  $$
(the additional term $\operatorname{div}_\mu X$ ensures that $\widehat X$
be a formally self-adjoint operator on~$L^2(\XX,\mu)$). One~then has the
commutation relations
$$ [\widehat X,\widehat Y]=-\frac{ih}{2\pi} \widehat {[X,Y]},
\qquad [\widehat X,\widehat f]=-\frac{ih}{2\pi} \widehat {Xf},
\qquad [\widehat f,\widehat g]=0,  $$
which clearly generalize~(\ref{tag:CCR}).

A~method using infinite dimensional diffeomorphism groups, obtained from local
current algebras on the physical space, was suggested by
Goldin, et al. ~\cite{bib:gol1,bib:-golmensh2,bib:-golmensh1}.
The relation to diffeomorphism groups of the configuration space was also
noticed by Segal, who in fact in
the same paper~\cite{bib:Segal} lifted the theory to the cotangent bundle
$T^{*}\bfrakQ$ and thereby anticipated the theory of geometric quantization.
Segal also pointed out that the number of inequivalent such
quantizations was related to the first cohomology group of $\bfrakQ$.

\subsection{Segal quantization}\label{subsec-segquant}

Let us elaborate a bit on the technique suggested by Segal. The configuration
space $\bfrakQ$ of the system is, in general, an $n$-dimensional
$C^\infty$-manifold. Since in the case when $\bfrakQ = {\mathbb R}^n$,
canonical quantization represents the classical position observables $q_i$ as
the operators $\widehat{q}_i$ of multiplication by the corresponding position
variable, on the  Hilbert space $\HH = L^2 ({\mathbb R}^n , d\bx )$, Segal
generalized this idea and  defined an entire class of observables of
position  using the smooth functions $f: \bfrakQ \rightarrow {\mathbb R}$.
Similarly, since canonical quantization on $\bfrakQ = {\mathbb R}^n$
replaces the classical observables of momentum, $p_i$, by derivatives with
respect to these variables, in Segal's scheme an entire family of quantized
momentum observables is obtained by using the vector fields $X$ of the manifold
$\bfrakQ$.

With this idea in mind, starting with a general configuration space manifold,
one first has to choose a Hilbert space. If the manifold is orientable, its
volume form determines a measure, $\nu$, which is locally equivalent to the
Lebesgue measure:
\be
   d\nu (\bx) = \rho (\bx)\; dx_1\;dx_2 \ldots dx_n ,
 \qquad \bx \in \bfrakQ \; . \label{volmeas}
\en
where $\rho$ is a positive, non-vanishing function. The quantum mechanical
Hilbert space is then taken to be $\HH = L^2 (\bfrakQ , d\nu )$.
In local coordinates we shall write the vector fields of $\bfrakQ$ as
$$ X = \sum_{i=1}^{n}a_i(\bx)\frac {\partial}{\partial x_i},   $$
for $C^{\infty}$-functions $a_i: \bfrakQ \rightarrow {\mathbb R}$.
The generalized quantum observables of position
are then defined by the mappings, $f \mapsto \widehat{q}(f)$,
such that on some  suitable dense set of vectors $\psi \in \HH$,
\be
     (\widehat{q}(f)\psi )(\bx) = f(\bx)\psi (\bx).
\label{quposops}
\en
Ignoring technicalities involving domains of these operators, they are easily
seen to be self-adjoint ($f$ is real). In order to obtain a set of quantized
momentum observables, we first notice that quite generally the natural action
of the vector field $X$, $\phi \mapsto X(\phi)$, on a suitably chosen set of
smooth functions $\phi \in \HH$, defines an operator on the Hilbert space. This
operator may not be bounded and may not be self adjoint. However, denoting by
$X^*$ the adjoint of the operator $X$, the combination,
\be
   \widehat{p}(X) = \frac \hbar{2i}[X - X^*],
\label{qumomops}
\en
does define a self-adjoint operator (if again we ignore domain related
technicalities), and we take this to be the generalized momentum operator
corresponding to the vector field $X$. An easy computation then leads to the
explicit expression,
\be
      \widehat{p}(X) = -i\hbar (X + K_{X}),
\label{segalquant1}
\en
where $K_X$ is the operator of multiplication by the function
\be
 k_X (\bx) = \frac 12\;\mbox{\rm div}_{\nu}\;(X)(\bx)
= \frac 12\;\left[X(\log \rho )(\bx) +
      \sum_{i=i}^N \frac {\partial a_i (\bx)}{\partial x_i}\right]\;.
  \label{segalquant2}
\en
In terms of the Lie bracket $[X,Y] = X\circ Y - Y\circ X$ of
the vector fields, one then obtains for the quantized operators the following
commutation relations, which clearly generalize the canonical commutation
relations:
\be \begin{aligned}[0pt]
[\widehat{p}(X), \widehat{p}(Y)]  &=  -i\hbar\; \widehat{p}([X,Y]) \\
[\widehat{q}(f), \widehat{p}(X)]  &=  i\hbar\; \widehat{q}(X(f)) \\
    [\widehat{q}(f), \widehat{q}(g)] &= 0.
\end{aligned}  \label{segalquant3} \en

It~ought to be pointed out here that the above commutation relations constitute
an  infinite dimensional Lie algebra, $\bfrakX_c (\bfrakQ )\oplus\CinfRQ$. This
is the Lie algebra of the
(infinite-dimensional) group, $\bfrakX_c (\bfrakQ )\rtimes \mbox{\rm Diff}
(\bfrakQ )$, the semi-direct product of the (additive) linear group of all
complete vector fields of $\bfrakQ$ with the group (under composition) of
diffeomorphisms of $\bfrakQ$ (generated by the elements of
$\bfrakX_c(\bfrakQ))$. The product of two elements $(f_1 , \phi_1 )$ and $(f_2
, \phi_2 )$ of this group is defined as:
$$ (f_1 , \phi_1 )(f_2 , \phi_2 ) = (f_1 + \phi_1 (f_2 ) ,\;
 \phi_1 \circ \phi_2 ) .$$
The Lie algebra generated by the first set of commutation relations (for the
momentum operators) in (\ref{segalquant3}) is called a {\em current algebra}.
When modelled on the physical space, rather than the configuration space, the
relations (\ref{segalquant3}) are precisely the {\em non-relativistic current algebra}
introduced by Dashen and Sharpe \cite{bib:DasShar}. The corresponding semi-direct product group
was obtained in this context by Goldin \cite{bib:gol1}.

   Next note that if $\theta$ is a fixed
one-form of $\bfrakQ$, then replacing $\widehat{p}(X)$ by
\be
 \widehat{p}(X)' = \widehat{p}(X) + X\rfloor\theta \;,
\label{equivrep}
\en
in (\ref{segalquant1}) does not change the commutation relations in
(\ref{segalquant3}). Indeed, by choosing such one-forms appropriately, one can
generate inequivalent families of representations of the Lie algebra
$\bfrakX_c (\bfrakQ )\oplus \CinfRQ$. In particular, if $\theta$ is
logarithmically exact, i.e., if $\theta = \frac {dF}F$, for some smooth
function $F$, then the representations  generated by the two sets of operators,
$\{\widehat{p}(X), \; \widehat{q}(f)\}$ and
$\{\widehat{p}(X)', \; \widehat{q}(f)\}$ are unitarily equivalent. In~other
words, there exists a unitary operator $V$ on $\HH$ which commutes with all
the $\widehat{q}(f), \;\; f \in \CinfRQ$, and such that
$$ V \widehat{p}(X)V^* = \widehat{p}(X)' ,
 \qquad X \in \bfrakX_c (\bfrakQ )\; . $$

\subsubsection*{Some simple examples}

The obvious example illustrating the above technique is provided by
taking  $\bfrakQ = {\mathbb R}^3  , \;\; \HH = L^2 ({\mathbb R}^3 , \; d\bx )$.
Consider the functions and vector fields,
\be
  f_i (\bx ) = x_i\; , \qquad X_i = \frac \partial{\partial x_i} \; , \qquad
  J_i = \varepsilon_{ijk}\; x_j\; \frac \partial{\partial x_k}\;,  \qquad
  i,j,k = 1,2,3, \;
\label{funcandvects}
\en
where $\varepsilon_{ijk}$ is the well-known completely antisymmetric tensor
(in the indices $i,j,k$) and summation being implied over repeated indices.
Quantizing these according to the above procedure we get the usual position,
momentum and angular momentum operators,
\be
  \widehat{q}_i := \widehat{q}(f_i ) = x_i , \quad
  \widehat{p}_i = \widehat{p}(X_i) =
               -i\hbar\; \frac {\partial}{\partial x_i}\;,
   \quad \widehat{J}_i = \widehat{p}(J_i) =
- i\hbar\; \varepsilon_{ijk}\; x_j\; \frac \partial{\partial x_k}\; .
\label{segalquant7}
\en
Computing the commutation relations between these operators, following
(\ref{segalquant3}), we get the well-known results,
\be \begin{aligned}[0pt]
    [\widehat{q}_i \; , \widehat{q}_j ] =  [\widehat{p}_i \; ,
     \widehat{p}_j ] & =  0 \; , \\
[\widehat{q}_i \; , \widehat{p}_j ]  =  i\hbar\; \delta_{ij} \; I\; , \qquad
[\widehat{q}_i \; , \widehat{J}_j ] & =  i\hbar\; \varepsilon_{ijk}\;
\widehat{q}_k \; , \\
[\widehat{p}_i , \widehat{J}_j ] = i\hbar \varepsilon_{ijk}\widehat{p}_k \;
, \qquad [\widehat{J}_i \; , \widehat{J}_j ]& =  i\hbar\; \varepsilon_{ijk}\;
\widehat{J}_k     \; .
\end{aligned} \label{segalquant8} \en
Note that these are just the commutation relations between the infinitesimal
generators of the orthochronous Galilei group ${\mathbf G}_{\rm orth}$
in a space of three dimensions and hence
they define its Lie algebra,
which now emerges as a subalgebra of the Lie algebra
$\bfrakX_c (\bfrakQ )\oplus \CinfRQ$.

Now let $\bA (\bx ) = (A_1 (\bx ) , A_2 (\bx ),  A_3 (\bx ))$ be a magnetic
vector potential, $\bB = \nabla \times \bA$ the corresponding magnetic field.
Consider the one form
$$\theta = -\frac ec \sum_{i=1}^3 A_i \;dx_i$$
($e = $ charge of the electron and $c = $ velocity of light). The set of
quantized operators
\be
  \widehat{q}(f) \quad \mbox{\rm and} \quad  \widehat{p}(X)' = -i\hbar X +
  \frac 12\sum_{i=1}^3 \left[\widehat{p}_i - \frac {2e}c A_i\right]a_i , \quad
 \text{where } X(\bx ) = \sum_{i=1}^3 a_i (\bx )\frac \partial{\partial x_i}\;,
\label{quantizedops}
\en
realize a quantization of a nonrelativistic charged particle in a magnetic
field. (For a ``current algebraic'' description, see Menikoff and Sharp
\cite{bib:MenSh}.) In~particular, if $d\theta = 0$  (i.e., $\nabla \times \bA = \bB = 0$),
then $\theta$ is closed, hence exact, and there is no magnetic field. Hence,
from a physical point of view, the quantizations corresponding to different
such $\theta$ must all be unitarily equivalent and indeed, as noted above,
this is also true mathematically. This point is illustrated by taking vector
potential  $\bA (\bx ) = \mu (x_2 , x_1 , 0)$ where $\mu$ is a constant.
Then $\nabla \times \bA = 0$ and the one-form
$\theta = -\frac {e\mu}c\; [x_2\;dx_1 + x_1 \; dx_2]$
is logarithmically exact:
$$ \theta = \frac {dF}F, \quad \mbox{\rm with} \quad F =
     \exp[-\frac {e\mu}c\; x_1 x_2 ] \; . $$

     On the other hand, consider the case where $\bA (\bx ) = \frac B2\;
(-x_2 , x_1 , 0), \;\; B > 0$.
This is the case of a constant magnetic field $\bB = (0, 0, B)$ of strength $B$
along the third axis. The corresponding one-form
$\theta = \frac {eB}{2c}\; [x_2\;dx_1  -  x_1 \; dx_2]$ is not closed and for
each different value of $B$ we get an inequivalent quantization.

\bigskip

As the next example, let $\bfrakQ = {\mathbb R}^3\backslash\{\mathbb R\}$, the
three dimensional Euclidean space with the third axis removed. We take the
measure $d\nu (\bx ) = d\bx$ and the Hilbert space $\HH = L^2 (\bfrakQ, d\nu)$.
Consider the vector potential,
$$ \bA (\bx ) =  -\frac \mu {r^2}(-x_2 , \; x_1 , \; 0 )\quad \mu > 0 ,
   \quad  r^2 =  (x_1)^2 +  (x_2)^2 . $$
Then $\nabla \times \bB = 0$ and the one-form
\be
   \theta (\bx ) = \frac{\mu e}{c r^2}\;[x_2\;dx_1 -  x_1\;dx_2 ]
\label{closedoneform}
\en
is closed. However, $\theta$ is not exact, since we may write $\theta = dF$,
with
\be
   F = -\frac{\mu e}c\; \tan^{-1} \left(\frac {x_2}{x_1}\right),
\label{aharonovpot}
\en
which is a multivalued function on $\bfrakQ$. Since $\bB = 0$, physically the
classical systems with $\bA = 0$ and $\bA$ given as above should be equivalent.
However, the quantizations for the two cases (which can be easily computed
using (\ref{equivrep})) are inequivalent. This is an example of the
Aharonov-Bohm effect (see \cite{bib:-ahabohm}).

Finally, for the same configuration space
${\mathbb R}^3\backslash\{\mathbb R\}$,
consider the case in which the magnetic field itself is given by
$$ \bB (\bx )  = \frac {2I}{c r^2}\; ( -x_2 , x_1 , 0 ) , \qquad   r^2 =
               (x_1)^2 +  (x_2)^2 . $$
This is the magnetic field generated by an infinite current bearing wire
(of current strength  $I$)
placed along the $x_3$-axis.  The vector potential, given locally by
$$
  \bA (\bx ) = \frac {2I}c (0, 0, \phi ) , \qquad
  -\frac {\pi}2 <  \phi = \tan^{-1} \left(\frac {x_2}{x_1}\right)
       < \frac {\pi}2 \; , $$
does not give rise to a closed form  and for each value of $I$ one gets a
different quantization.

\bigskip

As mentioned earlier, Segal actually suggested going over to the group of
diffeomorphisms $\mbox{\rm Diff} (\bfrakQ )$ and its unitary representations,
to attend to domain questions associated to $\widehat{q}(f), \widehat{p}(X)$,
and then suggested a classification scheme for possible unitarily inequivalent
quantizations in these terms. Note also, that the Segal quantization method is
based on configuration space, rather than on phase space. As such, the primary
preoccupation here is to generalize the method of canonical quantization.
On~the other hand, as we said before, Segal also extended the theory to phase
space and in that sense, Segal's method leads to similar results as other
methods that we shall study, on the representations of the Poisson algebra
on Hilbert space.

  At this point we should also mention that Goldin, Sharp and their collaborators
proposed to describe quantum theory by means of unitary representations of groups of
diffeomorphisms of the physical space \cite{bib:gol1,bib:GGPS,bib:GolSh1}.
Deriving the current algebra from second quantized canonical fields, their programme
has succeeded in predicting unusual possibilities, including the statistics
of {\em anyons} in two space dimensions \cite{bib:GoMeSh,bib:-golmensh1,bib:GoSh2,bib:LeiMyr}.
Diffeomorphisms of the physical space act naturally on the configuration space
$\bfrakQ$ and thus form a subgroup. In fact, the unitary representations of this group
are sufficient to characterize the quantum theory, so that the results of Goldin,
{\em et al.\/}, carry over to the quantization framework described in the next section.
In particular, the unitarily inequivalent representations describing particle
statistics were first obtained by Goldin, Menikoff and Sharp
\cite{bib:GoMeSh2,bib:GoMeSh,bib:-golmensh1}. For an extended review of these ideas,
see \cite{bib:gol2}.

\subsection{Borel quantization}\label{sec-synborquant}

We pass on to the related, and certainly more assiduously studied, method of
{\em Borel quantization}. This method focuses on both the geometric and measure
theoretic properties of the configuration space manifold $\bfrakQ$ as well as
attempting to incorporate internal symmetries by lifting $\bfrakQ$ to a complex
Hermitian vector bundle with connection and curvature, compatible with the
Hermitian structure.

Consider a one-parameter family of diffeomorphisms $s \mapsto \phi_{s}$ of
${\mathbb R}^n$, which are sufficiently well behaved in the parameter
$s \in \mathbb R$, in an appropriate sense. Then,
\be
   \frac d{ds} f\circ \phi_{s}\vert_{s=0} = X(f),
\label{diffeomorph8}
\en
where $f$ is an arbitrary smooth function, defines a vector field $X$.
Its quantized form $\widehat{p}(X)$, according to Segal's procedure will be a
general momentum observable acting on $\psi \in  L^{2}({\mathbb R}^n, d\bx )$
in the manner
\be
    (\widehat{p}(X)\psi )(\bx ) = -i\hbar (X\psi )(\bx ) -
      \frac {i\hbar}{2}\frac {\partial a_i}{\partial x_i}(\bx ) \psi (\bx ),
      \quad \mbox{\rm where} \quad X(\bx ) =
      \sum_{i=1}^{n}a_i(\bx ) \frac {\partial}{\partial x_i},
\label{genmomobs1}
\en
and together, the set of all such momentum observables
then form an algebra under the bracket
operation (see (\ref{segalquant3})):
\be
   [\widehat{p}(X), \; \widehat{p}(Y)] = -i\hbar \widehat{p}([X, \;Y]).
\label{genmomobs2}
\en
We~write $\phi_{s} = \phi_{s}^{X}$, to indicate the generator, and define the
transformed sets
\be
   \phi_{s}^{X}(\Delta ) = \{\phi_{s}^{X}(\bx ) \;\vert\; \bx \in \Delta\},
\label{transfset1}
\en
for each Borel set $\Delta$ in ${\mathbb R}^n$.

Next, denote the $\sigma$-algebra of the Borel sets of
$\bfrakQ = {\mathbb R}^n$ by ${\mathcal B}({\mathbb R}^n)$. Corresponding
to each $\Delta \in {\mathcal B}({\mathbb R}^n)$, define an operator
$P(\Delta )$ on $\HH$:
\be
  (P(\Delta )\bpsi )(\bx ) = \chi_{\Delta} (\bx )\bpsi (\bx ), \qquad
  \chi_{\Delta} (\bx ) = \left\{ \begin{array}{ll} 1, & \mbox{\rm if} \;\;
  \bx \in \Delta,\\
  0, & \mbox{\rm otherwise.}    \end{array}\right.
\label{PVmeas1}
\en
This is a projection operator, $P(\Delta ) =  P(\Delta )^{*} = P(\Delta )^{2}$,
and has the following measure theoretic properties:
\bea
   P(\emptyset ) = 0 , & \quad & P({\mathbb R}^n) = I \nonumber \\
   P(\cup_{i\in J}\Delta_{i}) = \sum_{i\in J}P(\Delta_{i}) & \qquad &
   \text{if } \Delta_{i}\cap\Delta_{j} = \emptyset, \;\; i\neq j,
\label{PVmeas2}
\ena
where $J$ is a discrete index set and the convergence of the sum is meant in
the weak sense. Such a set of projection operators $P(\Delta ), \;\; \Delta \in
{\mathcal B}({\mathbb R}^n)$, is called a (normalized) {\em projection valued
measure} (or~{\em PV-measure} for short) on~$\RR^n$. Note that, for any
$\bpsi \in \HH$,
\bea
  \mu_{\bpsi}(\Delta) & = & \langle\bpsi\vert P(\Delta )\bpsi\rangle ,
       \nonumber\\
                      & = & \int_{\Delta}\Vert\bpsi (\bx )\Vert^{2}\;d\bx ,
       \qquad  \Delta \in {\mathcal B}({\mathbb R}^n),
\label{realmeas1}
\ena
defines a real measure, absolutely continuous with respect to the Lebesgue
measure.

It is then easily checked that for each $s \in \mathbb R$,
\be
   V(\phi_{s}^{X}) = \exp [-i\hbar s\widehat{p}(X)],
\label{genimprimitiv1}
\en
defines a unitary operator on $\HH$, such that $\{V,P\}$ is a {\em system of
imprimitivity} in the sense:
\be
     V(\phi_{s}^{X})P(\Delta )V(\phi_{-s}^{X})  = P( \phi_{s}^{X}(\Delta )).
\label{genimprimitiv2}
\en
Now considering {\em all} such one-parameter diffeomorphism groups and their
associated systems of imprimitivity, we find that the collective system is
certainly irreducibly realized on $\HH= L^{2}({\mathbb R}^n, d\bx )$.

Suppose now that the system which we wish to quantize has some {\em internal
degrees of freedom}, such as the spin of a particle. Thus there is some group
$G$ of internal symmetries, and for any UIR of $G$ on some (auxiliary) Hilbert
space~$\mathfrak K$, we~want to work on the Hilbert space $\HH=\mathfrak K
\otimes L^2(\RR^n,d\bx)$ instead of just $L^2(\RR^n,d\bx)$; and we would like
(\ref{genimprimitiv2}) to be irreducibly realized on this~$\HH$. For~instance,
for the free particle in~$\RR^3$, to~accommodate for its spin we need to
replace\footnote{From a purely mathematical point of view, this amounts to
replacing the original configuration space $\RR^3$ by its Cartesian product
with a discrete set consisting of~$2j+1$ points.} $L^2(\RR^3,d\bx)$ by
$\HH=\CC^{2j+1}\otimes L^2(\RR^3,d\bx)$, with $\CC^{2j+1}$ carrying the $j$-th
spinor representation of $SU(2)$, $j=0,\frac12,1,\frac32,\dots$.

The aim of Borel quantization is to construct such irreducible systems on
arbitrary configuration space manifolds $\bfrakQ$. It is clear that the problem
is related to that of finding irreducible representations of the diffeomorphism
group, $\mbox{\rm Diff}(\bfrakQ )$, which admit systems of imprimitivity based
on the Borel sets of $\bfrakQ$.

Let $\bfrakQ$ be a configuration space manifold, of dimension $n$, $\mu$ a
smooth measure on $\bfrakQ$ (i.e., locally equivalent to the Lebesgue measure
on ${\mathbb R}^n$) and let
$\widetilde\HH = {\mathbb C}^k \otimes L^2 (\bfrakQ , d\mu )$, where $k \geq 1$
is an integer.

Let $\widetilde{P}(E)$ be the projection valued measure on $\htil$:
\be
  (\widetilde{P}(E)\widetilde{\psi})(x) = \chi_E (x)\widetilde{\psi}(x), \qquad
  \widetilde\psi \in \htil , \qquad E \in {\mathcal B}(\bfrakQ ),
\label{manifimpri1}
\en
$\chi_E$ being the characteristic function of the set $E$ and ${\mathcal
B}(\bfrakQ )$ denoting the set of all Borel sets of $\bfrakQ$. Now let $\HH$
be another Hilbert space and $P$ a PV-measure on it (also defined over
${\mathcal B}(\bfrakQ )$).

\bedefin
The pair $\{\HH , P\}$ is called a $k$-homogeneous localized quantum system if
and only if it is unitarily equivalent to $\{\htil , \widetilde{P}\}$, i.e.,
iff~there exists a unitary map $W: \HH \longrightarrow \htil$ such that
\be
  WP(E)W^{-1} = \widetilde{P}(E), \qquad E \in {\mathcal B}(\bfrakQ ).
\label{khomloc}
\en
\end{defi}

Let $f \in \CinfRQ =$ (space of infinitely differentiable,
real-valued functions on $\bfrakQ$).

\bedefin
Let $\{\HH,P\}$ be a $k$-homogeneous localized quantum system. The self-adjoint
operator,
\be
   \widehat q (f) = \int_{\bfrakQ}f(x)\;dP_x ,
\label{genposop}
\en
defined on the domain,
$$
  {\mathcal D}(\widehat q (f)) = \{ \psi \in \HH \; \vert\;
    \int_{\bfrakQ}\vert f(x)\vert^2
    \;d\langle \psi\vert P_x \psi\rangle < \infty \} ,
$$
is called a generalized position operator.
\end{defi}

Note that under the isometry (\ref{khomloc}), $\widehat q (f)$ becomes the
operator of multiplication by $f$ on $\htil$. The following properties of these
operators are easily verified:

\begin{enumerate}

\item $\widehat q (f)$ is a bounded operator if and only if $f$ is a bounded
function.

\item $\widehat q (f) = 0$ if and only if $f = 0$.

\item $\widehat q (\alpha f) = \alpha \widehat q (f)$, for
$\alpha \in \mathbb R$.

\item $\widehat q (f + g )   \supseteq \widehat q (f)  +\widehat q (g)$ and
      ${\mathcal D}(\widehat q (f) + \widehat q (g))
 = {\mathcal D}(\widehat q (f)) \bigcap {\mathcal D}(\widehat q (g))$.

\item $\widehat q (f\cdot g)) \supseteq \widehat q (f)\;\widehat q (g)$ and
      ${\mathcal D}(\widehat q (f)\;\widehat q (g)) =
      {\mathcal D}(\widehat q (f\cdot g ))\bigcap{\mathcal D}(\widehat q (f))$.

\end{enumerate}

We had mentioned earlier the notion of a {\em shift} on the manifold $\bfrakQ$.
This is a one parameter group of diffeomorphisms:
$\phi_s : \bfrakQ \longrightarrow \bfrakQ, \;\;
\phi_{s_2}\circ\phi_{s_1} = \phi_{s_1 + s_2}, \;\; s, s_1 , s_2 \in \mathbb R$,
$\phi_0$ being the identity map. Each such shift defines a {\em complete vector
field\/}, $X$ via,
\be
   X(f) : = \frac d{ds} f\circ \phi_s \vert_{s=0} ,
\label{vectfield1}
\en
$f$ being an arbitrary smooth function on the manifold and conversely,
every such vector field $X$ gives rise to a shift $\phi^X_s$, called the
{\em flow} of the vector field:
\be
   \pi(\phi^X_{-s} ) = e^{sX(x)},
\label{vectfield2}
\en
where $\pi(\phi^X_{-s} )$ is a linear operator on the space of smooth functions
$f$ on the manifold:
\be
  (\pi(\phi^X_{-s} )f)(x) = f(\phi^X_{s}(x)), \qquad x \in \bfrakQ\; .
\label{morevectfield}
\en
There is a natural action of the shifts on Borel sets $E \subset \bfrakQ$,
\be
   E \longmapsto \phi^X_s (E) = \{ \phi^X_s (x) \;\vert\; x \in E\} .
\label{shift}
\en
Since $\phi^X_s$ is smooth, the resulting set $\phi^X_s (E)$ is also a Borel
set. We want to represent the shifts $\phi^X_s$ on $\htil$ as one-parameter
unitary groups on Hilbert spaces $\HH$. Let ${\mathcal U}(\HH )$ denote the set
of all unitary operators on $\HH$ and, as~before, ${\bfrakX}_c (\bfrakQ )$ the
set of all complete vector fields on the manifold~$\bfrakQ$.

\bedefin
  Let $\{\HH ,P\}$ be a quantum system localized on $\bfrakQ$. A map
\be
   V: \phi^X_s \longmapsto  V( \phi^X_s ) \in {\mathcal U}(\HH ),
\label{shift2}
\en
is called a shift of the localized quantum system if, for all
$X \in {\bfrakX}_c (\bfrakQ )$, the map $s \longmapsto V( \phi^X_s )$ gives a
strongly continuous representation of the additive group of $\mathbb R$ and
$\{V( \phi^X_s ) , P \}$ is a system of imprimitivity with respect to the group
of real numbers $\mathbb R$ and the Borel $\mathbb R$-space $\bfrakQ$ with
group action $\phi^X_s$, i.e.,
\be
   V( \phi^X_s )P(E)V( \phi^X_{-s} ) = P(\phi^X_s (E)).
\label{shift3}
\en
The triple $\{ \HH , P , V \}$ is called a {\sl localized quantum system with
shifts.}  \end{defi}

Two localized quantum systems with shifts, $\{ \HH_j , P_j , V_j \}, \;\;
j=1,2$, are said to be unitarily equivalent if there exists a unitary map $W:
\HH_1 \longrightarrow \HH_2$, such that $W P_1 (E)W^{-1} = P_2 (E), \;\; E \in
{\mathcal B}(\bfrakQ )$ and $WV_1(\phi^X_s )W^{-1} = V_2(\phi^X_s ), \;\; x \in
{\bfrakX}_c (\bfrakQ ), \;\; s \in \mathbb R$. The map $\widehat{p} :
{\bfrakX}_c (\bfrakQ ) \longrightarrow  {\mathcal S}(\HH )$
(the~set of all self-adjoint operators on~$\HH$), where $\widehat{p}(X)$ is
defined via Stone's theorem as the infinitesimal generator~of
\be
     V( \phi^X_s ) = \exp [\frac i\hbar s\widehat{p}(X) ],
\label{kinmom}
\en
is called the {\em kinematical momentum} of $\{ \HH , P, V\}$.

The imprimitivity relation (\ref{shift3}) has the following important
consequences.

\belem
Let $\{ \HH , P , V \}$ be a $k$-homogeneous localized quantum system with
shifts. Then
\be
 V( \phi^X_s )\widehat{q}(f)V( \phi^X_{-s} ) = \widehat{q}(f\circ\phi^X_s).
\label{shift4}
\en
A~$k$-homogeneous quantum system with shifts $\{ \HH , P , V \}$ is unitarily
equivalent to $\{ \htil , \widetilde{P} , \widetilde{V} \}$, with $\htil$ and
$\widetilde{P}$ as in (\ref{manifimpri1}).
\enlem

The representation $\widetilde{V}$ acquires a very specific form. To understand
it we need the concept of a cocycle. Let $G$ be a locally compact group, $H$ a
standard Borel group, $X$ a Borel $G$-space with group action $x\longmapsto gx$
and $[\nu]$ a $G$-invariant measure class on $X$. (This means that if $\nu$ is
any measure in the class, then so also is $\nu_g$, where $\nu_g(E)=\nu(gE)$,
for all $E \in {\mathcal B}(\bfrakQ )$.)

A~Borel measurable map $\xi :G \times X \longrightarrow H$ is called a {\em
cocycle} of $G$, relative to the measure class $[\nu ]$ on $X$, with values
in~$H$,~if
\be \begin{aligned}[0pt]
  \xi (e, x ) & =  1, \\
  \xi (g_1 g_2 , x ) & =  \xi (g_1 , g_2 x )\;\xi(g_2 , x ),
\end{aligned}  \label{cocycle1}  \en
for $[\nu ]-$almost all $x \in X$ and almost all (with respect to the Haar
measure) $g_1 , g_2 \in G$ ($e$ is the identity element of $G$). Two cocycles
$\xi_1$ and $\xi_2$ are said to be {\em cohomologous} or {\em equivalent} if
there exists a Borel function $\zeta : X \longrightarrow H$, such that,
$$  \xi_2 (g, x) = \zeta (gx)\;\xi_1 (g, x)\;\zeta (x)^{-1} $$
for almost all $g \in G$ and $x \in X$. The  equivalence classes $[\xi ]$ are
called {\em cohomology classes of cocycles}. The following classification
theorem for localized quantum systems then holds.

\betheo
Any localized $k$-homogeneous quantum system $\{\HH , P, V\}$ on $\bfrakQ$,
with shifts, is unitarily equivalent to a canonical representation
$\{ \htil , \widetilde{P} , \widetilde{V}\}$, with $\htil =
{\mathbb C}\otimes L^2 (\bfrakQ , d\mu )$, for some smooth measure $\mu$ on
$\bfrakQ$,
$$ (\widetilde{P}(E)\widetilde\psi )(x) = \chi_E (x )\widetilde\psi (x), $$
for all $\widetilde\phi \in \htil$ and all $E \in {\mathcal B}(\bfrakQ )$, and
\be
  (V(\phi^X_s )\widetilde\psi )(x)= \xi^X (s, \phi^X_{-s}(x))\;
  \sqrt{\lambda (\phi^X_s , \phi^X_{-s}(x))}\; \widetilde\psi (\phi^X_{-s}(x)),
\label{grouprep1}
\en
for all $\widetilde\psi \in \htil$ and all $X \in \bfrakX_c (\bfrakQ )$, where
$\xi^X$ is a cocycle of the Abelian group $\mathbb R$ (relative to the class of
smooth measures on $\bfrakQ$), having values in ${\mathcal U}(k)$ (the group of
$k\times k$ unitary matrices) and $\lambda$ is the unique smooth Radon-Nikodym
derivative,
$$ \lambda (\phi^X_s , x ) = \frac {d\mu_{\phi^X_s}}{d\mu} (x) . $$
Moreover, equivalence classes of $k$-homogeneous localized quantum systems are
in one-to-one correspondence with equivalence classes of cocycle sets
$[\{\xi^X\}_{X \in {\mathfrak X}_c (\mathfrak Q )}]$, where
$$ \{\xi^X_1\}_{X \in {\mathfrak X}_c (\mathfrak Q )} \sim
    \{\xi^X_2\}_{X \in {\mathfrak X}_c (\mathfrak Q )} $$
if there exists a Borel function $\zeta:\bfrakQ\longrightarrow{\mathcal U}(k)$,
such that, for all $X \in {\mathfrak X}_c (\mathfrak Q ),\;\; s \in \mathbb R$
and  $x \in \bfrakQ$,
$$ \xi^X_2 (s,x ) = \zeta (\phi^X_s (x))\;\xi^X_1 (s,x)\;\zeta(x)^{-1}. $$
\entheo

Differentiating (\ref{grouprep1}) with respect to $s$, using (\ref{kinmom}),
and then setting $s=0$, we obtain,
\be
 \widehat{p} (X)\widetilde\psi = -i\hbar\;\bfrakL_X \widetilde\psi -
      \frac {i\hbar}2 \;\mbox{\rm div}_\nu (X) \widetilde\psi + \omega
(X)\widetilde\psi,
\label{borelquant1}
\en
where $\bfrakL_X \widetilde\psi$ is the Lie derivative of $\widetilde\psi$
along $X$ and,
\be \begin{aligned}[0pt]
 \frac 12 \mbox{\rm div}_\nu (X)(x) &= \frac d{ds}\;
         \sqrt{\lambda (\phi^X_s, \phi^X_{-s} (x))} \vert_{s=0}   \\
\alpha (X)(x) &= -i\hbar\; \frac d{ds}\;\xi^X (s, \phi^X_{-s} (x))\vert_{s=0}.
\end{aligned}  \label{borelquant2}  \en
The first two terms in (\ref{borelquant1}) are linear in $X$. It is now
possible to show that the following commutation relations hold:
\be \begin{aligned}[0pt]
  [\widehat{q}(f), \widehat{q}(g)] &= 0,  \\
\lbrack\widehat{p}(X),\widehat{q}(f)\rbrack  &=
  -i\hbar\;\widehat{q}(\bfrakL_X f),     \\
  \lbrack\widehat{p}(X),\widehat{p}(Y)\rbrack &= -i\hbar\;\widehat{p}([X,Y])
      -i\hbar\;\Omega (X,Y),
\end{aligned} \label{borelquant3} \en
for all $f, g \in \CinfRQ,\;\; X, Y \in \bfrakX_c (\bfrakQ )$, and where,
\be
  \Omega (X,Y) = -i\hbar\; [\alpha (X) , \alpha (Y)] + \bfrakL_X \alpha (Y) -
     \bfrakL_Y \alpha (X) -\alpha ([X,Y]) .
\label{borelquant4}
\en
The two-form $\Omega$ and the one-form $\alpha$ on $\bfrakQ$
are related in the same way as the curvature two-form
$\displaystyle{\frac 1\hbar}\;\Omega$ of a ${\mathbb C}^1$-bundle and its
connection one-form $\displaystyle{\frac 1\hbar}\;\alpha (X)$. Indeed, one can
show that if $D$ is the covariant derivative defined by the connection, then
$D\Omega = 0$, which is the {\em Bianchi identity\/}.

\bedefin
Let $\{\HH , P, V\}$ be a $k$-homogeneous localized quantum system with shifts
on $\bfrakQ$ and $\Omega$ a differential two-form on $\bfrakQ$ with values in
the set of all $k\times k$ Hermitian matrices. The kinematical momentum
$\widehat{p}$ is called $\Omega$-compatible if in a canonical representation
$\{\htil , \widetilde{P}, \widetilde{V}\}$, the associated kinematical momenta
$\widetilde{p}$ satisfy
\be
   [\widetilde{p}(X), \widetilde{p}(Y)]\widetilde{\psi}
   = -i\hbar \;(\widetilde{p}([X,Y])
    \widetilde{\psi} + \Omega (X,Y)\widetilde{\psi}).
\label{omegacomp}
\en
In this case, the quadruple $\{\HH , \widehat{q}, \widehat{p}, \Omega \}$ is
called an $\Omega$-compatible $k$-Borel kinematics.
\end{defi}

In order to arrive at a classification theory of localized quantum systems, we
first impose some additional smoothness conditions. An $\Omega$-compatible
$k$-quantum Borel kinematics $\{ \HH , \widehat{q}, \widehat{p}, \Omega\}$ is
said to be {\em differentiable} if it is equivalent to $\{\htil, \widetilde{q},
\widetilde{p}, \widetilde{\Omega}\}$, where

\begin{enumerate}

\item $\htil = L^2 (\mathbb E , \langle\cdot\vert\cdot\rangle , d\nu )$ for a
      ${\mathbb C}^k$-bundle $\mathbb E$ over $\bfrakQ$, with Hermitian metric
      $\langle\cdot\vert\cdot\rangle$ and a smooth measure $\nu$ on $\bfrakQ$.

\item $\widetilde{\Omega}$ is a two-form with (self-adjoint) values in the
      endomorphism bundle $L_{\mathbb E} = {\mathbb E}\otimes {\mathbb E}^*$.

\item $(\widetilde{q}(f)\sigma )(x) = f(x)\sigma (x)$, for all $f \in
      \CinfRQ$ and smooth sections $\sigma \in \Gamma_0$
      (= smooth sections of compact support).

\item $\widetilde{p} (X)\Gamma_0 \subset \Gamma_0$, for all $X \in \bfrakX_c
      (\bfrakQ )$.

\end{enumerate}

We then have the following canonical representation of a differentiable quantum
Borel kinematics:

\betheo
Let $\{ \HH , \widehat{q}, \widehat{p}, \Omega\}$ be a localized differentiable
quantum Borel kinematics on $\bfrakQ$ in canonical representation. Then there
is a Hermitian connection $\nabla$ with curvature ${\displaystyle \frac
1\hbar}\;\Omega$ on $\mathbb E$ and a covariantly constant self-adjoint section
$\Phi$ of $L_{\mathbb E} = {\mathbb E}\otimes {\mathbb E}^*$, the bundle of
endomorphisms of $\mathbb E$, such that for all $X \in \bfrakX_c (\bfrakQ )$
and all $\sigma \in \Gamma_0$,
\be
\widehat{p}(X)\sigma = -i\hbar\;\nabla_X\sigma
    + (-\frac {i\hbar}2\;\mathbb I + \Phi )
    \mbox{\rm div}_\nu (X)\sigma .
\label{diffquant}
\en
\entheo

For an elementary quantum Borel kinematics, i.e., when the
${\mathbb C}^k$-bundle is a line bundle, one can give a complete classification
of the possible equivalence classes of quantum Borel kinematics. Indeed, for
Hermitian line bundles, one has the classification theorem:

\betheo
Let $\bfrakQ$ be a connected differentiable manifold and $B$ a closed two-form
on $\bfrakQ$ (i.e., $dB = 0$). Then there exists a Hermitian complex line
bundle $(\mathbb E , \langle\cdot \vert\cdot\rangle , \nabla )$, with
compatible connection and curvature $\frac 1\hbar\;B$, if and only if $B$
satisfies the integrality condition
\be
  \frac 1{2\pi\hbar}\; \int_{\Sigma}B \in \mathbb Z ,
\label{integrcond1}
\en
for all closed two-surfaces $\Sigma$ in $\bfrakQ$. Furthermore, the various
equivalence classes of  $(\mathbb E , \langle\cdot\vert\cdot\rangle , \nabla )$
(for fixed curvature $\frac 1\hbar\;B$) are parameterized by $H^1 (\bfrakQ,
{\mathcal U}(1))\simeq \pi_1 (\bfrakQ )^*$, where $\pi_1 (\bfrakQ )^*$ denotes
the group of characters of the first fundamental group of $\bfrakQ$.
\entheo

The classification of the associated elementary quantum Borel kinematics is
then spelled out in the following theorem.

\betheo\label{classiftheo}
The equivalence classes of elementary localized
differentiable quantum Borel kinematics are in one-to-one correspondence with
$I^2 (\bfrakQ )\times \pi_1 (\bfrakQ )^* \times {\mathbb R}$, where $I^2
(\bfrakQ )$ denotes the set of all closed real two-forms on $\bfrakQ$,
satisfying the integrality condition (\ref{integrcond1}).
\entheo

For ${\mathbb C}^k$-bundles only a weaker result, for $\Omega = 0$, is known:

\betheo
The equivalence classes of ($\Omega = 0$)-compatible differentiable and
localized $k$-quantum Borel kinematics are in one-to-one correspondence with
the equivalence classes $\{(D, A)\}$ of pairs of unitary representations
$D \in \mbox{\rm Hom}(\pi_1 (\bfrakQ ), {\mathcal U} (k))$ and self adjoint
complex $k\times k$ matrices $A \in {\mathcal S}({\mathbb C}^k )\bigcap D'$,
where $D'$ is the commutant of the representation $D$, i.e.,
$D' = \{ M \in {\mathcal L} ({\mathbb C}^k )\;\vert\; [M, D(g)] = 0, \;\;
\forall g \in \pi_1 (\bfrakQ )\}$. Here two pairs $(D_1, A_1 )$ and
$(D_2 , A_2 )$ are equivalent if there is a unitary matrix $U$ such that
$D_2 = UD_1 U^{-1}$ and $A_2 = UA_1 U^{-1}$.
\entheo

Instead of enlarging the space of quantizable observables to include the
Hamiltonian, the~Borel quantization method then proceeds in a different way to
treat the time evolution of the quantized system, leading ultimately to a
nonlinear Schr\"odinger equation; see Ali \cite{bib:AliSurv}, Doebner and
Nattermann \cite{bib:DoebNa}, Angermann, Doebner and Tolar \cite{bib:ADT},
Angermann \cite{bib:Angm}, Tolar \cite{bib:Tolar}, Pasemann \cite{bib:Pasem}
and Mueller~\cite{bib:Muell} for the details. For~a comparison with geometric
quantization (to~be discussed in the next section) see Zhao~\cite{bib:ZhaoQ}.

\section{Geometric quantization}\label{sec2}
We pass on to a treatment of {\em geometric quantization\/}, which in addition
to being a physical theory has also emerged as a branch of mathematics.
The~starting point here is  a real symplectic manifold $\Omg$ (the~phase space)
of dimension $2n$, with symplectic form~$\omega$. For a function $f$ on $\Omg$,
the corresponding Hamiltonian vector field $X_f$ is given by
$\omega(\cdot,X_f)=df$. The Poisson bracket of two functions is defined by
\begin{equation}  \{f,g\} = -\omega(X_f,X_g).  \label{tag:PBG}  \end{equation}
Starting with such a manifold as the arena of classical mechanics, the goal of
geometric quantization is to assign to each such manifold $(\Omg,\omega)$ a
separable Hilbert space $\HH$ and a mapping $Q:f\mapsto Q_f$ from a subspace
$\Obs$ (as large as possible) of real-valued functions on~$\Omg$, which is a
Lie algebra under the Poisson bracket, into self-adjoint linear operators
on~$\HH$ in such a way that
\begin{enumerate}
\item[(Q1)] $Q_\jedna=I$, where $\jedna$ is the function constant one and $I$
the identity operator on~$\HH$;
\item[(Q2)] the mapping $f\mapsto Q_f$ is linear;
\item[(Q3)] $[Q_f,Q_g]=\frac{ih}{2\pi} Q_{\{f,g\}}$, \;\; $\forall f,g\in\Obs$;
\item[(Q4)] the procedure is {\sl functorial\/} in the sense that for two
symplectic manifolds $(\Omg^{(1)},\omega^{(1)})$, $(\Omg^{(2)},\omega^{(2)})$
and a diffeomorphism $\phi$ of $\Omg^{(1)}$ onto $\Omg^{(2)}$ which sends
$\omega^{(1)}$ into $\omega^{(2)}$, the composition with $\phi$ should map
$\Obs^{(2)}$ into $\Obs^{(1)}$ and there should be a unitary operator $U_\phi$
from $\HH^{(1)}$ onto $\HH^{(2)}$ such that
\begin{equation}  Q^{(1)}_{f\circ\phi} = U_\phi^* Q^{(2)}_f U_\phi \qquad
\forall f\in\Obs^{(2)} ;  \label{tag:COV}  \end{equation}
\item[(Q5)] for $(\Omg,\omega)=\RR^{2n}$ with the standard symplectic form, we
should recover the operators $Q_{q_j}$, $Q_{p_j}$ in~(\ref{tag:Schro}).
\end{enumerate}

\begin{remark*} The requirements (Q4) and (Q5) are, in some way, a substitute
for the irreducibility condition (\ref{tag:IRE}), which may be difficult to
interpret on a general symplectic manifold (i.e. in the absence of a global
separation of coordinates into the $q$ and $p$ variables). Another, frequently
used, possibility is to require that for some ``distinguished'' set of
observables $f$ the corresponding quantum operators $Q_f$ should act
irreducibly on~$\HH$; however, there seems to be no general recipe how one
should choose such ``distinguished'' sets. The requirement that there be no
nontrivial subspace in $\HH$ invariant for all $Q_f$, $f\in\Obs$, is
\underbar{not} the correct substitute; see Tuynman \cite{bib:TuyIrr} for a
thorough discussion of this point. Also we gave up the von Neumann rule~(q3),
but it turns out that this is usually recovered to some extent,
cf.~\cite{bib:GGTuy}.   \qed  \end{remark*}

\begin{remark*} Observe that if there is a group $G$ of symplectomorphisms
acting on $(\Omg,\omega)$, then the covariance axiom~(Q4) implies (taking
$\Omg_1=\Omg_2=\Omg$) that the quantization map $f\mapsto Q_f$ is (essentially)
$G$-invariant.  \qed  \end{remark*}

\medskip

The solution to the above problem was first given by Kostant \cite{bib:Kost}
and Souriau \cite{bib:SouSD}. It~is accomplished in two steps: prequantization
and polarization. {\sl Prequantization\/} starts with introducing a complex
Hermitian line bundle $L$ over $\Omg$ with a connection $\nabla$ whose
curvature form satisfies $\operatorname{curv} \nabla=2\pi\omega/h$.
(For~$(L,\nabla)$ to exist it is necessary that the cohomology class of
$\omega/h$ in $H^2(\Omg,\RR)$ be integral; this is known as the
{\sl prequantization condition.\/}) One then defines for each
$f\in C^\infty(\Omg)$ the differential operator
\begin{equation}  Q_f = -\frac{ih}{2\pi} \nabla_{X_f} + f  \label{tag:GQF}
\end{equation}
where the last $f$ stands for the operator of multiplication by~$f$. Plainly
these operators satisfy (Q1),(Q2) and (Q4), and a short computation reveals
that they also satisfy~(Q3).

Unfortunately, (Q5) is manifestly violated for the operators~(\ref{tag:GQF});
in fact, for $\Omg=\RR^{2n}$ these operators act not on $L^2(\RR^n)$ but on
$L^2(\RR^{2n})$, so we need somehow to throw away half of the variables. More
precisely, one checks that for $\Omg=\RR^{2n}$ the operators (\ref{tag:GQF})
are given by
$$ Q_f = -\frac{ih}{2\pi} \sum_j \bigg(
\frac{\partial f}{\partial p_j} \frac\partial{\partial q_j} -
\frac{\partial f}{\partial q_j} \frac\partial{\partial p_j} \bigg)
+ \bigg(f-\sum_j p_j \frac{\partial f}{\partial p_j} \bigg) ,  $$
so restricting $Q_f$ to the space of functions depending only on $q$ and
square-integrable over $q\in\RR^n$ one recovers the desired
operators~(\ref{tag:Schro}). For a general symplectic manifold $(\Omg,\omega)$,
making sense of ``functions depending on and square-integrable over only half
of the variables'' is achieved by {\sl polarization.\/} The latter amounts,
roughly speaking, to choosing a subbundle $\PP$ of complex dimension $n$ in the
complexified tangent bundle $T\Omg^\CC$ in a certain way and then restricting
to functions on $\Omg$ which are constant along the directions
in~$\PP$~\footnote{If~$\Omg$ is a cotangent bundle, i.e.~$\Omg=T^*\bfrakQ$ for
some configuration space~$\bfrakQ$, one can polarize simply by restricting to
functions depending on $q$ only; however, for general symplectic manifolds the
global separation into position and momentum coordinates is usually impossible.
A~well-known example of a physical system whose phase space is not a cotangent
bundle is the phase space of classical spin (discussed extensively in
Souriau~\cite{bib:SouSD}),
which can be identified with the Riemann sphere~$\bold S^2$.}. This settles the
``dependence on half of the variables''. As~for the ``square-integrability'',
the simplest solution is the use of half-densities, which however does not give
the correct quantization for the harmonic oscillator; one therefore has to
apply the {\sl metaplectic correction,\/} which amounts to using not
half-densities but half-forms
and gives the right answer for the
harmonic oscillator (but not in some other cases, cf.~\cite{bib:TuyWis}).
Finally, for functions $f$ which leave $\PP$ invariant, i.e.~$[X_f,\PP]\subset
\PP$, the corresponding operator given (essentially) by (\ref{tag:GQF}) maps a
function constant along $\PP$ into another such function, and thus one arrives
at the desired quantum operators.

Since geometric quantization is still probably the most widely used
quantization method, we~will now discuss all the above ingredients in some
more detail prior to embarking on the discussion of other approaches.

\subsection{Prequantization} \label{sec21}
The aim of prequantization is to construct a mapping $f\mapsto Q_f$ satisfying
all the required axioms except~(Q5). For simplicity, let us start with the case
when $\Omg$ is a cotangent bundle: $\Omg=T^*\bfrakQ$. One can then define
globally a real one-form $\theta$ (the {\sl symplectic potential\/}) satisfying
\begin{equation}  d\theta=\omega.  \label{tag:SYMPOT}  \end{equation}
Actually, if $m\in \Omg$ and $\xi\in T_m\Omg$, then one sets
$$ \theta(\xi):= m(\pi_*\xi)  $$
where $\pi:\Omg\to\bfrakQ$ denotes the cotangent bundle projection and
$\pi_*:T\Omg\to T\bfrakQ$ is the derivative map of~$\pi$. In~terms of local
coordinates $q_j$ on $\bfrakQ$ and $(p_j,q_j)$ on~$\Omg$, one~has
\begin{equation}  \theta=\sum_{j=1}^n p_j\,dq_j, \qquad
\omega=\sum_{j=1}^n dp_j\wedge dq_j.  \label{tag:PDQ}  \end{equation}
The Hamiltonian field $X_f$ of a function $f$ on $\Omg$ is in these coordinates
given~by
\begin{equation}  X_f = \sum_{j=1}^n \Big( \frac{\partial f}{\partial p_j}
\frac{\partial}{\partial q_j} - \frac{\partial f}{\partial q_j}
\frac{\partial}{\partial p_j} \Big),   \label{tag:HAMI}  \end{equation}
and the Poisson bracket $\{f,g\}=- \omega(X_f,X_g)=X_f g$ of two functions
$f,g$ is again expressed by~(\ref{tag:POAB}).

A~simple computation shows that $[X_f,X_g]= - X_{\{f,g\}}$, thus
$Q_f=-\frac{ih}{2\pi} X_f$ satisfies the conditions (Q2), (Q3) and (Q4).
Unfortunately, (Q1) fails, since $X_1=0$. Let us try correcting this by taking
$$ Q_f = -\frac{ih}{2\pi} X_f +f  $$
(where the latter $f$ is to be taken as the operator of multiplication by the
function~$f$). Then $Q_1=I$, as desired, but
$$ [Q_f,Q_g] = \frac{ih}{2\pi} (Q_{\{f,g\}} + \{f,g\})  $$
so now (Q3) is violated. Observe, however, that
$$ X_f(\theta(X_g))-X_g(\theta(X_f)) = - \theta(X_{\{f,g\}})+\{f,g\}  $$
by a straightforward computation using (\ref{tag:HAMI}) and~(\ref{tag:PDQ}).
Thus taking
\begin{equation}  Q_f = -\frac{ih}{2\pi} X_f - \theta(X_f) + f
\label{tag:QFA}  \end{equation}
it follows that all of (Q1)~--~(Q4) will be satisfied.

Having settled the case of the cotangent bundle, let us now turn to general
symplectic manifolds $(\Omg,\omega)$. By~a theorem of Darboux, one can always
cover $\Omg$ by local coordinate patches $(p_j,q_j)$ such that the second
formula in (\ref{tag:PDQ}) (and, hence, also~(\ref{tag:HAMI})) holds; however,
the corresponding symplectic potentials need not agree on the intersections of
two coordinate patches. Let us therefore examine what is the influence of a
different choice of potential on the operator~(\ref{tag:QFA}).
If~$\omega=d\theta=d\theta'$, then $\theta'=\theta+du$ (locally) for some real
function~$u$; then $\theta'(X_f)- \theta(X_f)=X_f u=-e^u X_f e^{-u}$, whence
\begin{equation}  e^{\frac{2\pi}{ih}u} Q'_f \phi = Q_f e^{\frac{2\pi}{ih}u}
\phi\; , \qquad \forall \phi\in C^\infty. \label{tag:QFB}  \end{equation}

Recall now that, quite generally, a complex {\sl line bundle\/} $L$ over a
manifold $\Omg$ is given by the following data:
\begin{enumerate}
\item[(1)] a~covering (atlas) $\{U_\alpha\}_{\alpha\in\Cal I}$ of $\Omg$ by
coordinate patches,
\item[(2)] a~family of transition functions $\{g_{\alpha\beta}\}_{\alpha,\beta
\in\Cal I}$, each $g_{\alpha\beta}$ being a nonvanishing $C^\infty$ function in
$U_\alpha\cap U_\beta$, satisfying the cocycle condition
\begin{equation}  g_{\alpha\beta} g_{\beta\gamma} = g_{\alpha\gamma}
\qquad\text{in } U_\alpha\cap U_\beta\cap U_\gamma  \label{tag:QFC}
\end{equation}
(${}\implies g_{\alpha\alpha}=1$, $g_{\beta\alpha}=1/g_{\alpha\beta}$).
\end{enumerate}
A~{\sl section\/} $\phi$ of $L$ is a family of functions $\phi_\alpha:U_\alpha
\to\CC$ such that
\begin{equation}  \phi_\alpha = g_{\alpha\beta} \phi_\beta \qquad
\text{in } U_\alpha\cap U_\beta.   \label{tag:QFD}  \end{equation}
(Similarly, one defines {\sl vector bundles\/} by demanding that $f_\alpha$ be
mappings from $U_\alpha$ into a (fixed) vector space $\Bbb V$, and $g_{\alpha
\beta}\in GL(\Bbb V)$ be linear isomorphisms of~$\Bbb V$; more generally,
a~({\sl fiber\/}) {\sl bundle\/} with some object $\frak G$ as fiber is defined
by taking $f_\alpha$ to be mappings from $U_\alpha$ into $\frak G$, and
$g_{\alpha\beta}$ to be isomorphisms of the object~$\frak G$.)

For later use, we also recall that $L$ is said to be {\sl Hermitian\/} if, in
addition, there is given a family $e_\alpha$ of positive $C^\infty$ functions
on $U_\alpha$ such that
$$ e_\alpha = |g_{\alpha\beta}|^{-2} e_\beta \qquad
\text{in }U_\alpha\cap U_\beta.  $$
In~that case, for two sections $\phi,\psi$ one can define unambiguously their
``local'' scalar product --- a~function on~$\Omg$~---~by
$$ (\phi,\psi)_m = e_\alpha(m) \overline{\phi_\alpha(m)}\psi_\alpha(m),
\qquad\text{ if } m\in U_\alpha.  $$

Further, a mapping $(\xi,\phi)\mapsto\nabla_\xi \phi$ from $\bfrakX(\Omg)\times
\Gamma(L)$ into $\Gamma(L)$, where $\Gamma(L)$ denotes the space of all smooth
(i.e.~$C^\infty$) sections of $L$ and $\bfrakX(\Omg)$ the space of all smooth
vector fields on~$\Omg$, is~called a {\sl connection\/} on $L$ if it is linear
in both $\xi$ and $\phi$,
\begin{equation}  \nabla_{f\xi}\phi=f \nabla_\xi\phi  \label{tag:QFEa}
\end{equation}
and
\begin{equation}  \nabla_\xi (f\phi) = (\xi f)\phi + f\nabla_\xi \phi
\label{tag:QFEb}  \end{equation}
for any $f\in C^\infty(\Omg)$. The {\sl curvature\/} of this connection is the
2-form on $\Omg$ defined~by
\begin{equation}  \curv(\nabla)(\xi,\eta)\phi := i(\nabla_\xi
\nabla_\eta-\nabla_\eta \nabla_\xi -\nabla_{[\xi,\eta]} ) \phi, \qquad \forall
\xi,\eta\in \bfrakX(\Omg),\ \phi\in\Gamma(L).  \label{tag:QFEc} \end{equation}
Finally, a connection on a Hermitian line bundle is said to be {\sl
compatible\/} (with the Hermitian structure)~if
\begin{equation}  \xi (\phi,\psi) = (\nabla_{\overline\xi} \phi,\psi) +
(\phi,\nabla_\xi \psi)  \label{tag:QFF}  \end{equation}
for $\phi,\psi\in\Gamma(L)$ and complex vector fields $\xi\in V(\Omg)^\CC$.

Returning to our symplectic manifold $(\Omg,\omega)$, suppose now that we have
an open cover $\{U_\alpha\}_{\alpha\in\Cal I}$ of $\Omg$ and collections
$\{\theta_ \alpha\}_{\alpha\in\Cal I}$ and
$\{u_{\alpha\beta}\}_{\alpha,\beta\in\Cal I}$
such that $\theta_\alpha$ is a symplectic potential on $U_\alpha$ and $\theta_
\alpha=\theta_\beta+du_{\alpha\beta}$ on $U_\alpha\cap U_\beta$. Comparing
(\ref{tag:QFB}) and (\ref{tag:QFD}), we see that if we can take
\begin{equation}  g_{\alpha\beta}=\exp\Big(-\frac{2\pi}{ih}
u_{\alpha\beta}\Big)  \label{tag:QFG} \end{equation}
then the local operators $Q_f$ can be glued together into a well-defined global
operator on the sections of the corresponding line bundle~$L$.

    The functions defined by the last formula satisfy the consistency condition
(\ref{tag:QFC}) if and only if
$\exp(-\frac{2\pi}{ih}(u_{\alpha\beta}+u_{\beta\gamma}+
u_{\gamma\alpha}))=1$, that~is, if~and only if there exist integers
$n_{\alpha\beta\gamma}$ such that
$$ u_{\alpha\beta}+u_{\beta\gamma}+u_{\gamma\alpha}=n_{\alpha\beta\gamma}h  $$
for all $\alpha,\beta,\gamma$ such that $U_\alpha\cap U_\beta\cap U_\gamma$ is
nonempty. One~can show that this condition is independent of the choice of the
cover $\{U_\alpha\}$ etc.~and is, in~fact, a~condition on~$\omega$: it~means
that the de Rham cohomology class defined by $h^{-1}\omega$ in $H^2(\Omg,\RR)$
should be integral. This is known as the {\sl integrality condition\/} (or~{\sl
prequantization\/} condition), and we will assume it to be fulfilled throughout
the rest of this section (Section~\ref{sec2}). The bundle $L$ is called the
{\sl prequantization bundle.\/}

Observe that since the transition functions (\ref{tag:QFG}) are
unimodular\footnote{In~general, if the transition functions $g_{\alpha\beta}$
of a (fiber) bundle all belong to a group~$G$, $G$~is said to be the
{\sl structure group\/} of the bundle. Thus the line bundle $L$ above has
structure group~$U(1)$, and, similarly, the frame bundles $\FF^k\PP$ to be
constructed in the next subsection have structure groups $GL(k,\RR)$.}
(because $u_{\alpha\beta}$ are real), we~can equip the bundle $L$ with a
Hermitian structure simply by taking $e_\alpha=1$ $\forall\alpha$; that~is,
$$ (\phi,\psi)_m = \overline{\phi_\alpha(m)} \psi_\alpha(m).  $$
We~finish this subsection by exhibiting a compatible connection $\nabla$
on~$L$, in~terms of which the operators $Q_f$ assume a particularly simple
form. Namely, define, for $\xi\in \bfrakX(\Omg)$, $\psi\in\Gamma(L)$ and a
local chart $U_\alpha$,
\begin{equation}  (\nabla_\xi \psi)_\alpha := \xi\psi_\alpha + \frac{2\pi}{ih}
\theta_\alpha(\xi) \psi_\alpha.  \label{tag:QFH}  \end{equation}
One easily checks that this definition is consistent (i.e.~that $\phi:=\nabla
_\xi\psi$ satisfies the relations~(\ref{tag:QFD})) and that $\nabla$ satisfies
(\ref{tag:QFEa}), (\ref{tag:QFEb}) and (\ref{tag:QFF}), i.e.~defines a
compatible connection. Now comparing (\ref{tag:QFA}) and (\ref{tag:QFH}) we see
that the prequantum operators $Q_f$ can be rewritten simply~as
\begin{equation}  Q_f = -\frac{ih}{2\pi} \nabla_{X_f} + f.  \label{tag:QFI}
\end{equation}

To~summarize our progress, we have shown that on an arbitrary symplectic
manifold $(\Omg,\omega)$ such that $h^{-1}\omega$ satisfies the integrality
condition, there exists a Hermitian line bundle $L$ and operators $Q_f$ on
$\Gamma(L)$ (the space of smooth sections of~$L$) such that the correspondence
$f\mapsto Q_f$ satisfies the conditions (Q1)~--~(Q4). In~more detail --- there
is a compatible connection $\nabla$ on~$L$, and the operators $Q_f$ are given
by the formula~(\ref{tag:QFI}).

\begin{remark*} It~can be shown that the curvature of the connection
(\ref{tag:QFH}) is given~by
$$ \curv(\nabla) = \frac{2\pi}h \omega.  $$
The fact that, for a given symplectic manifold $(\Omg,\omega)$, there exists a
Hermitian line bundle $L$ with a compatible connection $\nabla$ satisfying
$\curv(\nabla)=2\pi\omega$ if and only if $\omega$ satisfies the integrality
condition, is the content of a theorem of A.~Weil \cite{bib:WeilVK} (see
also~\cite{bib:Kost}). Furthermore, the equivalence classes of such bundles
$(L,\nabla,(\cdot,\cdot))$ are then parameterized by the elements of the first
cohomology group $H^1(\Omg,\TT)$ with coefficients in the circle group~$\TT$.
This should be compared to the content of Theorem \ref{classiftheo}, which we
stated in the context of Borel quantization.
\qed  \end{remark*}

\begin{remark*} In~another guise, the integrability condition can be expressed
by saying that the integral of $\omega$ over any closed orientable
2-dimensional surface in~$\Omg$ should be an integer multiple of~$2\pi$.
This is reminiscent of the Bohr-Sommerfeld quantization condition,
familiar from the old quantum theory. \qed
\end{remark*}

\begin{remark*} It~is possible to give an alternative description of the whole
construction above in the language of {\it connection forms.\/} Namely, let
$L^\times$ denote the line bundle $L$ with the zero section removed. The {\it
fundamental vector field\/} on $L^\times$ corresponding to $c\in\CC$ is
defined~by
$$ (\eta_c f)(m,z)=\frac d{dt} f(e^{2\pi i c t}z) \big|_{t=0}, \qquad
\forall m\in \Omg,\ z\in L^\times_m,  $$
for any function $f$ on $L^\times$. A~{\sl connection form\/} is a one-form
$\alpha$ on $L^\times$ which is $\CC^\times$-invariant and satisfies $\alpha
(\eta_c)=c$ $\forall c\in\CC$; in~other words, it is locally given by $\alpha
=\pi^*\Theta+i\frac{dz}z$, with $\Theta$ a one-form on $\Omg$ and $z$ the
coordinate in the fiber $L^\times_m\simeq\CC^\times$. A~vector field $\zeta$ on
$L^\times$ is called {\it horizontal\/} (with respect to~$\alpha$) if $\alpha
(\zeta)=0$. It~can be shown that every vector field $\xi$ on $\Omg$ has a
unique {\sl horizontal lift\/} $\tilde\xi$ on~$L^\times$, defined by the
requirements that
$$ \pi_*\tilde\xi=\xi\qquad\text{and}\qquad \alpha(\tilde\xi)=0
\text{ (i.e.~$\tilde\xi$ is horizontal).}  $$
One can then easily verify that the recipe
$$ (\nabla_\xi \phi) := \tilde\xi \phi_\beta
\qquad\text{in a local chart }U_\beta, $$
or, equivalently,
$$ \nabla_\xi \phi = 2\pi i \phi^*\alpha(\xi) \phi,  $$
defines a connection on~$L^\times$. Our connection (\ref{tag:QFH}) corresponds
to the choice
$$ \alpha_\beta = \frac{2\pi}h \theta_\beta + i\frac{dz}z \qquad
\text{in a local chart } U_\beta\times\CC^\times.  $$
See Sniatycki~\cite{bib:SniaB}, Section~3.1 for the details.   \qed
\end{remark*}

\begin{remark*} Still another (equivalent) description may be based on the use
of connection one-forms in a principal $U(1)$-bundle over~$\Omg$ and the Reeb
vector field therein; see \cite{bib:TuyIrr} and the references therein.   \qed
\end{remark*}

\medskip

We~conclude by mentioning also an alternative characterization of the
prequantum operators~$Q_f$ when the Hamiltonian field $X_f$ of $f$ is complete.
In~that case, the field $X_f$ generates a one-parameter group (a~{\sl flow\/})
$\rho_t=\exp(t X_f)$ of canonical transformations (symplectomorphisms) of
$(\Omg,\omega)$. This flow lifts uniquely to a flow --- again denoted~$\rho_t$
--- of linear connection-preserving transformations on $\Gamma(L)$.
The~operator $Q_f$ is then given~by
$$ Q_f \phi = -\frac{ih}{2\pi} \frac d{dt} (\rho_t\phi) \big|_{t=0}.  $$
For the details we refer to Sniatycki~\cite{bib:SniaB}, Section~3.3.
In~particular, since the induced transformations $\rho_t$ on $\Gamma(L)$ are
unitary, it~follows by the Stone theorem that $Q_f$ are (essentially)
self-adjoint operators on the Hilbert space
$$ \HH_{\text{preq}}:=\text{the completion of } \Big\{\phi\in\Gamma(L): \;
\text{$\into (\phi,\phi)_m |\omega^n| <\infty$}  \Big\}  $$
of all square-integrable sections of~$L$. This is also akin to the construction
of the operators $\widehat{p}(X)$ in Borel quantization
(see~(\ref{borelquant1})).

\subsection{Real polarizations and half-densities}\label{sec22}
We~now discuss the second step of geometric quantization --- namely, making
sense of ``functions depending on'' and ``square-integrable over'' only half of
the variables. The simplest way of doing this is via real polarizations and
half-densities, which we now proceed to describe.

A~(real) {\sl distribution\/}\footnote{This is not to be confused with the
distributions (generalized functions) in the sense of L.~Schwartz!} $\DD$ on
$\Omg$ is a map which assigns to each point $m\in \Omg$ a linear subspace
$\DD_m$ of $T_m \Omg$ such that \begin{enumerate}
\item[(i)] $\dim\DD_m=k$ (a~constant independent of $m\in \Omg$)
\item[(ii)] $\forall m_0\in \Omg$ $\exists$ a neighbourhood $U$ of $m_0$ and
vector fields $X_1,\dots,X_k$ on $U$ such that $\forall m\in U$, $\DD_m$ is
spanned by $X_1|_m,\dots,X_k|_m$. \end{enumerate}
A~distribution is called {\sl involutive\/} if for any two vector fields
$X,Y\in\DD$ (i.e.~$X_m,Y_m\in\DD_m$ $\forall m$) implies that $[X,Y]\in\DD$ as
well; and {\sl integrable\/} if for each $m_0\in \Omg$ there exists a
submanifold $N$ of $\Omg$ passing through $m_0$ and such that $\forall m\in N:$
$\DD_m=T_m N$. A~theorem of Frobenius asserts that for real distributions, the
notions of integrability and involutiveness are equivalent. An~integrable
distribution is also called a {\sl foliation,\/} and the maximal connected
submanifolds $N$ as above are called its {\sl leaves.\/} A~foliation is called
{\sl reducible\/} (or~{\sl fibrating\/}) if the set of all leaves --- denoted
$\MD$ --- can be given a structure of a manifold in such a way that the natural
projection map $\pi:\Omg\to \MD$ is a (smooth) submersion.

So~far, all these definitions make sense for an arbitrary (smooth)
manifold~$\Omg$. If~$\Omg$ is symplectic, then we further define $\DD$ to
be {\sl isotropic\/} if $\omega(X,Y)=0$ $\forall X,Y\in\DD$; and
{\sl Lagrangian\/} if it is maximal isotropic,
i.e.~$\dim\DD_m=n:=\frac12\dim \Omg$ $\forall m\in \Omg$.
A~Lagrangian foliation is called a {\sl real polarization\/} on~$\Omg$.

One~can prove the following alternative characterization of real polarizations:
a~smooth distribution $\DD$ on $\Omg$ is a real polarization if and only if for
each $m_0\in \Omg$ there exists a neighbourhood $U$ of $m_0$ and $n$
independent functions $f_1,\dots,f_n$ on~$U$ (i.e.~$\forall m\in U:$
$df_1,\dots,df_n$ are independent in~$T^*_m \Omg$) such that:
\begin{equation}  \begin{array}{rl}
\text{(i) }&\text{$\forall m\in U$, $\DD_m$ is spanned by $X_{f_1}|_m,\dots,
X_{f_n}|_m$;}\\
\text{(ii) }&\text{$\{f_i,f_j\}=0$ on $U$, $\forall i,j=1,\dots,n$.}
\end{array}  \label{tag:REALPOL}  \end{equation}
(That is --- $\DD$ is locally spanned by commuting Hamiltonian vector-fields.)

Now~we say that a section $\phi$ of our prequantization bundle $L$ with
connection~$\nabla$ (constructed in the preceding subsection) is {\sl
covariantly constant\/} along $\DD$~if
$$ \nabla_X \phi=0\; ,  \qquad \forall X\in \DD.  $$
In~view of the compatibility relation (\ref{tag:QFF}), the ``local'' scalar
product $(\phi,\psi)$ of two covariantly constant sections is then a function
on $\Omg$ constant along~$\DD$ (i.e.~$X(\phi,\psi)=0$ $\forall X\in\DD$),
hence, defines a function on~$\MD$.

Let us now deal with the issue of ``integrating'' over~$\MD$.

The simplest solution would be to take the integral of $(\phi,\psi)_m$ with
respect to some measure on~$\MD$. That is, if $\mu$ is a (nonnegative regular
Borel) measure on~$\MD$, let $\HH$ be the Hilbert space of all sections
$\phi$ of $L$ such that $\phi$ is covariantly constant along $\DD$ and
$$ \int_{\MD} (\phi,\psi)_m \, d\mu(x) <\infty  $$
(where, for each $x\in \MD$, $m$ is an arbitrary point in the fiber
$\pi^{-1}(x)$ above~$x$). For~a real function $f$ on~$\Omg$, the quantum
operator could then be defined on $\HH$~by
\begin{equation}  Q_f \phi = -\frac{ih}{2\pi} \nabla_{X_f} \phi + f\phi,
\label{tag:QPP}  \end{equation}
granted this takes $\phi\in\HH$ again into a section covariantly constant
along~$\DD$. In~view of (\ref{tag:QFEb}) and (\ref{tag:QFEc}), the latter is
readily seen to be the case~if
\begin{equation}  [X_f,X]\in\DD \qquad\forall X\in\DD.   \label{tag:QQQ}
\end{equation}
Hence, proclaiming the set of all functions satisfying (\ref{tag:QQQ}) to be
the space $\Obs$ of quantizable observables, we have arrived at the desired
quantization recipe.

Unfortunately, there seems to be no canonical choice for the measure $\mu$ on
$\MD$ in general. For~this reason, it is better to incorporate the choice of
measure directly into the bundle~$L$: that is, to pass from the prequantum line
bundle $L$ of \S\ref{sec21} to the tensor product of $L$  with some ``bundle of
measures on~$\MD$''. In~order for this product to make sense, we must (first
of all define this ``bundle of measures'' over $\MD$, and second) turn the
latter bundle into a bundle over~$\Omg$ (instead of~$\MD$). Let us now explain
how all this is done.

Consider, quite generally, a manifold $\xX$ of dimension~$n$, and let
$\pi:\FF^n\xX\to\xX$ be the bundle of $n$-frames\footnote{The bundle $\FF^k\xX$
of $k$-frames, where $1\le k\le n$, is defined similarly; in~particular,
$\FF^1\xX$ is just the tangent bundle without the zero section.} over~$\xX$,
i.e.~the fiber $\FF^n_x\xX$ at $x\in\xX$ consists of all ordered $n$-tuples of
linearly independent vectors $(\xi_1,\dots,\xi_n)$ from $T_x\xX$. The group
$GL(n,\RR)$ of real nonsingular $n\times n$ matrices acts on $\FF^n\xX$ in a
natural way: if~$\xi_{jk}$ are the coordinates of $\xi_j$ with respect to some
local chart $U\times\RR^n$ of $T_x\xX$, then $g\in GL(n,\RR)$ acts~by
$$ (\xi\cdot g)_{jk} = \sum_{l=1}^n \xi_{jl} g_{lk}.  $$
Now recall that one possible definition of a complex $n$-form is that it is a
mapping $\eta:\FF^n\xX\to\CC$ assigning to a point $x\in\xX$ and an $n$-frame
$(\xi_1,\dots,\xi_n)\in\FF^n_x\xX$ a complex number $\eta_x(\xi_1,\dots,\xi_n)$
such that
$$ \eta_x(\xi\cdot g) = \eta_x(\xi)\cdot\det g \qquad\forall g\in GL(n,\RR). $$
By~analogy, we therefore define a {\sl density\/} on $\xX$ as a mapping $\nu$
from $\FF^n\xX$ into $\CC$ satisfying
$$ \nu_x(\xi\cdot g)=\nu_x(\xi)\cdot|\det g\,| \qquad\forall g\in GL(n,\RR), $$
and, more generally, an $r$-density, where $r$ is any (fixed) real number,~by
\begin{equation}  \nu_x(\xi\cdot g) = \nu_x(\xi)\cdot|\det g\,|^r
\qquad \forall g\in GL(n,\RR).    \label{tag:HDA}  \end{equation}

Similarly, one defines, for a distribution $\DD$ on a manifold, an {\sl
$r$-$\DD$-density\/} as a mapping from the bundle $\FF^n\DD$ ($n=\dim\DD$) of
$n$-frames of~$\DD$ (i.e.~the fiber $\FF^n_m\DD$ consists of all ordered bases
of~$\DD_m$) into $\CC$ which satisfies
\begin{equation}  \nu_m(\xi\cdot g)=\nu_x(\xi)\cdot|\det g\,|^r
\qquad\forall\xi\in\FF^n\DD, \ \forall g\in GL(n,\RR).  \label{tag:HDB}
\end{equation}

Let~us now apply this to the case of $\xX=\MD$ with $\DD$ a real polarization
as above. Thus, a $\frac12$-density on $\MD$ is a function $\phi$ which assigns
to any ordered $n$-tuple of independent tangent vectors $\xi_j\in T_x(\MD)$ a
complex number $\phi_x(\xi_1,\dots,\xi_n)$ such that (\ref{tag:HDA}) holds with
$r=\frac12$. We~now define a ``lift'' from $\frac12$-densities on $\MD$ to
\mhD-densities on $\Omg$ as follows. Let $m\in \Omg$ and let
$\xi_1,\dots,\xi_n$ be a frame of $T_{\pi(m)}(\MD)$, where $\pi:\Omg\to\MD$
denotes the canonical projection. Then there exists a unique dual basis
$c_1,\dots,c_n\in T^*_{\pi(m)}(\MD)$, defined by $c_j(\xi_k)=\delta_{jk}$. This
basis is mapped by $\pi^*$ onto $n$ independent vectors of $T^*_m \Omg$, and we
can therefore define tangent vectors $\tilde\xi_j\in T_m \Omg$ by the recipe
$$ \omega(\cdot,\tilde\xi_j)=\pi^*_m c_j.  $$
From the properties of the symplectic form $\omega$ one easily sees that
$\pi_*\tilde\xi_j=0$, that is, $\tilde\xi_1,\dots,\tilde\xi_n$ is, in fact,
a~basis of~$\DD_m$, and the correspondence $(\xi)\mapsto(\tilde\xi)$ between
the frames of $T_{\pi(m)}(\MD)$ and the frames of $\DD_m$ is bijective. For~a
half-density $\phi$ on~$\MD$, we can therefore define a function $\tilde\phi$
on $\FF^n\DD$~by
$$ \tilde\phi(\tilde\xi) := \phi(\xi).  $$
An~easy computation shows that
$$ \tilde\phi(\tilde\xi\cdot g) = \tilde\phi(\widetilde{\xi\cdot g^{-1T}})
= \tilde\phi(\tilde\xi) \cdot|\det g^{-1T}|^{1/2},  $$
where ${}^T$ stands for matrix transposition. Thus $\tilde\phi$ is a
\mhD-density on~$\Omg$.

Let~us denote by $\BB^\DD$ the complex fibre bundle of \mhD-densities
on~$\Omg$. (That~is: the fiber $\BB_m^\DD$ consists of all functions
$\nu_m:\FF^n\DD\to\CC$ satisfying~(\ref{tag:HDB}), and the sections of
$\BB^\DD$ are thus \mhD-densities on~$\Omg$.) The~map $\phi\mapsto\tilde\phi$
above thus defines a lifting from $\Delta^{1/2}(\MD)$, the (similarly defined)
line bundle of $\tfrac12$-densities on~$\MD$, into~$\BB^\DD$. It~turns out that
the image of this lifting consists precisely of the sections of $\BB^\DD$ which
are ``covariantly constant'' along~$\DD$. Namely, for any $\zeta\in\DD$ one can
define a mapping $\nabla_\zeta$ on $\BB^\DD$ as follows: if~$\nu$ is a
\mhD-density, then
\begin{equation}  (\nabla_\zeta \nu)_m (\eta_\sharp) := \zeta(\nu(\eta))|_m
\qquad \forall m\in \Omg,   \label{tag:HDC}  \end{equation}
where $\eta_\sharp$ is an arbitrary frame in $\DD_m$ and
$\eta=(\eta_1,\dots,\eta_n)$, where $\eta_j$ are $n$ linearly independent
locally Hamiltonian vector fields on $\Omg$ which span $\DD$ in a neighbourhood
of $m$ and such that $\eta|_m=\eta_\sharp$ (such vector fields exist because
$\DD$ is a polarization, cf.~(\ref{tag:REALPOL})). It~is not difficult to
verify that $\nabla_\zeta\nu$ is independent of the choice of~$\eta$, and that
$\nabla$ satisfies the axioms (\ref{tag:QFEa}) and~(\ref{tag:QFEb}), and is
thus a well-defined partial connection on~$\BB^\DD$. (The term ``partial''
refers to the fact that it is defined for $\zeta\in\DD$ only.)
From (\ref{tag:HDC}) it also follows that $\nabla$ is {\sl flat,\/}~i.e.
$$ \nabla_\xi\nabla_\zeta-\nabla_\zeta\nabla_\xi=\nabla_{[\xi,\zeta]}
\qquad \forall \xi,\zeta\in\DD.  $$
Now it can be proved that a \mhD-density $\nu$ on $\Omg$ is a lift of a
$\frac12$-density $\phi$ on $\MD$, i.e.~$\nu=\tilde\phi$, if and only~if
$$ \nabla_\zeta\nu =0 \qquad\forall \zeta\in\DD,  $$
i.e.~if and only if $\nu$ is covariantly constant along~$\DD$.

Coming back to our quantization business, consider now the tensor product
\begin{equation}  QB := L\otimes \BB^\DD  \label{tag:QBU}  \end{equation}
(the {\sl quantum bundle\/}) with the (partial) connection given~by
\begin{equation}  \nabla_\zeta(s\otimes\nu)=\nabla_\zeta s\otimes\nu +
s\otimes\nabla_\zeta\nu \qquad (\zeta\in\DD, s\in\Gamma(L),
\nu\in\Gamma(\BB^\DD)).  \label{tag:HDCC}  \end{equation}
Collecting all the ingredients above, it transpires that for any two sections
$\phi=s\otimes\nu$ and $\psi=r\otimes\mu$ of $QB$ which are covariantly
constant along~$\DD$ (i.e.~$\nabla_\zeta\phi=\nabla_\zeta\psi=0$, \;
$\forall\zeta\in\DD$), we can unambiguously define a half-density $(\phi,\psi)$
on $\MD$ by the formula
$$ (\phi,\psi)_{\pi(m)}(\pi_*\xi) := (s,r)_m \overline{\nu_m(\zeta)}
\mu_m(\zeta) |\epsilon_\omega (\zeta,\xi)|,  $$
where $(\zeta,\xi)$ is an arbitrary basis of $T_m \Omg$ such that $(\zeta)$ is
a basis of~$\DD_m$, and
\begin{equation}  \epsilon_\omega = \frac{(-1)^{n(n-1)/2}}{n!} \, \omega^n
\label{tag:HDEE}  \end{equation}
is the symplectic volume on~$\Omg$. Now~introduce the Hilbert space
$$ \HH=\text{the completion of } \Big\{ \psi\in\Gamma(QB):\,\nabla_\zeta \psi=0
\ \forall\zeta\in\DD \text{ and } \int_{\MD} (\psi,\psi) <\infty \Big\}   $$
of all square-integrable sections of $QB$ covariantly constant along~$\DD$,
with the obvious scalar product.

Finally, for a vector field $\zeta$ on~$\Omg$, let $\rho_t=\exp(t\zeta)$ be
again the associated flow of diffeomorphisms of~$\Omg$. The derived map
$\rho_{t*}$ on the tangent vectors defines a flow $\tilde\rho_t$
on~$\FF^n \Omg$:
$$ \tilde\rho_t (m,(\xi_j)) := (\rho_t m,(\rho_{t*}\xi_j)).  $$
One~can prove that~if
\begin{equation}  [\zeta,\DD]\subset\DD \qquad
(\text{i.e.~} [\zeta,\eta]\in\DD\ \forall\eta\in\DD)  \label{tag:HDD}
\end{equation}
then $\tilde\rho_t$ maps the subbundle $\FF^n\DD\subset\FF^n \Omg$ into itself,
and we can therefore define a lift $\tilde\zeta$ of $\zeta$ to $\FF^n\DD$ by
the recipe
$$ \tilde\zeta (m,(\xi)) := \frac d{dt} \tilde\rho_t(m,(\xi)) \Big|_{t=0}.  $$
Now if $\nu$ is a \mhD-density then it is a function on $\FF^n\DD$, hence we
can apply $\tilde\zeta$ to~it, and the result $\tilde\zeta\nu:=\LL_\zeta\nu$
will again be a \mhD-density. Further, $\LL_\zeta\nu$ is linear in~$\nu$;
\begin{equation}  \LL_\zeta(g\nu) = g\LL_\zeta\nu+(\zeta g)\nu;
\label{tag:LIEA}  \end{equation}
if $\eta$ is another vector field for which $[\eta,\DD]\subset\DD$, then
\begin{equation}  \LL_\zeta\LL_\eta-\LL_\eta\LL_\zeta = \LL_{[\zeta,\eta]};
\label{tag:LIEB}  \end{equation}
and if $\zeta$ is a locally Hamiltonian vector field in~$\DD$, then
$\LL_\zeta\nu=\nabla_\zeta\nu$ coincides with the partial connection
$\nabla_\zeta$ constructed above.

Now we are ready to define (at~last!) the quantum operators. Namely, if
$f:\Omg\to\RR$ is a smooth function whose Hamiltonian vector field $X_f$
satisfies~(\ref{tag:HDD}),~i.e.
\begin{equation}  [X_f,\DD]\subset\DD,   \label{tag:HDE}  \end{equation}
then the quantum operator $Q_f$ is defined on sections of~$QB$ as follows:
\begin{equation}  Q_f(s\otimes\nu) := \Big( -\frac{ih}{2\pi} \nabla_{X_f} s +
fs\Big) \otimes\nu + s\otimes\Big(-\frac{ih}{2\pi} \LL_{X_f} \nu \Big).
\label{tag:HDF}  \end{equation}
From the properties of $\LL$ and $\nabla$ it transpires that if $s\otimes\nu$
is covariantly constant along $\DD$ then so is~$Q_f(s\otimes\nu)$, and so $Q_f$
gives rise to a well-defined operator (denoted again by~$Q_f$) on the Hilbert
space~$\HH$ introduced above; it~can be shown that if $X_f$ is complete then
$Q_f$ is (essentially) self-adjoint.

The space of all real functions $f\in C^\infty(\Omg)$ satisfying
(\ref{tag:HDE})~is, by~definition, the space $\Obs$ of quantizable observables.

Unfortunately, it turns out that, no matter how elegant, the quantization
procedure described in this section gives sometimes incorrect answers: namely,
for the one-dimensional harmonic oscillator (corresponding to the observable
$f=\frac12(p^2+q^2)$ on the phase space $\Omg=\RR^2$ with the usual symplectic
form $\omega=dp\wedge dq$), one has first of all to modify the whole procedure
further by allowing ``distribution valued'' sections\footnote{This time the
distributions \underbar{are} those of L.~Schwartz (not subbundles of~$T\Omg$).}
of~$QB$ (see~\S\ref{sec261} below), and even then the energy levels come out as
$nh/2\pi$, $n=1,2,\dots$, instead of the correct answer $(n-\frac12)h/2\pi$.
It~turns out that the reason for this failure is the use of half-densities
above instead of the so-called half-forms; in~order to describe how the
situation can still be saved, we need to introduce complex tangent spaces and
complex polarizations. We~therefore proceed to describe this extended setup
in the next subsection, and then describe the necessary modifications
in~\S\ref{sec24}.\footnote{Another reason for allowing complex polarizations is
that there are symplectic manifolds on which no real polarizations exist ---
for instance, the sphere $\Bbb S^2$.}

\subsection{Complex polarizations}\label{sec23}
From now on, we start using complex objects such as the complexified tangent
bundle $T\Omg^\CC$, complex vector fields $\xi\in \bfrakX(\Omg)^\CC$, etc., and
the bar $\overline{\phantom X}$ will denote complex conjugation. A~{\sl complex
polarization\/} $\PP$ on the manifold $\Omg$ is a complex distribution on
$\Omg$ such that \begin{enumerate}
\item[(i)] $\PP$ is involutive (i.e.~$X,Y\in\PP\implies[X,Y]\in\PP$)
\item[(ii)] $\PP$ is Lagrangian (i.e.~$\dim_\CC\PP=n\equiv\tfrac12\dim_\RR\Omg$
and $\omega(X,Y)=0$ $\forall X,Y\in\PP$)
\item[(iii)] $\dim_\CC \PP_m\cap\overline\PP_m=:k$ is constant on~$\Omg$
(i.e.~independent of~$m$)
\item[(iv)] $\PP+\overline\PP$ is involutive.  \end{enumerate}
Again, one can prove an alternative characterization of complex polarizations
along the lines of~(\ref{tag:REALPOL}): namely, a~complex distribution $\PP$ on
$\Omg$ is a complex polarization if and only if $\forall m_0\in \Omg$ there is
a neighbourhood $U$ of $m_0$ and $n$ independent complex $C^\infty$ functions
$z_1,\dots,z_n$ on $U$ such that
\begin{equation}   \begin{array}{rl}
\text{(i) }&\text{$\forall m\in U$, $\PP_m$ is spanned (over~$\CC$) by the
Hamiltonian vector fields}\\
&\text{$X_{z_1}|_m,\dots,X_{z_n}|_m$;} \\
\text{(ii) }&\text{$\{z_j,z_k\}=0$ on $U$ $\forall j,k=1,\dots,n$;} \\
\text{(iii) }&\text{$\dim_\CC \PP_m\cap\overline\PP_m=:k$ is constant on~$\Omg$
(i.e.~independent of~$m$ and~$U$);} \\
\text{(iv) }&\text{the functions $z_1,\dots,z_k$ are real and $\forall m\in U,$
$\PP_m\cap\overline\PP_m$ is spanned} \\
&\text{by $X_{z_1}|_m,\dots,X_{z_k}|_m$.}
\end{array}   \label{tag:CPLXPOL}  \end{equation}
To~each complex polarization there are associated two \underbar{real}
involutive (and, hence, integrable) distributions $\DD,\EE$ on~$\Omg$~by
\begin{alignat*} 2
\DD &= \PP\cap\overline\PP\cap T\Omg \qquad &&\text{(so
$\DD^\CC=\PP\cap\overline\PP$, $\dim_\RR\DD=k$)} \\
\EE &= (\PP+\overline\PP)\cap T\Omg \qquad &&\text{(so
$\EE^\CC=\PP+\overline\PP$, $\dim_\RR\EE=2n-k$)}.  \end{alignat*}
One~has $\EE=\DD^\perp$, $\DD=\EE^\perp$ (the orthogonal complements with
respect to~$\omega$), so~that, in~particular, $X_f\in\EE\iff f$ is constant
along~$\DD$ (i.e.~$\xi f=0$ $\forall\xi\in\DD$), and similarly $X_f\in\DD\iff
f$ is constant along~$\EE$.

A~complex polarization is called {\sl admissible\/} if the space of leaves
$\MD$ admits a structure of a manifold such that $\pi:\Omg\to\MD$ is a
submersion. In~that case, $\tilde\EE:=\pi_*\EE$ defines a real integrable
distribution of dimension $2(n-k)$ on~$\MD$, and using the Newlander-Nirenberg
theorem one can show that the mapping $\Cal J: T_x\Bbb L\to T_x\Bbb L$ defined
on each leaf $\Bbb L$ of $\tilde\EE$ in $\MD$~by
$$ \Cal J(\pi_*\operatorname{Re}w) = \pi_*\operatorname{Im}w  $$
is~an integrable complex structure on $\Bbb L$ and if $X_{z_1},\dots,X_{z_k}$
are local Hamiltonian vector fields as in~(\ref{tag:CPLXPOL}) then the
functions $z_{k+1},\dots,z_n$ form, when restricted to~$\Bbb L$, a~local system
of complex coordinates which makes $\Bbb L$ a complex manifold. In~particular,
if $z$ is a complex function on an open set $U\subset \Omg$, then $X_z\in\PP$
if and only if locally $z=\tilde z\circ\pi$ where $\tilde z:\pi^{-1}(U)\subset
\MD\to\CC$ is holomorphic when restricted to any leaf of~$\tilde\EE$.

Throughout the rest of this section, unless explicitly stated otherwise, we
will consider only admissible complex polarizations.

Let~us now proceed to define the quantum Hilbert space $\HH$ and the quantum
operators $Q_f$ in this new setting. For real polarizations $\DD$, we~did this
by identifying functions on $\MD$ with sections on $\Omg$ covariantly constant
along~$\DD$, and then solving the problem of integration by lifting the
half-densities on $\MD$ to \mhD-densities on~$\Omg$. For~complex polarizations,
the ``quotient'' $\Omg/\PP$ does not make sense; and if we use $\MD$ instead,
then, since $\dim\MD$ can be smaller than $n$ in general, the passage from
half-densities on $\MD$ to ``\mhD-densities'' on $\Omg$ breaks down.
What we do is, then, that we trust our good luck and
just carry out the final quantization procedure as described for real
polarizations, and see if it works --- and it does!

Let~us start by defining $\FnPC$ to be the bundle of all complex frames
of~$\PP$.\footnote{The superscript $\CC$ is just to remind us that this is a
complex object; there is no such thing as~$\FF^n\PP^\RR$!} There is a natural
action of $GL(n,\CC)$, written as $(\eta)\mapsto(\eta)\cdot g$, on the fibers
of~$\FnPC$, and we define a \mhP-density $\nu$ on $\Omg$ as a complex function
on $\FnPC$ such that
\begin{equation}  \nu_m((\eta)\cdot g) = \nu_m((\eta))\cdot |\det g\,|^{-1/2}
\qquad \forall (\eta)\in\FnPC,\ \forall g\in GL(n,\CC),  \label{tag:CPA}
\end{equation}
and denote the (complex line) bundle of all \mhP-densities on $\Omg$ by
$\BB^\PP$. Next we define $\nabla_\zeta\nu$, for~$\zeta\in\PP$,~by
\begin{equation}  (\nabla_\zeta \nu)_m ((\eta)|_m) = \frac{\zeta[\nu((\eta))
\cdot |\epsilon_{\omega,k}(\eta_{k+1},\dots,\eta_n,\overline\eta_{k+1},\dots,
\overline\eta_n)|^{1/4}]}
{|\epsilon_{\omega,k}(\eta_{k+1},\dots,\eta_n,\overline\eta_{k+1},\dots,
\overline\eta_n)|^{1/4}} \ \bigg|_m  \label{tag:CPEPS}  \end{equation}
where $(\eta_1,\dots,\eta_n)$ are any vector fields which span $\PP$ in a
neighbourhood of $m$ such that $\eta_1,\dots,\eta_k$ are real Hamiltonian
vector fields spanning~$\DD$, and $\epsilon_{\omega,k}$ is the $2(n-k)$-form
defined~by
\begin{equation}  \epsilon_{\omega,k} = \frac{(-1)^{(n-k)(n-k-1)/2}}{(n-k)!}
\,\omega^{n-k}  \label{tag:CPX}  \end{equation}
(so that, in particular, $\epsilon_{\omega,0}=\epsilon_\omega$ is the volume
form~(\ref{tag:HDEE})). It~again turns out that $\nabla_\zeta\nu$ is a
\mhP-density if $\nu$~is\footnote{The factor $|\epsilon_{\omega,k}|^{1/4}$ in
(\ref{tag:CPEPS}) needs some explanation. The~reason for it is that if we
defined $\nabla_\zeta\nu$ simply by the same formula (\ref{tag:HDC}) as for the
real polarizations, then $\nabla_\zeta\nu$ might fail to be a \mhP-density:
it~would have satisfied the relation (\ref{tag:CPA}) only if there were no
absolute value around $\det g$ there. (That~is, if $(\hat\eta)=(\eta)\cdot g$
is another frame satisfying the conditions imposed on~$\eta$, then we have
$\zeta(\det g)=0$, which need not imply $\zeta|\!\det g|=0$.) This difficulty
does not arise for real polarizations (since then $\det g$ is locally of
constant sign), nor for the half-forms discussed in the next subsection (where
there is no absolute value around the determinant). On~the other hand,
(\ref{tag:CPEPS})~has the advantage that it defines $\nabla_\zeta$ consistently
not only for $\zeta\in\PP$, but even for $\zeta\in \EE^\CC=\PP+\overline\PP$;
however, we will not need this refinement in the sequel. \endgraf It~should be
noted that the correction factor $|\epsilon_{\omega,k}|^{1/4}$ is such that the
combination
$\nu(\eta)\cdot|\epsilon_{\omega,k}(\eta_{k+1},\dots,\overline\eta_n)|^{1/4}$
depends only on the vectors $\eta_1,\dots,\eta_k$ spanning~$\DD$, and defines
thus a \mhD-density on~$\Omg$.}, and defines thus a flat partial connection
on~$\BB^\PP$. The~formula (\ref{tag:HDCC}) then defines a partial connection on
the quantum bundle $QB:=L\otimes\BB^\PP$ ($L$~being, as before, the prequantum
bundle from~\S\ref{sec21}). Now~if $\phi=s\otimes\nu$, $\psi=r\otimes\mu$ are
two arbitrary (smooth) sections of~$QB$, then we~set
\begin{equation}  \begin{aligned}
& (\phi,\psi)_m (\pi_*(\zeta_{k+1},\dots,\zeta_n,\xi_1,\dots,\xi_n)) :=
(s,r)_m\,\overline{\nu_m(\zeta_1,\dots,\zeta_n)}\,\mu_m(\zeta_1,\dots,\zeta_n)
\cdot{} \\ & \qquad\qquad\qquad\qquad {} \cdot
|\epsilon_{\omega,k}(\zeta_{k+1},\dots,\zeta_n,\overline\zeta_{k+1},\dots,
\overline\zeta_n)|^{1/2} \cdot |\epsilon_\omega(\zeta_1,\dots,\zeta_n,\xi_1,
\dots,\xi_n)|  \end{aligned}   \label{tag:CPB}  \end{equation}
where $\zeta_1,\dots,\zeta_n,\xi_1,\dots,\xi_n$ is any basis of $T_m\Omg^\CC$
such that $\zeta_1,\dots,\zeta_k$ is a basis of $\DD_m^\CC=\PP_m\cap\overline
\PP_m$ and $\zeta_1,\dots,\zeta_n$ is a basis of~$\PP_m$, and $\epsilon_
{\omega,k}$ and $\epsilon_\omega$ are the forms given by (\ref{tag:CPX})
and~(\ref{tag:HDEE}), respectively. This time not every basis of $T_{\pi(m)}
(\MD)^\CC$ arises as $\pi_*(\zeta_{k+1},\dots,\zeta_n,\xi_1,\dots,\xi_n)$ with
$\zeta,\xi$ as above, but it is easily seen that the values of $(\phi,\psi)_m$
on different frames are related in the correct way and thus $(\phi,\psi)_m$
extends to define consistently a unique density on $\FF^{2n-k}_{\pi(m)}(\MD)^
\CC$ (the fiber at $\pi(m)$ of the bundle of all complex $(2n-k)$-frames
on~$\MD$). From the proof of the Frobenius theorem one can show that for any
local Hamiltonian vector fields $X_{z_1},\dots,X_{z_n}$ as in
(\ref{tag:CPLXPOL}) there exist vector fields $Y_1,\dots,Y_k$ (possibly on a
subneighbourhood of~$U$) such that
$\pi_*(X_{z_{k+1}},\dots,X_{z_n},X_{\overline z_{k+1}},\dots,X_{\overline z_n},
Y_1,\dots,Y_k)$ is a basis of $T_{\pi(m)}(\MD)^\CC$ which depends only on
$\pi(m)$, and $\epsilon_\omega(X_{z_1},\dots,X_{z_n},X_{\overline z_{k+1}},
\dots,X_{\overline z_n},Y_1,\dots,Y_k)$ is a function constant on the leaves
of~$\DD$. Taking these vector fields for the $\zeta_j$ and $\xi_j$
in~(\ref{tag:CPB}), it can be proved in the same way as for the real
polarizations that
$$ \eta(\phi,\psi)_m(\pi_*(X_z,X_{\overline z},Y)) =
(\nabla_{\overline\eta} \phi,\psi)_m(\pi_*(X_z,X_{\overline z},Y)) +
(\phi,\nabla_\eta \psi)_m (\pi_*(X_z,X_{\overline z},Y))  $$
for any $\eta\in\DD_m$. Thus, in~particular, if $\phi,\psi$ are covariantly
constant along~$\DD$, then $(\phi,\psi)_m$ depends only on $\pi(m)$ and defines
thus a density on~$\MD$.

We~can therefore define, as before, the Hilbert space
\begin{equation}  \HH = \text{the completion of } \Big\{
\psi\in\Gamma(QB):\,\nabla_\zeta\psi= 0\ \forall\zeta\in\PP\text{ and }
\int_{\MD}(\psi,\psi)<\infty\Big\}\label{tag:CPC} \end{equation}
of square-integrable sections of $QB$ covariantly constant along~$\PP$ (with
the obvious inner product).

Finally, if $\zeta$ is a real vector field on~$\Omg$ satisfying $[\zeta,\PP]
\subset\PP$, with the associated flow $\rho_t$, and $\nu$ a \mhP-density
on~$\Omg$, then we may again define $\LL_\zeta\nu$~by
\begin{equation}  (\LL_\zeta\nu)_m(\eta) = \frac d{dt} \nu_{\rho_t m}
(\tilde\rho_t(\eta))
\Big|_{t=0}, \qquad (\eta\in\FnPC)  \label{tag:CPL}  \end{equation}
and show that $\LL_\zeta\nu$ is again a \mhP-density and that $\LL_\zeta$ has
all the properties of a ``flat partial Lie derivative''
((\ref{tag:LIEA})~and~(\ref{tag:LIEB})) and that $\LL_{X_f}=\nabla_{X_f}$
whenever $f$ is a real function for which $X_f\in\PP$ (hence $X_f\in\DD$). Now
the operator
\begin{equation}  Q_f (s\otimes\nu) := \Big( -\frac{ih}{2\pi} \nabla_{X_f} s +
fs\Big) \otimes\nu + s\otimes\Big(-\frac{ih}{2\pi} \LL_{X_f} \nu \Big),
\label{tag:CPD}  \end{equation}
defined for any real function $f$ such that
\begin{equation}  [X_f,\PP]\subset\PP,  \label{tag:POLPRES}  \end{equation}
maps sections covariantly constant along $\PP$ again into such sections, and
thus defines an operator on~$\HH$, which can be shown to be self-adjoint if
$X_f$ is complete.

Having extended the method of \S\ref{sec22} to complex polarizations, we now
describe the modification needed to obtain the correct energy levels for the
harmonic oscillator: the metalinear correction.

\subsection{Half-forms and the metalinear correction} \label{sec24}
What this correction amounts to is throwing away the absolute value in the
formula~(\ref{tag:CPA}); that~is, to~pass from half-densities to {\sl
half-forms.\/} To~do that we obviously need to have the square root of the
determinant in (\ref{tag:CPA}) defined in a consistent manner; this is achieved
by passing from $GL(n,\CC)$ to the {\sl metalinear group\/} $ML(n,\CC)$, and
from the frame bundle $\FnPC$ to the bundle $\hatFnPC$ of {\sl metalinear
$\PP$-frames.\/}

The~group $ML(n,\CC)$ consists, by definition, of all pairs $(g,z)\in GL(n,\CC)
\times\CC^\times$ satisfying
$$ z^2=\det g $$
with the group law
$$ (g_1,z_1)\cdot(g_2,z_2):=(g_1 g_2,z_1 z_2).  $$
We~will denote by $p$ and $\lambda$ the canonical projections
\begin{alignat*} 2
p &: ML(n,\CC)\to GL(n,\CC) &&:\quad (g,z)\mapsto g, \\
\lambda &: ML(n,\CC)\to \CC^\times &&:\quad (g,z)\mapsto z, \end{alignat*}
respectively. To~define the bundle $\hatFnPC$, suppose that $\{U_\alpha\}$ is
a trivializing cover of $\FnPC$ (i.e.~$U_\alpha$ are local patches on $\Omg$
such that the restrictions $\FnPC|U_\alpha$ are isomorphic to Cartesian
products $U_\alpha\times GL(n,\CC)$) with the corresponding transition
functions $g_{\alpha\beta}:U_\alpha\cap U_\beta\to GL(n,\CC)$. Suppose
furthermore that there exist (continuous) lifts $\tilde g_{\alpha\beta}:
U_\alpha\cap U_\beta\to ML(n,\CC)$ such that $p\tilde g_{\alpha \beta}=
g_{\alpha\beta}$ and that the cocycle conditions $\tilde  g_{\alpha \beta}
\tilde g_{\beta\gamma} = \tilde g_{\alpha\gamma}$ are satisfied. Then the cover
$\{U_\alpha,\tilde g_{\alpha\beta}\}$ defines the desired bundle $\hatFnPC$.
It~turns out that such lifts $\tilde g_{\alpha\beta}$ exist (possibly after
refining the cover $\{U_\alpha\}$ if necessary) if and only if the cohomology
class determined by the bundle $\FnPC$ in $H^2(\Omg,\ZZ_2)$ vanishes; from now
on, we will assume that this condition is satisfied.

The~mapping $\tilde p:\hatFnPC\to\FnPC$, obtained upon applying $p$ in each
fiber, yields then a 2-to-1 covering of $\FnPC$ by $\hatFnPC$.

A~{\sl \mhP-form\/} on $\Omg$ is, by definition, a function
$\tilde\nu:\hatFnPC \to\CC$ satisfying
$$ \tilde\nu_m(\tilde\xi\cdot\tilde g) = \tilde\nu_m(\tilde\xi) \cdot
\lambda(\tilde g)^{-1} \qquad \forall \tilde\xi\in\hatFnPC,
\forall \tilde g\in ML(n,\CC).  $$
The~complex line bundle of all \mhP-forms will be denoted by $\tBP$.

Next we define the (partial) connection $\nabla$ on~$\tBP$. Let~$\eta_1,\dots,
\eta_n$ be local Hamiltonian vector fields spanning $\PP$ in a neighbourhood of
a point $m_0\in \Omg$ (cf.~(\ref{tag:CPLXPOL})). Since $\tilde p$ is a local
homeomorphism, there exists a local lifting $(\tilde\eta_1,\dots,\tilde\eta_n)
\in\hatFnPC$ (possibly defined on a smaller neighbourhood of~$m_0$) such that
$\tilde p(\tilde \eta_j)=\eta_j$. We~can also arrange that $(\tilde\eta_1,
\dots,\tilde\eta_n)|_{m_0}$ coincides with any given metaframe $\tilde f_0\in
\hat\FF^n_{m_0}\PP^\CC$. For $\zeta\in\PP$, we then define
$$ (\nabla_\zeta\tilde\nu)_{m_0}(\tilde f_0) := \zeta \tilde\nu
(\tilde\eta_1,\dots,\tilde\eta_n) \big|_{m_0}.  $$
One~checks as usual that this definition is consistent (i.e.~independent of the
choice of the Hamiltonian metaframe $\tilde\eta$ satisfying $\tilde\eta|_{m_0}
=\tilde f_0$) and defines again a \mhP-form on~$\Omg$; further, the resulting
map $\nabla$ is again a flat partial connection on~$\tBP$. Denoting by $QB$ the
tensor product (quantum bundle)
$$ QB:= L \otimes \tBP  $$
(with $L$ the prequantization bundle from~\S\ref{sec21}), we then have the
corresponding partial connection (\ref{tag:HDCC}) in~$QB$.

For~arbitrary two sections $\phi=s\otimes\tilde\nu$ and
$\psi=r\otimes\tilde\mu$ of~$QB$, $m\in \Omg$ and
$\tilde f\in\hat\FF^n_m\PP^\CC$ a metaframe at~$m$, denote
$(\zeta_1,\dots,\zeta_n)=\tilde p(\tilde f)$ and
choose $\xi_1,\dots,\xi_n\in T_m \Omg^\CC$ such that
$\zeta_1,\dots,\zeta_n,\xi_1,\dots,\xi_n$ is a basis of~$T_m \Omg^\CC$.
Assume that $\zeta_1,\dots,\zeta_k$ is a basis of $\DD_m^\CC$. Then a function
$(\phi,\psi)_m$ can be defined on $\FF^{2n-k}_{\pi(m)}(\MD)^\CC$~by
\begin{equation}  \begin{aligned}
& (\phi,\psi)_m (\pi_*(\zeta_{k+1},\dots,\zeta_n,\xi_1,\dots,\xi_n)) :=
(s,r)_m \, \overline{\tilde\nu_m(\tilde f)} \, \tilde\mu_m(\tilde f)
\cdot{} \\ & \qquad\qquad\qquad\qquad {} \cdot
|\epsilon_{\omega,k}(\zeta_{k+1},\dots,\zeta_n,\overline\zeta_{k+1},\dots,
\overline\zeta_n)|^{1/2} \cdot |\epsilon_\omega(\zeta_1,\dots,\zeta_n,\xi_1,
\dots,\xi_n)| .  \end{aligned}   \label{tag:MLA}  \end{equation}
Although $(\phi,\psi)_m$ is again defined only on a certain subset of
$\FF^{2n-k}_{\pi(m)}(\MD)^\CC$, one can check as before that it extends
consistently to a (unique) density on $\FF^{2n-k}_{\pi(m)}(\MD)^\CC$, and,
further, if $\phi$ and $\psi$ are covariantly constant along $\PP$ then
$(\phi,\psi)_m$ depends only on $\pi(m)$, and thus defines a (unique) density
on~$\MD$.

Finally, if $\zeta$ is a real vector field on $\Omg$ preserving $\PP$
(i.e.~$[\zeta,\PP]\subset\PP$), then the associated flow $\rho_t$ (which
satisfies $\tilde\rho_{t*}\PP_m\subset\PP_{\rho_t m}$) induces a flow
$\tilde\rho_t$ on $\PP$-frames which, for $t$ small enough, lifts uniquely to
a flow $\Tilde{\Tilde\rho}_t$ on~the metaframes such that $\tilde p\Tilde
{\Tilde\rho}_t=\tilde\rho_t\tilde p$. Using this action we define
\begin{equation}  (\LL_\zeta\tilde\nu) (\tilde f) := \frac d{dt}
\tilde\nu_{\rho_t m} (\Tilde{\Tilde\rho}_t \tilde f) \Big|_{t=0}, \qquad
\tilde f \in \hat\FF^n_m\PP^\CC.  \label{tag:MLB}  \end{equation}
As~before, it is easily seen that $\LL_\zeta\tilde\nu$ is again a \mhP-form,
for any \mhP-form~$\tilde\nu$, that $\LL_\zeta$ satisfies the axioms
(\ref{tag:LIEA}) and (\ref{tag:LIEB}) of a ``flat partial Lie derivative'', and
that $\LL_{X_f}=\nabla_{X_f}$ if $f$ is a real function with $X_f\in\PP$.

Introducing the Hilbert space $\HH$ as before,
$$ \HH = \text{the completion of } \Big\{ \psi\in\Gamma(QB):\,\nabla_\zeta\psi=
0\ \forall\zeta\in\PP\text{ and }\int_{\MD}(\psi,\psi)<\infty \Big\} ,   $$
a~straightforward modification of the corresponding arguments for
\mhP-densities shows that the operators defined by~(\ref{tag:CPD}),~i.e.
\begin{equation}  Q_f (s\otimes\nu) := \Big( -\frac{ih}{2\pi} \nabla_{X_f} s +
fs\Big) \otimes\nu + s\otimes\Big(-\frac{ih}{2\pi} \LL_{X_f} \nu \Big)
\label{tag:MLC}  \end{equation}
(but now with the Lie derivative (\ref{tag:CPL}) replaced
by~(\ref{tag:MLB})~etc.!), for $f:\Omg\to\RR$ such that (\ref{tag:POLPRES})
holds, are densely defined operators of $\HH$ into itself; and if $X_f$ is
complete, they are self-adjoint.

We~have thus arrived at the final recipe of the original geometric quantization
of Kostant and Souriau: that~is, starting with a phase space --- a~symplectic
manifold $(\Omg,\omega)$ --- satisfying the integrality condition:
$$ h^{-1}[\omega] \text{ is an integral class in } H^2(\Omg,\RR),  $$
and with a complex polarization $\PP$ on $\Omg$ satisfying the condition for
the existence of the metaplectic structure:
$$ \text{the class of $\FnPC$ in $H^2(\Omg,\ZZ_2)$ vanishes,}  $$
we have constructed the Hilbert space $\HH$ as (the completion of) the space of
all sections of the quantum bundle $QB=L\otimes\tBP$ which are covariantly
constant along $\PP$ and square-integrable over~$\MD$; and for a function~$f$
belonging to the space
\begin{equation}  \Obs=\{f:\Omg\to\RR; \, [X_f,\PP]\subset\PP \}
\label{tag:MLO} \end{equation}
(the {\sl space of quantizable observables\/}) we have defined by
(\ref{tag:MLC}) the corresponding quantum operator $Q_f$ on~$\HH$, which is
self-adjoint if the Hamiltonian field $X_f$ of $f$ is complete, and such that
the correspondence $f\mapsto Q_f$ satisfies the axioms (Q1)~--~(Q5) we have set
ourselves in the beginning.\footnote{In~(Q4), one of course takes the
polarizations on the two manifolds which correspond to each other under the
given diffeomorphism.}

\subsection{Blattner-Kostant-Sternberg pairing} \label{sec25}
The~space (\ref{tag:MLO}) of quantizable observables is often rather small: for
instance, for $\Omg=\RR^{2n}$ (with the standard symplectic form) and the
vertical polarization $\partial/\partial p_1,\dots,\partial/\partial p_n$, the
space $\Obs$ essentially coincides with functions at most linear in~$p$, thus
excluding, for instance, the kinetic energy~$\frac12\|\bp\|^2$. There is a
method of extending the quantization map $Q$ to a larger space of
functions\footnote{However, on the extended domain $Q$ does in general no
longer satisfy the axiom~(Q3); see the discussion in \S\ref{sec27} below.}
so~that $Q_f$ is still given by (\ref{tag:MLC}) if $f$ satisfies
(\ref{tag:POLPRES}), while giving the correct answer
$Q_f=-\frac{h^2}{8\pi}\Delta$ for the kinetic energy
$f(\bp,\bq)=\frac12\|\bp\|^2$. The method is based on a pairing of half-forms,
due to Blattner, Kostant and Sternberg \cite{bib:BlattPSPM}, which we now
proceed to describe.

Suppose $\PP$ and $\GG$ are two (complex) polarizations for which there exist
two real foliations $\hatDD$ and $\hatEE$ (of constant dimensions $k$ and
$2n-k$, respectively) such that
\begin{equation}  \begin{aligned}
& \overline\PP\cap\GG = \hatDD^\CC, \\
& \overline\PP+\GG = \hatEE^\CC,  \\
& \Omg/\hatDD\text{ has a manifold structure and }\pi:\Omg\to \Omg/\hatDD\text{
is a submersion.}   \end{aligned}  \label{tag:TUA}  \end{equation}
Pairs of polarizations satisfying the first and the third condition are called
{\sl regular\/}\footnote{This definition of regularity slightly differs from
the original one in \cite{bib:BlattLN}, where it is additionally required that
the Blattner obstruction (\ref{tag:BLOB}) vanish.}; if~in addition
$\hatDD=\{0\}$ (which implies that the second condition also holds, with
$\hatEE=T\Omg$), they are called {\sl transversal.\/} If~the polarizations
$\PP$ and $\GG$ are {\sl positive,\/} which means that
\begin{equation}  i\,\omega(x,\overline x)\ge0  \qquad\forall x\in\PP,
\label{tag:POSIT}  \end{equation}
and similarly for $\GG$, then $\overline\PP\cap\GG$ is automatically
involutive, so the first condition in (\ref{tag:TUA}) is equivalent to the
(weaker) property that $\overline\PP\cap\GG$ be of constant rank.

For $m\in \Omg$, choose a basis $\xi_1,\dots,\xi_n,\overline\xi'_{k+1},\dots,
\overline\xi'_n,t_1,\dots,t_k$ of $T_m \Omg^\CC$ such that $\xi_1,\dots,\xi_k$
span $\hatDD_m$, $\xi_1,\dots,\xi_n$ span $\PP_m$ and $\xi_1,\dots,\xi_k,
\xi'_{k+1},\dots,\xi'_n$ span~$\GG_m$. Now if $\phi=s\otimes\nu$ and
$\psi=r\otimes\mu$ are (local) sections of $L\otimes\tBP$ and $L\otimes\tilde
\BB^{\GG}$, respectively, then we can ``define'' a function on
$\FF^{2n-k}(\Omg/\hatDD)^\CC$~by
\begin{equation}  \begin{aligned}
& \spr\phi\psi _m (\pi_*(\xi_{k+1},\dots,\xi_n,\overline\xi'_{k+1},\dots,
\overline\xi'_n,t_1,\dots,t_k)) =
(s,r)_m \, \overline{\nu_m((\xi_1,\dots,\xi_n)\sptilde\,)} \cdot {} \\
& \hskip10em {}\cdot
\mu_m((\xi_1,\dots,\xi_k,\xi'_{k+1},\dots,\xi'_n)\sptilde\,)
\cdot {} \\
& \hskip10em {} \cdot     %h^{-n/2} \,
\sqrt{\epsilon_{\omega,k}
(\xi_{k+1},\dots,\xi_n,\overline\xi'_{k+1},\dots,\overline\xi'_n)} \cdot
|\epsilon_\omega(\xi_1,\dots,t_k)|.  \end{aligned}   \label{tag:TUB}
\end{equation}
(Here $\epsilon_{\omega,k}$ is given by~(\ref{tag:CPX})). Moreover, if~$\phi$
and $\psi$ are covariantly constant along $\PP$ and $\GG$, respectively, then
this expression is independent of the choice of $m$ in the fiber above
$\pi(m)$, and thus defines a density --- denoted $(\phi,\psi)_{\pi(m)}$ ---
on~$\Omg/\hatDD$. However, there are two problems with~(\ref{tag:TUB}): first,
we need to specify which metaframes $(\xi_1,\dots,\xi_n)\sptilde$ above
$(\xi_1, \dots,\xi_n)$ and $(\xi_1,\dots,\xi'_n)\sptilde$ above
$(\xi_1,\dots,\xi'_n)$ to choose; and, second, we must specify the choice of
the branch of the square root of $\epsilon_{\omega,k}$.

Both problems are solved by introducing the {\sl metaplectic frame bundle\/}
on~$\Omg$, which, basically, amounts to a recipe for choosing metalinear lifts
$\tBP$ of $\BB^\PP$ for all complex polarizations $\PP$ on $\Omg$ \underbar
{simultaneously}.\footnote{More precisely, for all positive complex
polarizations (see the definition below). In~other words, the choice of a
metaplectic frame bundle uniquely determines a metalinear frame bundle for each
positive complex polarization.}

\begin{remark*} On an abstract level, the basic idea behind  the half-form
pairing can be visualized as follows (Rawnsley~\cite{bib:RawnCMP}). Let
$\PP^\perp\subset T^*\Omg^\CC$ denote the bundle of one-forms vanishing
on~$\PP$; in~view of the Lagrangianity of~$\PP$, the mapping $\xi\mapsto\omega
(\xi,\cdot)$ is an isomorphism of $\PP$ onto~$\PP^\perp$. The~exterior power
$\bigwedge^n\PP^\perp=:K^\PP$ is a line bundle called the canonical bundle
of~$\PP$. If~the polarization $\PP$ is positive, then the Chern class of
$K^\PP$ is determined by~$\omega$, so~that $K^\PP$ and $K^\GG$ are isomorphic
for any two positive polarizations $\PP$ and~$\GG$. In~this case the bundle
$K^\PP\otimes\overline{K^\GG}$ is trivial, and a choice of trivialization will
yield the pairing. In~particular, if $\overline\PP\cap\GG=\{0\}$, then exterior
multiplication defines an isomorphism of $K^\PP\otimes\overline{K^\GG}$ with
$\bigwedge^{2n} T^*\Omg^\CC$, and the latter is trivialized by the volume
form~$\epsilon_\omega$; hence one can define $\spr\nu\mu$~by
$$ i^n \spr\nu\mu \epsilon_\omega = \mu\wedge\overline\nu, \qquad
\mu\in\Gamma(K^\PP), \nu\in\Gamma(K^\GG).  $$
If~$\overline\PP\cap\GG$ has only constant rank, then the positivity of $\PP$
and $\GG$ implies that the first two conditions in (\ref{tag:TUA}) hold, for
some real foliation $\hatDD$ and some real distribution (but not necessarily a
foliation) $\hatEE=\hatDD^\perp$ on~$\Omg$. Then $\omega$ induces a nonsingular
skew form $\omega_\hatDD$ on $\hatEE/\hatDD$, and $\PP$ and $\GG$ project to
Lagrangian subbundles $\PP/\hatDD$ and $\GG/\hatDD$ of $(\hatEE/\hatDD)^\CC$
such that $\overline{\PP/\hatDD}\cap(\GG/\hatDD)=\{0\}$. Thus $K^{\PP/\hatDD}$
and $K^{\GG/\hatDD}$ can be paired by exterior multiplication as above. To~lift
this pairing back to $K^\PP$ and $K^\GG$, consider $m\in \Omg$ and a frame
$\xi_1, \dots,\xi_n$ of $\PP_m$ such that $\xi_1,\dots,\xi_k$ is a frame
of~$\hatDD_m$. Then any $\nu\in K^\PP_m$ is of the form
$$ \nu = a\,\omega(\xi_1,\cdot\,)\wedge\omega(\xi_2,\cdot\,)\wedge\dots\wedge
\omega(\xi_n,\cdot\,)  $$
for some $a\in\CC$. The~projections $\tilde\xi_j$ of $\xi_j\in\hatDD_m^{\perp
\CC}$ onto $(\hatEE/\hatDD)_m^\CC$, $j=k+1,\dots,n$, then form a frame for
$(\PP/\hatDD)_m$, and we~set
$$ \tilde\nu(\tilde\xi_1,\dots,\tilde\xi_k) := a \,\omega_\hatDD(\tilde\xi
_{k+1},\cdot\,)\wedge \dots \wedge \omega_\hatDD(\tilde\xi_n,\cdot\,) \in
K_m^{\PP/\hatDD}.  $$
Projecting $\mu\in K^\GG_m$ in the same fashion, we then put $\spr\nu\mu_m:=
\spr{\tilde\nu}{\tilde\mu}_m$. Thus in any case we end up with a
$-2$-$\hatDD$-density on~$\Omg$, which defines, using the volume density
$|\epsilon_\omega|$, a 2-density on $T\Omg$ normal to $\hatDD$ (i.e.~vanishing
if any of its arguments is in~$\hatDD$). Thus if $\spr\nu\mu_m$ is covariantly
constant along the leaves, we can project down to a 2-density on
$T(\Omg/\hatDD)$. Now~if the Chern class of $K^\PP$ is even --- in~which case
$(\Omg,\omega)$ is called {\sl metaplectic\/} --- then the symplectic frame
bundle of $\Omg$ has a double covering, by~means of which one can canonically
construct a square root $Q^\PP$ of~$K^\PP$, for any positive
polarization~$\PP$. (Sections of $Q^\PP$ are called half-forms {\sl normal\/}
to~$\PP$.) Further, these square roots still have the property that
$Q^\PP\otimes\overline{Q^\GG}$ is trivial. Applying the ``square root'' to the
above construction, one thus ends up with a density on~$\Omg/\hatDD$.
Integrating this density gives a complex number, and we thus finally arrive at
the desired pairing
$$ \Gamma(Q^\PP) \times \overline{\Gamma(Q^\GG)} \to\CC.   $$

In~particular, choosing $\GG=\PP$ (i.e.~pairing a polarization with itself),
passing from $Q^\PP$ to the tensor product $L\otimes Q^\PP$ with the prequantum
bundle, and using again Lie differentiation to define a partial connection
along $\hatDD$ in the densities on $T\Omg$ normal to $\hatDD$, we can also
continue as before and recover in this way in an equivalent guise the Hilbert
space $\HH$ and the quantum operators~$Q_f$ from the preceding subsection(s).
\qed   \end{remark*}

\medskip

We~now give some details about the construction of the metaplectic frame
bundle. As~this is a somewhat technical matter, we will confine ourselves to
the simplest case of transversal polarizations, i.e.~such that (\ref{tag:TUA})
holds with $\hatDD=\{0\}$ (and, hence, $\hatEE=T\Omg$); the general case can be
found in~\cite{bib:SniaB}, Chapter~5, or~\cite{bib:BlattLN}. We~will also
assume throughout that the polarizations are positive,
i.e.~(\ref{tag:POSIT})~holds.

A~{\sl symplectic frame\/} at $m\in \Omg$ is an (ordered) basis
$(u_1,\dots,u_n, v_1,\dots,v_n)\equiv(u,v)$ of $T_m\Omg$ such that
$$ \omega(u_j,u_k)=\omega(v_j,v_k)=0,\qquad \omega(u_j,v_k)=\delta_{jk}.  $$
The~collection of all such frames forms a right principal $Sp(n,\RR)$ bundle
$\FwM$, the {\sl symplectic bundle\/}; here $Sp(n,\RR)$, the $n\times n$ {\sl
symplectic group,\/} consists of all $g\in GL(2n,\RR)$ which preserve~$\omega$
(i.e.~$\omega(g\xi,g\eta)=\omega(\xi,\eta)$). The~group $Sp(n,\RR)$ can be
realized as the subgroup of $2n\times 2n$ real matrices $g$ satisfying $g^tJg=
J$, where $J$ is the block matrix $\Big[\begin{matrix} 0 & -I\\I
&0\end{matrix}\Big]$. The fundamental group of $Sp(n,\RR)$ is infinite cyclic,
hence there exists a unique double cover $Mp(n,\RR)$, called the {\sl
metaplectic group.\/} We~denote by $p$ the covering homomorphism. The~{\sl
metaplectic frame bundle\/} $\tFwM$ is a right principal $Mp(n,\RR)$ bundle
over $\Omg$ together with a map $\tau: \tFwM\to\FwM$ such that
$\tau(\tilde\xi\cdot \tilde g)=\tau (\tilde\xi)\cdot p(\tilde g)$, for all
$\tilde\xi\in\tFwM$ and $\tilde g\in Mp(n,\RR)$. The~existence of $\tFwM$ is
equivalent to the characteristic class of $\FwM$ in $H^2(\Omg,\ZZ)$ being even
(cf.~the construction of the metalinear frame bundle~$\hatFnPC$).

A~{\sl positive Lagrangian frame\/} at $m\in \Omg$ is a frame $(w_1,\dots,w_n)
\equiv w\in T_m \Omg^\CC$ such that
\begin{equation}  \omega(w_j,w_k)=0 \qquad \forall j,k=1,\dots,n,
\label{tag:SMA}  \end{equation}
and
\begin{equation}  i\,\omega(w_j,\overline w_j)\ge0 \qquad \forall j=1,\dots,n.
\label{tag:SMB}  \end{equation}
The corresponding {\sl bundle of positive Lagrangian frames\/} is denoted
by~$\Lag$.

In~terms of a given symplectic frame $(u,v)$, a positive Lagrangian frame can
be uniquely expressed as
\begin{equation}  w=(u,v) \bigg[ \begin{matrix} U \\V \end{matrix} \bigg]
\label{tag:SMC}  \end{equation}
where $U,V$ are $n\times n$ matrices satisfying
\begin{equation}  \rank\bigg[\begin{matrix} U\\V\end{matrix}\bigg]=n,\qquad
U^tV=V^t U, \label{tag:SNA}  \end{equation}
in~view of (\ref{tag:SMA}), and
\begin{equation}  i(V^*U-UV^*)\text{ is positive semidefinite}  \label{tag:SNB}
\end{equation}
in~view of (\ref{tag:SMB}). This sets up a bijection between the set of all
positive Lagrangian frames at a point $m\in \Omg$ and the set $\Pi$ of all
matrices  $U,V$ satisfying (\ref{tag:SNA}) and~(\ref{tag:SNB}). The~action of
$Sp(n,\RR)$ on $\Pi$ by left matrix multiplication defines thus an action on
$\Lag$ and a positive Lagrangian frame $w$ at $m$ can be identified with the
function $w^\sharp:\FwM\to\Pi$ satisfying
$$ w^\sharp( (u,v)\cdot g) = g^{-1} w^\sharp (u,v)
\qquad \forall g\in Sp(n,\RR) $$
by the recipe
\begin{equation}  w= (u,v) w^\sharp(u,v).  \label{tag:SND}  \end{equation}

From (\ref{tag:SNA}) it follows that the matrix $C$ defined~by
$$ C:= U-iV  $$
is nonsingular, and that the matrix $W$ defined~by
$$ W=(U+iV)C^{-1}  $$
is symmetric ($W^t=W$). From (\ref{tag:SNB}) it then follows that $\|W\|\le1$,
i.e.~$W$ belongs to the closed unit ball
$$ \Bbb B := \{W\in\CC^{n\times n}: W^t=W,\,\|W\|\le 1\}  $$
of symmetric complex $n\times n$ matrices.

Since
\begin{equation}  U=\frac{(I+W)C}2, \qquad V=\frac{i(I-W)C}2,  \label{tag:SNC}
\end{equation}
the mapping $\bigg[\begin{matrix} U\\V\end{matrix}\bigg] \mapsto (W,C)$ sets up
a bijection between $\Pi$ and $\Bbb B\times GL(n,\CC)$. The~action of
$Sp(n,\RR)$ on $\Pi$ translates into
$$ g\cdot(W,C) =: (g_\sharp(W),\alpha(g,W)C),  $$
where $g_\sharp$ is a certain (fractional linear) mapping from $\Bbb B$ into
itself and $\alpha$ is a certain (polynomial) mapping from $Sp(n,\RR)\times
\Bbb B$ into $GL(n,\CC)$. Since $\Bbb B$ is contractible, there exists a unique
lift $\tilde\alpha: Mp(n,\RR)\times\Bbb B\to ML(n,\CC)$ of $\alpha$ such that
$$ \tilde\alpha(\tilde e,W)=\tilde I  \qquad \forall W\in\Bbb B,  $$
where $\tilde e$ and $\tilde I$ stand for the identities in $Mp(n,\RR)$ and
$ML(n,\CC)$, respectively,~and
$$ p(\tilde\alpha(\tilde g,W)) = \alpha(p(\tilde g), W) \qquad
\forall \tilde g\in Mp(n,\RR),\ \forall W\in\Bbb B ,   $$
where $p$ also denotes (on~the left-hand side), as before, the canonical
projection of $ML(n,\CC)$ onto $GL(n,\CC)$. Let $\tilde\Pi=\Bbb B\times
ML(n,\CC)$; then there is a left action of $Mp(n,\RR)$ on $\tilde\Pi$
defined~by
$$ \tilde g\cdot (W,\tilde C) := (p(\tilde g)_\sharp(W),
\tilde\alpha(\tilde g,W)\tilde C), \qquad \tilde g\in Mp(n,\RR),  $$
and $\tilde\Pi$ is a double cover of $\Pi$ with the covering map $\tau:\tilde
\Pi\to\Pi$ given by (\ref{tag:SNC}) with $C$ replaced by~$p(\tilde C)$.
In~analogy with~(\ref{tag:SND}), we~now define a {\sl positive metalinear
Lagrangian frame\/} as a function $\tilde w^\sharp:\tFwM\to\tilde\Pi$ such that
$$ \tilde w^\sharp (\widetilde{(u,v)}\cdot\tilde g) = \tilde g^{-1} \cdot
\tilde w^\sharp (\widetilde{(u,v)}) \qquad \forall \widetilde{(u,v)} \in
\tilde\FF^\omega_m \Omg,\ \forall \tilde g\in Mp(n,\RR),  $$
and let $\tildeLag$ be the corresponding bundle of all such frames.
The~covering map $\tau:\tilde\Pi\to\Pi$ gives rise to the similar map $\tilde
\tau:\tildeLag\to\Lag$, showing that the former is a double cover of the
latter. Finally, the obvious right action of $GL(n,\CC)$ on $\Lag$ lifts
uniquely to a right action of $ML(n,\CC)$ on $\tildeLag$.

Let~now $\PP$ be a positive polarization on $(\Omg,\omega)$. Then the bundle
$\FnPC$ of $\PP$-frames is a subbundle of $\Lag$ invariant under the action of
$GL(n,\CC)$ just mentioned. The~inverse image of $\FnPC$ under $\tilde\tau$ is
a subbundle $\tildeFnPC$ of $\tildeLag$ invariant under the action of
$ML(n,\CC)$, and $\tilde\tau$ restricted to $\tildeFnPC$ defines a double
covering $\tilde\tau:\tildeFnPC\to\FnPC$. It~follows that $\tildeFnPC$ is a
metalinear frame bundle of~$\PP$, which we will call the metalinear frame
bundle {\sl induced\/} by~$\tildeLag$.

Finally, notice that for two positive polarizations $\PP$ and $\GG$ satisfying
the transversality condition
\begin{equation}  \overline\PP\cap\GG=\{0\}  \label{tag:TUD}  \end{equation}
and frames $(\xi_1,\dots,\xi_n)\equiv\xi$ and $(\xi'_1,\dots,\xi'_n)\equiv\xi'$
of $\PP$ and $\GG$, respectively, at~some point $m\in \Omg$ as in
(\ref{tag:TUB}) (with~$k=0$), if~we identify $\xi$ and $\xi'$ with the matrices
$\bigg[\begin{matrix} U\\V\end{matrix}\bigg]$ and $\bigg[\begin{matrix}
U'\\V'\end{matrix}\bigg]$ as in~(\ref{tag:SMC}) with respect to some choice of
a symplectic frame $(u,v)$ at~$m$, then the expression $\epsilon_{\omega,k}
(\dots)$ in (\ref{tag:TUB}) reduces~to
\begin{equation}
\det\bigg[\frac{\omega(\xi_j,\overline{\xi'_l})}{i}\bigg]_{j,l=1}^n =
\det\Big[\frac12 C^{\prime*} (I-W^{\prime*}W)C \Big],  \label{tag:TUE}
\end{equation}
with $(W,C)$ and $(W',C')$ as~in~(\ref{tag:SNC}). The~transversality hypothesis
implies that the matrix on the left-hand side is invertible, hence so must be
$I-W^{\prime*}W$. Since the subset $\Bbb B_0$ of all matrices in $\Bbb B$ for
which $1$ is not an eigenvalue is contractible, there exists a unique map
$\tilde\gamma:\Bbb B_0\to ML(n,\CC)$ such that
$$ p(\tilde\gamma(S)) = I-S \qquad \forall S\in\Bbb B_0,
\qquad \text{and } \tilde\gamma_0=\tilde I.  $$
(Note that $\tilde\gamma$ is independent of the polarizations $\PP$
and~$\GG$~!) Consequently, the function
$$ \lambda\Big( \frac12 \widetilde{C'}^* \tilde\gamma(I-W^{\prime*}W)
\tilde C \Big),  $$
with $\lambda$ having the same meaning as in~\S\ref{sec24}, gives the sought
definition of the square root of~(\ref{tag:TUE}) which makes the right-hand
side of (\ref{tag:TUB}) well-defined and independent of the choice of the
metalinear frames $\tilde\xi, \tilde\xi'$ above $\xi$ and~$\xi'$.

Finally, integrating the density (\ref{tag:TUB}) over $\Omg/\hatDD$, we obtain
the sesquilinear pairing
\begin{equation}  \phi,\psi \mapsto \spr\phi\psi \in\CC   \label{tag:BKS}
\end{equation}
between sections $\phi$ and $\psi$ of $L\otimes\tBP$ and
$L\otimes\tilde\BB^\GG$ covariantly constant along $\PP$ and $\GG$,
respectively. This is the {\sl Blattner-Kostant-Sternberg  pairing\/} (or~just
BKS-pairing for short) originally introduced in~\cite{bib:BlattPSPM}.

Unfortunately, there seems to be no known general criterion for the existence
of $\spr\phi\psi$, i.e.~for the integrability of the density~(\ref{tag:TUB}).
All one can say in general is that $\spr\phi\psi$ exists if both $\phi$ and
$\psi$ are compactly supported. In~many concrete situations, however,
(\ref{tag:BKS})~extends continuously to the whole Hilbert spaces $\HH_\PP$ and
$\HH_\GG$ defined by (\ref{tag:CPC}) for the polarizations $\PP$ and $\GG$,
respectively, and, further, the operator $H_{\PP\GG}:\HH_\PP\to\HH_\GG$
defined~by
$$ (\psi,H_{\PP\GG}\phi)_{\HH_\GG} = \spr\psi\phi
\qquad\forall \phi\in\HH_\PP, \psi\in\HH_\GG,  $$
turns out to be, in fact, unitary. For instance, for $\PP=\GG$, $H_{\PP\GG}$ is
just the identity operator (so~that the BKS pairing coincides with the inner
product in~$\HH_\PP$), and for $\Omg=\RR^{2n}$ and $\PP$ and $\GG$ the
polarizations spanned by the $\partial/\partial p_j$ and the $\partial/\partial
q_j$, respectively, $H_{\PP\GG}$ is the Fourier transform. It~may happen,
though, that $H_{\PP\GG}$ is bounded and boundedly invertible but not
unitary~\cite{bib:RawnKep}; no~example is currently known where $H_{\PP\GG}$
would be unbounded.

Turning finally to our original objective --- the extension of the quantization
map $f\mapsto Q_f$ --- let now $f$ be a real function on $\Omg$ such that $X_f$
does not necessarily preserve the polarization~$\PP$. The flow $\rho_t=\exp
(tX_f)$ generated by $X_f$ then takes $\PP$ into a polarization $\tilde\rho_t
\PP=:\PP_t$, which may be different from~$\PP$. The flow $\rho_t$ further
induces the corresponding flows on the spaces $\Gamma(L)$ of sections of the
prequantum bundle~$L$, as~well as from sections of the metalinear bundle $\tBP$
into the sections of~$\tilde\BB^{\PP_t}$; hence, it gives rise to a (unitary)
mapping, denoted~$\rho^\sharp_t$, from the quantum Hilbert space $\HH_{\PP_t}
=:\HH_t$ into~$\HH$. Assume now that for all sufficiently small positive~$t$,
the polarizations $\PP_t$ and $\PP$ are such that the BKS pairing between them
is defined~on (or~extends by continuity~to) all of $\HH_t\times\HH$ and the
corresponding operator $H_{\PP_t\PP}=:H_t$ is unitary. Then the promised
quantum operator given by the BKS pairing~is
\begin{equation}  Q_f \phi = -\frac{ih}{2\pi} \,\frac d{dt} \big(
H_t\circ\rho_t^\sharp \big) \Big| _{t=0}.   \label{tag:BKQ}  \end{equation}

In~view of the remarks in the penultimate paragraph, in~practice it may be
difficult to verify the (existence and) unitarity of $H_t$, but one may still
use (\ref{tag:BKQ}) to compute $Q_f$ on a dense subdomain and investigate the
existence of a self-adjoint extension afterwards.

Observe also that for $f\in\Obs$, i.e.~for functions preserving the
polarization ($[X_f,\PP]\subset\PP$), one has $\PP_t=\PP$ and $H_t=I$ $\forall
t>0$, and, hence, it can easily be seen that (\ref{tag:BKQ}) reduces just to
our original prescription~(\ref{tag:MLC}). In~particular, if $f$ is constant
along~$\PP$ (i.e.~$X_f\in\PP$), then $Q_f$ is just the operator of
multiplication by~$f$.

If~the polarization $\PP=\DD$ is real and its leaves are simply connected, it
is possible to give an explicit local expression for the
operator~(\ref{tag:BKQ}). Namely, let $V$ be a contractible coordinate patch
on~$\MD$ such that on $\pi^{-1}(V)$ (where, as before, $\pi:\Omg\to\MD$ is the
canonical submersion) there exist real functions $q_1,\dots,q_n$, whose
Hamiltonian vector fields span $\PP|_{\pi^{-1}(V)}$, and functions
$p_1,\dots,p_n$ such that $\omega|_{\pi^{-1}(V)} = \sum_{j=1}^n dp_j\wedge
dq_j$. Using a suitable reference section on $\pi^{-1}(V)$ covariantly constant
along~$\PP$, the subspace in $\HH_\PP$ of sections supported in $\pi^{-1}(V)$
can be identified with $L^2(V,dq_1\dots dq_n)$. If~$\psi$ is such a section,
then under this identification, the operator (\ref{tag:BKQ}) is given~by
\begin{equation}  Q_f\psi = \frac{ih}{2\pi} \, \frac{d\psi_t}{dt} \bigg|_{t=0},
\label{tag:SNATa} \end{equation}
where
\begin{equation} \begin{aligned} \psi_t(q_1,\dots,q_n) &=
\bigg(\frac{2\pi}{ih}\bigg)^{n/2} \int
\exp\Big[-\frac{2\pi}{ih}\int_0^t (\theta(X_f)-f)\circ\rho_{-s}\,ds\Big]
\times {} \\ & \qquad\quad {}\times
\sqrt{\det\big[ \omega(X_{q_j},\rho_t X_{q_k}) \big]_{j,k=1}^n}
\,\psi(q_1\circ\rho_{-t},\dots,q_n\circ\rho_{-t}) \, dp_1 \dots dp_n,
\end{aligned}  \label{tag:SNATb}  \end{equation}
where $\theta=\sum_{j=1}^n p_j\,dq_j$ and $\rho_t$ is, as~usual, the flow
generated by~$X_f$. See~\cite{bib:SniaB}, Section~6.3.

The~conditions (\ref{tag:TUA}), under which the BKS pairing was constructed
here, can be somewhat weakened, see
Blattner~\cite{bib:BlattLN}.\footnote{Originally, the pairing was defined in
Blattner's paper \cite{bib:BlattPSPM} for a pair of transversal real
polarizations; the transversality hypothesis was then replaced by regularity in
\cite{bib:BlattGSPM}, and finally regular pairs of positive complex
polarizations were admitted in~\cite{bib:BlattLN}.} In~particular, for positive
polarizations the pairing can still be defined even if the middle condition
in (\ref{tag:TUA}) is omitted. In~that case, a new complication can arise: it
may happen that for two sections $\phi$ and $\psi$ which are covariantly
constant along $\PP$ and $\GG$, respectively, their ``local scalar product''
$\spr\phi\psi_m$ is not covariantly constant along~$\hatDD$ (i.e.~does not
depend only on~$\pi(m)$). More precisely: $\spr\phi\psi_m$ is covariantly
constant whenever $\phi$ and $\psi$ are if and only if the
one-form~$\chi_{\PP\GG}$ (the {\sl Blattner obstruction\/})
defined on $\overline\PP\cap\GG$~by
\begin{equation}  \chi_{\PP\GG} := \sum_{j=1}^{n-k} \omega([v_j,w_j],\cdot\,)
\label{tag:BLOB}  \end{equation}
vanishes. Here $k=\dim\overline\PP\cap\GG$ and $v_1,\dots,v_{n-k},w_1,\dots,
w_{n-k}$ are (arbitrary) vector fields in $\overline\PP+\GG$ such that
$$ \omega(v_i,v_j)=\omega(w_i,w_j)=0, \qquad \omega(v_i,w_j)=\delta_{ij}.  $$
The simplest example when $\chi_{\PP\GG}\neq0$ is $\Omg=\RR^4$ (with the usual
symplectic form) and $\PP$ and $\GG$ spanned by $\partial/\partial p_1,
\partial/\partial p_2$ and $p_1\partial/\partial p_1+p_2 \partial/\partial p_2,
p_2 \partial/\partial q_1-p_1 \partial/\partial q_2$, respectively.

We~remark that so far there are no known ways of defining the BKS pairing if
the dimension of $\overline\PP\cap\GG$ varies, or if the intersection is not of
the form $\hatDD^\CC$ for a real distribution~$\hatDD$ which is fibrating.
Robinson~\cite{bib:RobiTAMS} showed how to define the ``local'' product
$\spr\phi\psi_m$ for a completely arbitrary pair of polarizations $\PP$
and~$\GG$, however his pairing takes values not in a bundle of densities but in
a certain line bundle over~$\Omg$ (coming from higher cohomology groups) which
is not even trivial in general, so it is not possible to integrate the local
products into a global (\hbox{$\CC$-valued}) pairing. (For a regular pair of
positive polarizations, Robinson's bundle is canonically isomorphic to the
bundle of densities~on~$\Omg$.)

A~general study of the integral kernels mediating BKS-type pairings was
undertaken by Gaw\c{e}dzki \cite{bib:GaweD}~\cite{bib:GaweGSPM}; he also
obtained a kernel representation for the quantum operators~$Q_f$. His~kernels
seem actually very much akin to the reproducing kernels for vector bundles
investigated by Peetre~\cite{bib:Pee} and others, cf.~the discussion in
Section~\ref{sec4} below.

A~completely different method of extending the correspondence $f\mapsto Q_f$
was proposed by Kostant in~\cite{bib:KostSymp}. For~a set $\XXX$ of vector
fields on $\Omg$ and a polarization~$\PP$, denote by $(\ad\PP)\XXX$ the set
$\{[X,Y]; X\in\XXX, Y\in\PP\}$, and~let
$$ C_\PP^k := \{ f\in C^\infty(\Omg): (\ad\PP)^k\{X_f\}\subset\PP\},
\qquad k=0,1,2,\dots.   $$
Then, in~view of the involutivity of~$\PP$, $C^k_\PP\subset C^{k+1}_\PP$, and,
in~fact, $C^0_\PP$ is the space of functions constant along~$\PP$, and
$C^1_\PP=\Obs$; one can think of $C^k_\PP$ as the space of functions which are
``polynomial of degree at most $k$ in the directions transversal to~$\PP$''.
Kostant's method extends the domain of the mapping $f\mapsto Q_f$ to the union
$C^*_\PP:= \bigcup_{k\ge0} C^k_\PP$; though phrased in completely geometric
terms, in the end it essentially boils down just to choosing a particular
ordering of the operators $P_j$ and $Q_j$ (cf.~Section~\ref{sec5} below).
Namely, let $\GG$ be an auxiliary polarization on $\Omg$ such that locally near
any $m\in \Omg$, there exist functions $q_1,\dots,q_n$ and $p_1,\dots,p_n$ such
that $X_{q_j}$ span~$\PP$, $X_{p_j}$~span~$\GG$, and $\{q_j,p_k\}=\delta_{jk}$.
(Such polarizations are said to be {\sl Heisenberg related.\/}) Now~if $f$ is
locally of the form $p^m \phi(q)$ (any~function from $C^*_\PP$ is a sum of such
functions), then
$$ Q_f = \bigg(\frac{ih}{2\pi}\bigg)^{|\boldkey m|-1} \sum_{0\le|\boldkey k|\le
|\boldkey m|} \bigg(\frac{|\boldkey k|}2-1\bigg) \binom{\boldkey m}{\boldkey k}
\frac{\partial^{|\boldkey k|}\phi} {\partial q^{\boldkey k}} \, \frac
{\partial^{|\boldkey m-\boldkey k|}}{\partial q^{\boldkey m-\boldkey k}}.  $$
Here $\boldkey m=(m_1,\dots,m_n)$ is a multiindex, $|\boldkey m|=m_1+\dots+
m_n$, and similarly for~$\boldkey k$. Again, however, the axiom (Q3) is no
longer satisfied by these operators on the extended domain, and, further, the
operator $Q_f$ depends also on the auxiliary polarization~$\GG$: if~$f\in
C^k_\PP$, then $Q_f$ is a differential operator of order~$k$, and choosing a
different auxiliary polarization $\GG$ (Heisenberg related to~$\PP$) results in
an error term which is a differential operator of order~$k-2$. We~will say
nothing more about this method here.

\subsection{Further developments} \label{sec26}
In~spite of the sophistication of geometric quantization, there are still quite
a few things that can go wrong: the integrality condition may be violated,
polarizations or the metaplectic structure need not exist, the Hilbert space
$\HH$ may turn out to be trivial, there may be too few quantizable functions,
etc. We~will survey here various enhancements of the original approach that
have been invented in order to resolve some of these difficulties, and then
discuss the remaining ones in the next subsection.

\subsubsection{Bohr-Sommerfeld conditions and distributional sections}
\label{sec261}
An~example when the Hilbert space $\HH$ turns out to be trivial --- that~is,
when there are no square-integrable covariantly constant sections of $QB$
except the constant zero --- is that of $\Omg=\CC\setminus\{0\}$ ($\simeq
\RR^2$ with the origin deleted), with the standard symplectic form, and the
circular (real) polarization $\DD$ spanned by~$\partial/\partial\theta$, where
$(r,\theta)$ are the polar coordinates in~$\CC\simeq\RR^2$. The~leaves space
$\MD$ can be identified with $\RR^+$; upon employing a suitable reference
section, sections of the quantum bundle (\ref{tag:QBU}) can be identified with
functions on $\CC\setminus\{0\}$, and covariantly constant ones with those
satisfying $f(e^{i\theta}z)=e^{2\pi ir\theta/h} f(z)$. (See~\cite{bib:TuyCWI},
pp.~79-83.) However, as~the coordinate $\theta$ is cyclic, this forces the
support of $f$ to be contained in the union of the circles
\begin{equation}  r=\frac{kh}{2\pi}, \qquad k=1,2,\dots.  \label{tag:BSA}
\end{equation}
As~the latter is a set of zero measure, we get $\HH=\{0\}$.

A similar situation can arise whenever the leaves of $\DD$ are not simply
connected.

In~general, for any leaf $\Lambda$ of $\DD$, the partial connection $\nabla$ on
the quantum line bundle $QB$ induces a flat connection in the restriction
$QB|_\Lambda$ of $QB$ to~$\Lambda$. For~any closed loop $\gamma$ in~$\Lambda$,
a~point $m$ on $\gamma$ and $\phi\in QB_m\setminus\{0\}$, the parallel
transport with respect to the latter connection of $\phi$ along $\gamma$
transforms $\phi$ into $c\phi$, for some $c\in\CC^\times$; the set of all $c$
that arise in this way forms a group, the {\it holonomy group\/} $G_\Lambda$
of~$\Lambda$. Let $\sigma$ be the set of all leaves $\Lambda\in\MD$ whose
holonomy groups are trivial, i.e.~$G_\Lambda=\{1\}$. The~preimage $\Cal S=
\pi^{-1}(\sigma)\subset \Omg$ is called the {\sl Bohr-Sommerfeld variety,\/}
and it can be shown that any section of $QB$ covariantly constant along $\DD$
has support contained in~$\Cal S$. In~the example above, $\Cal S$ is the union
of the circles~(\ref{tag:BSA}).

For \underbar{real} polarizations $\PP$ such that all Hamiltonian vector fields
contained in $\PP$ are complete (the {\sl completeness condition\/}), the
problem can be solved by introducing distribution-valued sections of~$QB$.
See~\cite{bib:SniaB}, Section~4.5, and \cite{bib:WoodhOLD}, pp.~162--164.
In~the example above, this corresponds to taking $\HH$ to be the set of all
functions $\phi$ which are equal to $\phi_k$ on the circles (\ref{tag:BSA}) and
vanish everywhere else,~i.e.
\begin{equation}  \phi(r e^{i\theta}) = \begin{cases} \phi_k e^{ki\theta}
\qquad &\text{if }r=k\frac h{2\pi}, \quad k=1,2,\dots, \\ 0 &\text{otherwise,}
\end{cases}  \label{tag:PHI}  \end{equation}
with the inner product
$$ (\phi,\psi) = \sum_{k=1}^\infty \overline\phi_k \psi_k.  $$
For~real functions $f$ satisfying
\begin{equation}  [X_f,\PP]\subset\PP   \label{tag:BSB}  \end{equation}
(i.e.~preserving the polarization), the quantum operators $Q_f$ can then be
defined, essentially, in the same way as before, and extending the BKS pairing
to distribution-valued sections (see~\cite{bib:SniaB}, Section~5.1), one can
also extend the domain of the correspondence $f\mapsto Q_f$ to some functions
$f$ for which (\ref{tag:BSB}) fails.

For~complex polarizations, there exist some partial results
(e.g.~Mykytiuk~\cite{bib:Myky}), but the problem is so far unsolved in general.

\begin{remark*} It~turns out that in the situation from the penultimate
paragraph, the subspaces $\HH_\alpha\subset\HH$ consisting of sections
supported on a given connected component $\Cal S_\alpha$ of the Bohr-Sommerfeld
variety $\Cal S$ are invariant under all operators $Q_f$ (both if $f$ satisfies
(\ref{tag:BSB}) or if $Q_f$ is obtained by the BKS pairing); that~is, $\HH$ is
reducible under the corresponding set of quantum operators. One~speaks of the
so-called {\sl superselection rules\/} (\cite{bib:SniaB}, Section~6.4).  \qed
\end{remark*}

\subsubsection{Cohomological correction} \label{sec262}
Another way of attacking the problem of non-existence of square-integrable
covariantly constant sections is the use of higher cohomology groups.

Let $k\ge0$ be an integer and let $QB$ be the quantum bundle $L\otimes\BB^\PP$
or $L\otimes\tBP$ constructed in \S\S\ref{sec23} (or~\ref{sec22})
and~\ref{sec24}, respectively.
A~$k$-$\PP$-form with values in $QB$ is a $k$-linear and alternating map which
assigns a smooth section $\alpha(X_1,\dots,X_k)$ of $QB$ to any $k$-tuple of
vector fields $X_1,\dots,X_k\in\overline\PP$. We~denote the space of all such
forms by $\Lambda^k(\Omg,\PP)$; one has $\Lambda^0(\Omg,\PP)=\Gamma(QB)$, and,
more generally, any $\alpha\in\Lambda^k(\Omg,\PP)$ can be locally written as a
product $\alpha=\beta\tau$ where $\tau$ is a section of $QB$ covariantly
constant along $\overline\PP$ and $\beta$ is an ordinary complex $k$-form
on~$\Omg$, with two such products $\beta\tau$ and $\beta'\tau$ representing
the same $k$-$\PP$-form whenever $\beta-\beta'$ vanishes when restricted
to~$\overline\PP$.

The~operator $\opP:\Lambda^k(\Omg,\PP)\to\Lambda^{k+1}(\Omg,\PP)$ is defined~by
\begin{align*}
(\opP\alpha)(X_1,\dots,X_{k+1}) &= \sum_\sigma \Big( \nabla_{X_{\sigma(1)}}
(\alpha(X_{\sigma(2)},\dots,X_{\sigma(k+1)})) \\
& \qquad\qquad - \frac k2\, \alpha([X_{\sigma(1)},X_{\sigma(2)}],X_{\sigma(3)},
\dots, X_{\sigma(k+1)}) \Big)  \end{align*}
where the summation extends over all cyclic permutations $\sigma$ of the index
set $1,2,\dots,k+1$. It~can be checked that $\opP^2=0$; hence, we can define
the {\sl cohomology groups\/} $H^k(\Omg,\PP)$ as the quotients $\Ker(\opP|
\Lambda^k)/\Ran(\opP|\Lambda^{k-1})$ of the $\opP$-closed $k$-$\PP$-forms by
the $\opP$-exact ones.

Finally, for each real function $f$ satisfying (\ref{tag:BSB}) (i.e.~preserving
the polarization), one can extend the operator $Q_f$ given by (\ref{tag:MLC})
(or~(\ref{tag:HDF}) or~(\ref{tag:CPD})) to $\Lambda^k(\Omg,\PP)$ by setting
\begin{equation}  (Q_f \alpha)(X_1,\dots,X_k) := Q_f(\alpha(X_1,\dots,X_k)) +
\frac{ih}{2\pi} \sum_{j=1}^k \alpha(X_1,\dots,[X_f,X_j],\dots,X_k).
\label{tag:COHO}  \end{equation}
It~can be checked that $Q_f$ commutes with $\opP$, and thus induces an operator
--- also denoted~$Q_f$ --- on the cohomology groups~$H^k(\Omg,\PP)$.

Now~it may happen that even though $H^0(\Omg,\PP)$ contains no nonzero
covariantly constant sections, one of the higher cohomology groups
$H^k(\Omg,\PP)$ does, and one can then use it as a substitute for~$\HH$
(and~(\ref{tag:COHO}) as a substitute for~(\ref{tag:MLC})). For~instance,
in~the above example of $\Omg=\CC\setminus\{0\}$ with the circular
polarization, one can show that using $H^1(\Omg,\PP)$ essentially gives the
same quantization as the use of the distributional sections in~\S\ref{sec261}
(see~Simms~\cite{bib:SimmLN}). However, in general there are still some
difficulties left --- for instance, we need to define an inner product on
$H^k(\Omg,\PP)$ in order to make it into a Hilbert space, etc. The~details can
be found in Woodhouse~\cite{bib:WoodhOLD}, Section~6.4,
Rawnsley~\cite{bib:RawnTAMS}, or Puta~\cite{bib:Puta} and the references given
there.

\subsubsection{$Mp^\CC$-structures} \label{sec263}
One more place where the standard geometric quantization can break down is the
very beginning: namely, when the integrality condition $h^{-1}[\omega]\in
H^2(\Omg,\ZZ)$, or the condition for the existence of the metaplectic structure
$\frac12 c_1(\omega)\in H^2(\Omg,\ZZ)$, are not satisfied. This is the case,
for instance, for the odd-dimensional harmonic oscillator, whose phase space is
the complex projective space $\CC P^n$ with even~$n$. It~turns out that this
can be solved by extending the whole method of geometric quantization to the
case when the \underbar{sum} $h^{-1}[\omega] + \frac12 c_1(\omega)$, rather
then the two summands separately, is integral. This was first done by
Czyz~\cite{bib:Czyz} for compact K\"ahler manifolds, using an axiomatic
approach, and then by Hess~\cite{bib:Hess}, whose method was taken much further
by Rawnsley and Robinson \cite{bib:RawnRob} (see Robinson~\cite{bib:RobiSURV}
for a recent survey).

The~main idea is to replace the two ingredients just mentioned --- the
prequantum bundle and the metaplectic structure --- by~a single piece of data,
called the {\sl prequantized $Mp^\CC$ structure.\/} To~define it, consider,
quite generally, a real vector space $V$ of dimension $2n$ with a symplectic
form $\OMEGA$ and an irreducible unitary projective representation $W$ of $V$
on a separable complex Hilbert space $\Bbb H$ such that
$$ W(x) W(y) = e^{-\pi i \OMEGA(x,y)/h} W(x+y),  \qquad \forall x,y\in V.  $$
By~the Stone-von Neumann theorem, $W$ is unique up to unitary equivalence;
consequently, for any $g\in Sp(V,\OMEGA)$ there exists a unitary operator $U$
on $\Bbb H$ (unique up to multiplication by a unimodular complex number) such
that $W(gx)=UW(x)U^*$ for all $x\in V$. Denote by $Mp^\CC(V,\OMEGA)$ the group
of all such $U$'s as $g$ ranges over $Sp(V,\OMEGA)$, and let
$\sigma:Mp^\CC(V,\OMEGA)\to Sp(V,\OMEGA)$ be the mapping given by
$\sigma(U)=g$. The~kernel of $\sigma$ is just $U(1)$, identified with the
unitary scalar operators in~$\Bbb H$. There is a unique character $\eta:
Mp^\CC(V,\OMEGA)\to U(1)$ such that $\eta(\lambda I)=\lambda^2$ $\forall\lambda
\in U(1)$; the kernel of $\eta$ is our old friend, the metaplectic group
$Mp(V,\OMEGA)$. Let~now $Sp(\Omg,\omega)$ denote the symplectic frame bundle of
the manifold~$\Omg$, which we think of as being modelled fiberwise on
$(V,\OMEGA)$. An~{\sl $Mp^\CC$-structure\/} on $\Omg$ is a principal
$Mp^\CC(V,\OMEGA)$ bundle $P\overset{\pi}{\to}\Omg$ together with a
$\sigma$-equivariant bundle map $P\to Sp(\Omg,\omega)$. An~$Mp^\CC$ structure
is called {\sl prequantized\/} if, in~addition, there exists an
$Mp^\CC(V,\OMEGA)$-invariant $u(1)$-valued one-form $\gamma$ on $P$ such that
$d\gamma=\frac{2\pi}{ih} \pi^*\omega$ and $\gamma(\boldkey z)=\frac12\eta_*z$
for all $z$ in the Lie algebra of $Mp^\CC(V,\OMEGA)$; here $\boldkey z$ is the
fundamental vertical vector field corresponding to~$z$.

It~turns out that $Mp^\CC$ structures always exist on any symplectic manifold,
and prequantized ones exist if and only if the combined integrality condition
$$ \text{the class }h^{-1} [\omega] + \tfrac12 c_1(\omega)^\RR \in
H^2(\Omg,\RR) \text{ is integral}  $$
is fulfilled. In~that case, if $\PP$ is a positive polarization on~$\Omg$, one
can again consider partial connections and covariantly constant sections
of~$P$, and define the corresponding Hilbert spaces and quantum operators more
or less in the same way as before. Details can be found in Rawnsley and
Robinson~\cite{bib:RawnRob} and Blattner and Rawnsley~\cite{bib:BlattRaCo}.
It~is also possible to define the BKS pairing in this situation.

\subsection{Some shortcomings} \label{sec27}
Though the method of geometric quantization has been very successful, it has
also some drawbacks. One of them is the dependence on the various ingredients,
i.e. the choice of the prequantum bundle, metaplectic structure (or
prequantized $Mp^\CC$-structure), and polarization. The~(equivalence
classes~of) various possible choices of the prequantum bundle are parameterized
by the elements of the cohomology group $H^1(\Omg,\TT)$, and have very sound
physical interpretation (for instance, they allow for the difference between
the bosons and the fermions, see Souriau~\cite{bib:SouSD}). The~situation with
the choices for the metaplectic structure, which are parameterized by
$H^1(\Omg,\ZZ_2)$, is already less satisfactory (for instance, for the harmonic
oscillator, only one of the two choices gives the correct result for the energy
levels; see~\cite{bib:TuyCWI}, pp.~150--153). But~things get even worse
with the dependence on polarization. One would expect the Hilbert spaces
associated to two different polarizations of the same symplectic manifold to be
in some ``intrinsic'' way unitarily equivalent; more specifically, for any two
polarizations $\PP,\GG$ for which the BKS pairing exists, one would expect the
corresponding operator $H_{\PP\GG}$ to be unitary, and such that the
corresponding quantum operators satisfy $Q'_f H_{\PP\GG}= H_{\PP\GG} Q_f$ for
any real observable $f$ quantizable with respect to both $\PP$ and~$\GG$.
We~have already noted in \S\ref{sec25} that the former need not be the case
($H_{\PP\GG}$ can be a bounded invertible operator which is not unitary, nor
even a multiple of a unitary operator), and it can be shown that even if
$H_{\PP\GG}$ is unitary, the latter claim can fail too (cf.~\cite{bib:TuyGBK}).
Finally, it was shown by Gotay~\cite{bib:Gotay} that there are symplectic
manifolds on which there do not exist any polarizations
whatsoever.\footnote{It~should be noted that --- unlike the cohomology groups
$H^1(\Omg,\TT)$ for the choices of the prequantum bundle and $H^1(\Omg,\ZZ_2)$
for the choice of the metaplectic structure --- there seems to be, up to the
authors' knowledge, no known classifying space for the set of all polarizations
on a given symplectic manifold, nor even a criterion for their existence.}
Such~phase spaces are, of~course, ``unquantizable'' from the point of view of
conventional geometric quantization theory.

Another drawback, perhaps the most conspicuous one, is that the space of
quantizable observables is rather small; e.g.~for $\Omg=\RR^{2n}$ and
polarization given by the coordinates $q_1,\dots,q_n$, the space $\Obs$
consists of functions at most linear
in~$p$, thus excluding, for instance, the kinetic energy~$\frac12\|\bp\|^2$.
The~extension of the quantization map $f\mapsto Q_f$ by means of the BKS
pairing\footnote{Sometimes this is also called the {\sl method of infinitesimal
pairing\/}.}, described in~\S\ref{sec25}, (which gives the correct answer
$Q_f=-\frac{h^2}{8\pi}\Delta$ for the kinetic energy
$f(\bp,\bq)=\frac12\|\bp\|^2$) is~not entirely satisfactory, for the following
reasons. First of all, as we have already noted in \S\ref{sec25}, it~is
currently not known under what conditions the pairing extends from compactly
supported sections to the whole product $\HH_\PP\times\HH_{\PP_t}$ of the
corresponding quantum Hilbert spaces; and even if the pairing so extends, it is
not known under what conditions the derivative at $t=0$ in (\ref{tag:BKQ})
exists. (And neither is it even known under what conditions the polarizations
$\PP$ and $\PP_t$ are such that the pairing can be defined in the first place
--- e.g.~transversal etc.) Consequently, it is also unknown for which functions
$f$ the quantum operator $Q_f$ is defined at~all. For~instance, using the
formulas (\ref{tag:SNATa}) and~(\ref{tag:SNATb}), Bao and~Zhu
\cite{bib:BaoZnew} showed that for $\Omg=\RR^{2}$ (with the usual symplectic
form) and $f(p,q)=p^m$, $Q_f$ is undefined as soon as $m\ge3$ (the integral in
(\ref{tag:SNATb}) then diverges as $t\to0$). Second, even when $Q_f$ is defined
all right, then, as we have also already noted in~\S\ref{sec25}, owing to the
highly nonexplicit nature of the formula~(\ref{tag:BKQ}) it is not even
possible to tell beforehand whether this operator is at least formally
symmetric, not to say self-adjoint. Third, even if $Q_f$ are well defined and
self-adjoint, their properties are not entirely satisfactory: for instance,
in~another paper by Bao and~Zhu~\cite{bib:BaoZ} they showed that for
$\Omg=\RR^2$ and $f(p,q)=p^2 g(q)$, one can again compute from
(\ref{tag:SNATa})~--~(\ref{tag:SNATb}) that (upon identifying $\HH$ with
$L^2(\RR,dq)$ by means of a suitable reference section)
\begin{equation}  Q_f \psi = \Big(\frac{ih}{2\pi}\Big)^2 \Big[g\psi'' + g'\psi'
+ \Big( \frac{g''}4 - \frac{g^{\prime2}}{16g} \Big) \psi \Big] ,
\label{tag:BAOZ}  \end{equation}
so~that, in~particular, the dependence $f\mapsto Q_f$ is not even linear(!).
Finally, from the point of view of our axioms (Q1)~--~(Q5) set up in the
beginning, the most serious drawback of (\ref{tag:BKQ}) is that the operators
$Q_f$ so defined do not, in general, satisfy the commutator condition~(Q3)!

\begin{remark*} For~functions $f$ such that $X_f$ leaves $\PP+\overline\PP$
invariant, it was shown by Tuynman that $Q_f$ can be identified with a certain
Toeplitz-type operator; see~\cite{bib:TuyGBK}.  \qed  \end{remark*}

\medskip

For some further comments on why the standard theory of geometric quantization
may seem unsatisfactory, see Blattner \cite{bib:Blatt}, p.~42, or
Ali~\cite{bib:AliSurv}.

Finally, we should mention that in the case when $\Omg$ is a coadjoint orbit
of a Lie group~$G$, which operates on $\Omg$ by $\omega$-preserving
diffeomorphisms, the geometric quantization is intimately related to the
representation theory of~$G$ (the orbit method); see Kirillov \cite{bib:Kirr},
Chapter~14, and Vogan~ \cite{bib:Vogan} for more information.

For~further details on geometric quantization, the reader is advised to consult
the extensive bibliography on the subject. In~our exposition in
\S\S\ref{sec22}--\ref{sec26} we
have closely followed the beautiful CWI syllabus of Tuynman \cite{bib:TuyCWI},
as well as the classics by Woodhouse \cite{bib:WoodhOLD} (see also the new
edition~ \cite{bib:Woodh}) and Sniatycki \cite{bib:SniaB}; the books by
Guillemin and Sternberg \cite{bib:GuiSt} and Hurt \cite{bib:Hurt} are oriented
slightly more towards the theory of Fourier integral operators and the
representation theory, respectively. Other worthwhile sources include the
papers by Sniatycki \cite{bib:SniaA}, Blattner and Rawnsley \cite{bib:BlatRaw}
\cite{bib:BlattRaCo}, Czyz \cite{bib:Czyz}, Gawedzki \cite{bib:GaweD}, Hess
\cite{bib:Hess}, Rawnsley and Robinson \cite{bib:RawnRob}, Robinson
\cite{bib:RobiTAMS}, Blattner \cite{bib:BlattPSPM} \cite{bib:BlattGSPM}
\cite{bib:BlattLN}, Tuynman \cite{bib:TuyWis} \cite{bib:TuyIrr}
\cite{bib:TuyGBK} \cite{bib:TuyCm}, Rawnsley \cite{bib:RawnCMP},
Kostant\cite{bib:KostSymp} \cite{bib:KostOno} \cite{bib:Kost}, and Souriau
\cite{bib:SouSD}, the surveys by Blattner \cite{bib:Blatt}, Ali
\cite{bib:AliSurv}, Echeverria-Enriquez et~al.~\cite{bib:EEMLRRVM}, or Kirillov
\cite{bib:KiriEMS}, and the recent books by Bates and Weinstein
\cite{bib:BaWei} and Puta \cite{bib:Puta}, as well as the older one by Simms
and Woodhouse~\cite{bib:SimWo}.

\section{Deformation quantization} \label{sec3}
Deformation quantization tries to resolve the difficulties of geometric
quantization by relaxing the axiom (Q3)~to
\begin{equation}  [Q_f,Q_g] = -\frac{ih}{2\pi} Q_{\{f,g\}} + O(h^2).
\label{tag:Da}  \end{equation}
Motivated by the asymptotic expansion for the Moyal product (\ref{tag:MOY}),
one can try to produce this by first constructing a formal associative but
noncommutative product $*_h$ (a~{\sl star product\/}), depending on~$h$, such
that, in a suitable sense,
\begin{equation}  f *_h g = \sum_{j=0}^\infty  h^j C_j(f,g)  \label{tag:Db}
\end{equation}
as $h\to0$, where the bilinear operators $C_j$ satisfy
\begin{gather}
C_0(f,g)=fg, \qquad C_1(f,g)-C_1(g,f)=-\tfrac i{2\pi}\{f,g\},  \label{tag:Dc}
\\ C_j(f,\jedna)=C_j(\jedna,f)=0 \qquad \forall j\ge1.  \label{tag:Dca}
\end{gather}
Here ``formal'' means that $f*_h g$ is not required to actually exist for any
given value of~$h$, but we only require the coefficients $C_j:\Obs\times\Obs\to
\Obs$ to be well defined mappings for some function space $\Obs$ on $\Omg$ and
satisfy the relations which make $*_h$ formally associative. As~a second step,
one looks for an analogue of the Weyl calculus, i.e.~one wants the product
$*_h$ to be genuine (not only formal) bilinear mapping from $\Obs\times\Obs$
into $\Obs$ and seeks a linear assignment to each $f\in\Obs$ of an operator
$Q_f$ on a (fixed) separable Hilbert space $\HH$, self-adjoint if $f$ is
real-valued, such that\footnote{This is the condition which implies that $*_h$
must be associative (since composition of operators~is).}
\begin{equation}  Q_f Q_g = Q_{f*_h g}.  \label{tag:Dd}  \end{equation}
Further, we also want the construction to satisfy the functoriality
(=covariance) condition (Q4), which means that the star product should commute
with any symplectic diffeomorphism~$\phi$,
\begin{equation}  (f\circ\phi) *_h (g\circ\phi) = (f*_h g)\circ\phi.
\label{tag:Dda}  \end{equation}
Finally, for $\Omg=\RR^{2n}$ the star product should reduce to, or at least be
in some sense equivalent to, the Moyal product.

The first step above is the subject of {\sl formal deformation quantization,\/}
which was introduced by Bayen, Flato, Fronsdal, Lichnerowicz and Sternheimer~
\cite{bib:BFFLS}. Namely, one considers the ring $\AAA=C^\infty(\Omg)[[h]]$ of
all formal power series in $h$ with $C^\infty(\Omg)$ coefficients, and seeks an
associative $\CC[[h]]$-linear mapping $*:\AAA\times\AAA\to\AAA$ such that
(\ref{tag:Db}), (\ref{tag:Dc}) and (\ref{tag:Dca}) hold. This is a purely
algebraic problem which had been solved by Gerstenhaber \cite{bib:Gerst}, who
showed that the only obstruction for constructing $*$ are certain Hochschild
cohomology classes $c_n\in H^3(\AAA,\AAA)$ (the construction is possible if and
only if all $c_n$ vanish). Later Dewilde and Lecomte \cite{bib:DewLe} showed
that a formal star product exists on any symplectic manifold (thus the
cohomological obstructions in fact never occur). More geometric constructions
were subsequently given by Fedosov \cite{bib:Fedos} (see also his
book~\cite{bib:FedosBk}) and Omori, Maeda and Yoshioka~\cite{bib:OMY}, but the
question remained open whether the star product exists also for any Poisson
manifold (i.e.~for Poisson brackets given locally by $\{f,g\} = \omega^{ij}
(\partial_i f\cdot\partial_j g - \partial_j f\cdot \partial_i g)$ where the
2-form $\omega$ is allowed to be degenerate). This question was finally settled
in the affirmative by Kontsevich \cite{bib:Kon} on the basis of his ``formality
conjecture''. Yet another approach to formal deformation quantization on a
symplectic manifold can be found in Karasev and Maslov \cite{bib:KMbk}; star
products with some additional properties (admitting a {\sl formal trace\/}) are
discussed in Connes, Flato and Sternheimer \cite{bib:FScl} and Flato and
Sternheimer \cite{bib:FSbk}, and classification results are also
available~\cite{bib:BertCG},\cite{bib:Delig},\cite{bib:NTsy}.

A~formal star product is called {\sl local\/} if the coefficients $C_j$ are
differential operators. If~the manifold~$\Omg$ has a complex structure (for
instance, if $\Omg$ is K\"ahler), the star product is said to admit {\sl
separation of variables\/}\footnote{Or to be of {\sl Wick type\/};
{\sl anti-Wick type\/} is similarly obtained upon replacing $f*g$ by $g*f$.}
if~$f*g=fg$ (i.e.~$C_j(f,g)=0$ $\forall j\ge1$) whenever $f$ is holomorphic
or $g$ is anti-holomorphic. See Karabegov \cite{bib:KarA}, \cite{bib:KarB} for
a systematic treatment of these matters.

The second step\footnote{This is what we might call {\sl analytic deformation
quantization.\/}}, i.e.~associating the Hilbert space operators $Q_f$ to each
$f$, is more technical. In~the first place, this requires that $f*_h g$
actually exist as a function on $\Omg$ for some (arbitrarily small) values
of~$h$. Even this is frequently not easy to verify for the formal star products
discussed above. The usual approach is therefore, in fact, from the opposite
--- namely, one starts with some geometric construction of the operators $Q_f$,
and then checks that the operation $*$ defined by (\ref{tag:Dd}) is a star
product, i.e.~satisfies (\ref{tag:Db}),(\ref{tag:Dc}) and~(\ref{tag:Dca}).

In other words, one looks for an assignment $f\mapsto Q_f$, depending on the
Planck parameter~$h$, of operators $Q_f$ on a separable Hilbert space $\HH$ to
functions $f\in C^\infty(\Omg)$, such that as $h\to0$, there is an asymptotic
expansion
\begin{equation}  Q\h_f Q\h_g = \sum_{j=0}^\infty h^j Q\h_{C_j(f,g)}
\label{tag:Qa}  \end{equation}
for certain bilinear operators $C_j:C^\infty(\Omg)\times C^\infty(\Omg)\to
C^\infty(\Omg)$. Here (\ref{tag:Qa}) should be interpreted either in the weak
sense,~as
$$ \Big\langle a,\Big[ Q\h_f Q\h_g - \sum_{j=1}^N h^j Q\h_{C_j(f,g)} \Big]b
\Big\rangle = O(h^{N+1}) \qquad\forall a,b\in\HH,\quad\forall N=0,1,2,\dots, $$
where $\spr\cdot\cdot$ stands for the inner product in~$\HH$, or in the sense
of norms
$$ \Big\| Q\h_f Q\h_g - \sum_{j=1}^N h^j Q\h_{C_j(f,g)} \Big\| = O(h^{N+1})
\qquad \forall N=0,1,2,\dots,  $$
where $\|\cdot\|$ is the operator norm on~$\HH$. Further, $Q_f$ should satisfy
the covariance condition (Q4), should (in~some sense) reduce to the Weyl
operators $W_f$ for $\Omg=\RR^{2n}$, and preferably, the $C_j$ should be
local (i.e.~differential) operators.

For K\"ahler manifolds, these two problems are solved by the Berezin and
Berezin-Toeplitz quantizations, respectively, which will be described
in the next section. For a general symplectic (or even Poisson) manifold,
analogous constructions seem to be so far unknown. An~interesting method for
constructing non-formal star products on general symplectic manifolds,
using integration over certain two-dimensional surfaces (membranes) in the
complexification $\Omg^\CC \simeq \Omg\times \Omg$ of the phase space~$\Omg$,
has recently been proposed by Karasev~\cite{bib:Karas}.

A~systematic approach to such constructions\footnote{Sometimes referred to as
{\sl $C^*$-algebraic deformation quantization\/}
(Landsman~\cite{bib:+tgtgrp}).} has been pioneered by Rieffel \cite{bib:RiefA}
\cite{bib:RiefB}~\cite{bib:RiefM}. He~defines a {\sl strict deformation
quantization\/} as a dense $*$-subalgebra $\AAA$ of $C^\infty(\Omg)$ equipped,
for each sufficiently small positive~$h$, with a norm~$\|\cdot\|_h$,
an~involution~${}^{*_h}$ and an associative product~$\times_h$, continuous
with respect to~$\|\cdot\|_h$, such that
\begin{enumerate}
\item[$\bullet$] $h\mapsto\AAA_h:=\,$the completion of $(\AAA,{}^{*_h},
\times_h)$ with respect to $\|\cdot\|_h$, is a continuous field of
$C^*$-algebras;
\item[$\bullet$] ${}^{*_0},\ \times_0$ and $\|\cdot\|_0$ are the ordinary
complex conjugation, pointwise product and supremum norm on~$C^\infty(\Omg)$,
respectively;
\item[$\bullet$] $\lim_{h\to0} \|(f\times_h g-g\times_h f) + \tfrac{ih}{2\pi}
\{f,g\}\|_h =0$.
\end{enumerate}
Using the Gelfand-Naimark theorem, one can then represent the $C^*$-algebras
$\AAA_h$ as Hilbert space operators, and thus eventually arrive at the
desired quantization rule $f\mapsto Q_f$. (One~still needs to worry about the
covariance and irreducibility conditions (Q4) and (Q5), which are not directly
built into Rieffel's definition, but let us ignore these for a moment.) The
difficulty is that examples are scarce --- all of them make use of the Fourier
transform in some way and are thus limited to a setting where the latter makes
sense (for instance, one can recover the Moyal product in this way). In~fact,
the motivation behind the definition comes from operator algebras and Connes'
non-commutative differential geometry rather than quantization. A~broader
concept is a {\sl strict quantization\/} \cite{bib:Rief}: it is defined as a
family of $*$-morphisms $T_h$ from a dense $*$-subalgebra $\AAA$ of
$C^\infty(\Omg)$ into $C^*$-algebras $\AAA_h$, for $h$ in some subset of $\RR$
accumulating at~0, such that $\operatorname{Ran} T_h$ spans $\AAA_h$ for
each~$h$, $\AAA_0=C^\infty(\Omg)$ and $T_0$ is the inclusion map of $\AAA$ into
$\AAA_0$, the functions $h\mapsto\|T_h(f)\|_h$ are continuous for each
$f\in\AAA$, and
\begin{equation}  \begin{aligned}
& \|T_h(f) T_h(g) - T_h(fg)\|_h \to 0  , \\
& \|[T_h(f),T_h(g)] + \frac{ih}{2\pi} T_h(\{f,g\})\|_h \to 0 \end{aligned}
\label{tag:Rj}  \end{equation}
as $h\to0$, for each $f,g\in\AAA$. (Thus the main difference from strict
deformation quantization is that the product $T_h(f)T_h(g)$ is not required to
be in the range of~$T_h$.) Comparing the second condition with (\ref{tag:Da})
we see that $Q_f=T_h(f)$ gives the quantization rule we wanted. (We again
temporarily ignore (Q4) and~(Q5).) Though this seems not to have been treated
in Rieffel's papers, it is also obvious how to modify these definitions so as
to obtain the whole expansion (\ref{tag:Db}) instead of just~(\ref{tag:Da}).

Strict quantizations are already much easier to come by, see for instance
Landsman \cite{bib:LandsA} for coadjoint orbits of compact connected Lie
groups. However, even the notion of strict quantization is still unnecessarily
restrictive --- we shall see below that one can construct interesting
star-products even when (\ref{tag:Rj}) is satisfied only in a much weaker
sense. (The Berezin-Toeplitz quantization is a strict quantization but not
strict deformation quantization; the Berezin quantization is not even a strict
quantization.)

Recently, a number of advances in this ``operator-algebraic'' deformation
quantization %of Rieffel
have come from the theory of symplectic grupoids, see
Weinstein \cite{bib:WeinGrpd}, Zakrzewski \cite{bib:Zakr}, Landsman
\cite{bib:LandsO} \cite{bib:LandsJ}, and the books of Landsman
\cite{bib:LandsBk} and Weinstein and Cannas da Silva \cite{bib:CdSWei}.
A~discussion of deformation quantization of coadjoint orbits of a Lie group,
which again exhibits an intimate relationship to group representations and the
Kirillov orbit method, can be found e.g.~in Vogan~\cite{bib:Vogan}, Landsman
\cite{bib:LandsA}, Bar-Moshe and Marinov~\cite{bib:MarinA},
Lledo~\cite{bib:Lledo}, and Fioresi and Lledo~\cite{bib:FioL}.
A~gauge-invariant quantization method which, in the authors' words,
``synthesizes the geometric, deformation and Berezin quantization approaches'',
was proposed by Fradkin and Linetsky \cite{bib:FradLin} and
Fradkin~\cite{bib:Fradk}.

We remark that, in a sense, the second step in the deformation quantization is
not strictly necessary --- an~alternate route is to cast the von Neumann
formalism, interpreting $\spr{\Pi(Q_f)u}u$ (where $\Pi(Q_f)$ is the spectral
measure of~$Q_f$) as the probability distribution of the result of measuring
$f$ in the state~$u$, into a form involving only products of operators, and
then replace the latter by the corresponding star products. Thus, for instance,
instead of looking for eigenvalues of an operator~$Q_f$, i.e.~solving the
equation $Q_f u=\lambda u$, with $\|u\|=1$, one looks for solutions of $f*\pi=
\lambda\pi$, with $\pi=\overline\pi=\pi*\pi$ ($\pi$~corresponds to the
projection operator~$\spr\cdot u u$); or,~more generally, one defines the
{\sl $($star-$)$ spectrum\/} of $f$ as the support of the measure $\mu$ on
$\RR$ for which
$$ \operatorname{Exp}(tf) = \int_\RR e^{-2\pi i\lambda t/h} \,d\mu(\lambda)  $$
(in~the sense of distributions) where $\operatorname{Exp}(tf)$ is the {\sl star
exponential\/}
$$ \operatorname{Exp}(tf) := \sum_{m=0}^\infty \frac1{m!} \Big(\frac{2\pi
t}{ih}\Big)^m \,\underbrace{f*\dots*f}_{m\text{ times}} .  $$
See Bayen et~al.~\cite{bib:BFFLS}. In~this way, some authors even perceive
deformation quantization as a device for ``freeing'' the quantization of the
``burden'' of the Hilbert space.

Some other nice articles on deformation quantization are Sternheimer
\cite{bib:Stw}, Arnal, Cortet, Flato and Sternheimer \cite{bib:ACFS}, Weinstein
\cite{bib:WeinAst}, Fernandes \cite{bib:Fern}, and Blattner \cite{bib:Blatt};
two recent survey papers are Gutt \cite{bib:GuttV} and Dito and Sternheimer
\cite{bib:DiSt}. See~also Neumaier \cite{bib:Neumai}, Bordemann and Waldmann
\cite{bib:BoWa}, Karabegov \cite{bib:KarC} \cite{bib:KarD}, Duval, Gradechi
and Ovsienko \cite{bib:+DuGrOv}, and the above mentioned books by Fedosov
\cite{bib:FedosBk} and Landsman \cite{bib:LandsBk} and papers by Rieffel
\cite{bib:RiefB}~\cite{bib:Rief}.

\section{Berezin and Berezin-Toeplitz quantization on K\"ahler manifolds}
\label{sec4}
Recall that a Hilbert space $\HH$ whose elements are functions on a set $\Omg$
is called a {\sl reproducing kernel Hilbert space\/} (rkhs for short) if for
each $x\in\Omg$, the evaluation map $\phi\mapsto\phi(x)$ is continuous
on~$\HH$. By the Riesz-Fischer representation theorem, this means that there
exist vectors $K_x\in \HH$ such that
$$ \phi(x) = \spr {K_x}\phi \qquad \forall \phi\in \HH.  $$
The function
$$ K(x,y)=\spr{K_x}{K_y}, \qquad x,y\in\Omg $$
is called the {\sl reproducing kernel\/} of~$\HH$. Let us assume further that
the scalar product in $\HH$ is in fact the $L^2$ product with respect to some
measure $\mu$ on~$\Omg$. (Thus $\HH$ is a subspace of~$L^2(\Omg,\mu)$.) Then
any bounded linear operator $A$ on $\HH$ can be written as an integral
operator,
\begin{align*}   A\phi(x)
& = \spr{K_x}{A\phi} =\spr{A^*K_x} {\phi}= \into \phi(y) \overline{A^*K_x(y)}
\,d\mu(y) \\
& = \into \phi(y) \spr{A^*K_x}{K_y} \, d\mu(y) =
\into \phi(y) \spr {K_x}{AK_y} \,d\mu(y) ,  \end{align*}
with kernel $\spr {K_x}{AK_y}$. The~function
\begin{equation}  A(x,y) = \frac{\spr {K_x}{AK_y}}{\spr{K_x}{K_y}}
\label{tag:Bses}  \end{equation}
restricted to the diagonal is called the {\sl lower\/} (or {\sl covariant\/})
symbol $\tilde A$ of~$A$:
\begin{equation}  \tilde A(x) :=A(x,x) =
\frac{\spr{K_x}{AK_x}}{\spr{K_x}{K_x}}. \label{tag:Bsr}  \end{equation}
Clearly the correspondence $A\mapsto\tilde A$ is linear, preserves conjugation
(i.e.~$\widetilde{A^*}=\overline{\tilde A}$) and for the identity operator $I$
on $\HH$ one has $\tilde I=\jedna$.

For any function $f$ such that $f\HH\subset L^2(\Omg,\mu)$ --- for instance,
for any $f\in L^\infty(\Omg,\mu)$ --- the {\sl Toeplitz operator\/} on $\HH$ is
defined by $T_f(\phi) = P(f\phi)$, where $P$ is the orthogonal projection of
$L^2$ onto~$\HH$. In~other words,
\begin{equation}  T_f \phi(x) = \spr{K_x}{f\phi} = \into \phi(y) f(y) K(x,y)
\,d\mu(y). \label{tag:Bt}  \end{equation}
The function $f$ is called the {\sl upper\/} (or {\sl contravariant\/}\footnote
{The adjectives {\it upper\/} and {\it lower\/} seem preferable to the more
commonly used {\it contravariant\/} and {\it covariant,\/} as the latter have
quite different meanings in differential geometry. The terms {\sl active\/} and
{\sl passive\/} are also~used.}) symbol of the Toeplitz operator~$T_f$.
The operator connecting the upper and the lower symbol
\begin{equation}  f\mapsto \tilde T_f, \qquad \tilde T_f(x) = \into f(y) \,
\frac{|K(x,y)|^2} {K(x,x)} \,d\mu(y) =: Bf(x) ,  \label{tag:Bo}  \end{equation}
is called the {\sl Berezin transform.\/} (It~is defined only at points $x$
where $K(x,x)\neq0$.)

In~general, an operator $A$ need not be uniquely determined by its lower
symbol~$\tilde A$; however, this is always the case if $\Omg$ is a complex
manifold and the elements of $\HH$ are \underbar{holomorphic} functions. (This
is a consequence of the fact that $A(x,y)$ is then a meromorphic function of
the variables $y$ and $\overline x$, hence also of $u=y+\overline x$ and
$v=i(y-\overline x)$, and thus is uniquely determined by its restriction to the
real axes $u,v\in\RR^n$, i.e.~to~$x=y$.) In~that case the correspondence
$A\leftrightarrow\tilde A$ is a bijection from the space $\Cal B(\HH)$ of all
bounded linear operators on $\HH$ onto a certain subspace $\AAA_\HH\subset
C^\omega(\Omg)$ of real-analytic functions on~$\Omg$, and one can therefore
transfer the operator multiplication in $\Cal B(\HH)$ to a non-commutative and
associative product $*_\HH$ on~$\AAA_\HH$. Specifically, one~has
\begin{equation}  (f *_\HH g)(y) = \into f(y,x)g(x,y) \,
\frac{|K(x,y)|^2}{K(y,y)} \, d\mu(x), \qquad f,g\in\AAA_\HH,  \label{tag:Bpro}
\end{equation}
where $f(x,y),g(x,y)$ are functions on $\Omg\times\Omg$, holomorphic in $x$
and $\overline y$, such that $f(x,x)=f(x)$ and $g(x,x)=g(x)$
(cf.~(\ref{tag:Bses}) and~(\ref{tag:Bsr})).

In~particular, these considerations can be applied when $\HH$ is the Bergman
space $A^2(\Omg,\mu)$ of all holomorphic functions in the Lebesgue space
$L^2(\Omg,\mu)$ on a complex manifold $\Omg$ equipped with a measure $\mu$
such that $A^2(\Omg,\mu)\neq\{0\}$. Suppose now that we have in fact a family
$\mu_h$ of such measures, indexed by a small real parameter $h>0$. (It~suffices
that $h$ --- the Planck constant --- range over some subset of $\RR_+$ having 0
as an accumulation point.) Then one gets a family of Hilbert spaces
$\HH_h=A^2(\Omg,\mu_h)$ and of the corresponding products $*_{\HH_h}=:*_h$ on
the spaces $\AAA_{\HH_h}=:\AAA_h$. Berezin's idea (phrased in today's terms)
was to choose the measures $\mu_h$ in such a way that these products $*_h$
yield a star-product. More specifically, let $(\AAA,*)$ be the direct sum of
all algebras $(\AAA_h,*_h)$, and let $\tilde\AAA$ be a linear subset of $\AAA$
such that each $f=\{f_h(x)\}_h\in\tilde\AAA$ has an asymptotic expansion
\begin{equation}  f_h(x) = \sum_{j=0}^\infty h^j \, f_j(x) \qquad\text{as
}h\to0  \label{tag:Ba}  \end{equation}
with real-analytic functions $f_j(x)$ on~$\Omg$. We will say that $\tilde
\AAA$ is {\sl total\/} if for any $N>0$, $x\in\Omg$ and $F\in C^\omega(\Omg)
[[h]]$ there exists $f\in\tilde\AAA$ whose asymptotic expansion (\ref{tag:Ba})
coincides with $F(x)$ modulo~$O(h^N)$. Suppose that we can show that there
exists a total set $\tilde\AAA\subset\AAA$ such that for any
$f,g\in\tilde\AAA$, one has $f*g\in\tilde\AAA$ and
\begin{equation}  (f*g)_h(x) = \sum_{i,j,k\ge0} C_k(f_i,g_j)(x) \,h^{i+j+k}
\qquad \text{as } h\to0 ,  \label{tag:Bba}  \end{equation}
where $C_k:C^\omega(\Omg)\times C^\omega(\Omg)\to C^\omega(\Omg)$ are
some bilinear differential operators such that
\begin{equation}  C_0(\phi,\psi)=\phi\psi, \qquad C_1(\phi,\psi)-C_1(\psi,\phi)
= -\frac i{2\pi}\{\phi,\psi\}.  \label{tag:Bbd}  \end{equation}
Then the recipe
\begin{equation}  \Big( \sum_{i\ge0} f_i \, h^i\Big) * \Big(\sum_{j\ge0} g_j \,
h^j\Big) := \Big( \sum_{i,j,k\ge0} C_k(f_i,g_j) \, h^{i+j+k} \Big)
\label{tag:Bs}  \end{equation}
gives a star-product on $C^\infty(\Omg)[[h]]$ discussed in the preceding
section.  Moreover, this time it is not just a formal star product, since for
functions in the total set $\tilde\AAA$ it really exists as an element of
$C^\infty(\Omg)$, and, in fact, for each $h$ we can pass from $\AAA_h$ back
to $\Cal B(\HH_h)$ and thus represent $f_h(x)$ as an operator $\Oph f$ on the
Hilbert space~$\HH_h$. If~we can further find a linear and
conjugation-preserving ``lifting''
\begin{equation}  f\mapsto Lf  \label{tag:Bc}  \end{equation}
from $C^\infty(\Omg)$ (or a large subspace thereof) into $\tilde\AAA$ such
that $(L\phi)_0=\phi$, then the mapping
$$ \phi \mapsto \Oph (L\phi) =: Q_\phi  $$
will be the desired quantization rule, provided we can take care of the axioms
(Q4) (functoriality) and (Q5) (the case of~$\RR^{2n}\simeq\CC^n$). (It~is easy
to see that for real-valued $\phi$ the operators $\Oph(L\phi)$ are
self-adjoint.)

To see how to find measures $\mu_h$ satisfying
(\ref{tag:Bba})--(\ref{tag:Bbd}), consider first the case when there is a group
$G$ acting on $\Omg$ by biholomorphic transformations preserving the
symplectic form~$\omega$. In~accordance with our axiom (Q4), we then want the
product $*$ to be $G$-invariant, i.e.~to satisfy~(\ref{tag:Dda}).
An~examination of (\ref{tag:Bpro}) shows that for two Bergman spaces
$\HH=A^2(\Omg,\mu)$ and $\HH'=A^2(\Omg,\mu')$, the products $*_\HH$ and
$*_{\HH'}$ coincide if and only if
\begin{equation}  \frac{|K(x,y)|^2}{K(y,y)} \,d\mu(x) =
\frac{|K'(x,y)|^2}{K'(y,y)} \,d\mu'(x). \label{tag:Bd}  \end{equation}
In~particular, $d\mu'/d\mu$ has to be a squared modulus of an analytic
function; conversely, if $d\mu'=|F|^2\,d\mu$ with holomorphic ~$F$, then one
can easily check that $K(x,y)=\overline{F(x)}F(y) K'(x,y)$, and hence
(\ref{tag:Bd}) holds. Thus the requirement that $*_\HH$ be $G$-invariant means
that there exist analytic functions $\phi_g$, $g\in G$, such that
$$ d\mu(g(x)) = |\phi_g(x)|^2 \, d\mu(x).  $$
Assuming now that $\mu$ is absolutely continuous with respect to the
($G$-invariant) measure $\nu=\bigwedge^n\omega$ on~$\Omg$,
$$ d\mu(x) = w(x) \,d\nu(x),  $$
the last condition means that
$$ w(g(x)) = w(x) |\phi_g(x)|^2 .  $$
Hence the form $\partial\overline\partial\log w$ is $G$-invariant. But the
simplest examples of $G$-invariant forms (and if $G$ is sufficiently ``ample'',
the only ones) are clearly the constant multiples of the form~$\omega$. Thus if
$\omega$ lies in the range of $\partial\overline\partial$, i.e.~if $\omega$ is
not only symplectic but K\"ahler, we are led to take
\begin{equation}  d\mu_h(x) = e^{-\alpha \Phi(x)} \, d\nu(x)  \label{tag:Be}
\end{equation}
where $\alpha=\alpha(h)$ depends only on $h$ and $\Phi$ is a {\sl K\"ahler
potential\/} for the form~$\omega$
(i.e.~$\omega=\partial\overline\partial\Phi$).

In~his papers \cite{bib:Berez}, Berezin showed that for $\Omg=\CC^n$ with the
standard K\"ahler form $\omega= i \sum_j dz_j\wedge d\overline z_j$, as
well as for $(\Omg,\omega)$ a~bounded symmetric domain with the invariant
metric, choosing $\mu_h$ as in (\ref{tag:Be}) with $\alpha=1/h$ indeed yields
an (invariant) product $*$ satisfying (\ref{tag:Bba})--(\ref{tag:Bbd}), and
hence one obtains a star product. Berezin did not consider the
``lifting''~(\ref{tag:Bc}) (in~fact, he viewed his whole procedure as a means
of freeing the quantum mechanics from the Hilbert space!), but he established
an asymptotic formula for the Berezin transform $B=B_h$ in (\ref{tag:Bo}) as
$h\to0$ from which it follows that one can take as the lifting $Lf$ of $f\in
C^\infty(\Omg)$ the Toeplitz operators $T_f=T^{(h)}_f$ given by
(\ref{tag:Bt}). Finally, in the case of $\Omg=\CC^n\simeq\RR^{2n}$ one
obtains for $T_{\operatorname{Re}z_j}$ and $T_{\operatorname{Im}z_j}$ operators
which can be shown to be unitarily equivalent to the Schr\"odinger
representation (\ref{tag:Schro}). Thus we indeed obtain the desired
quantization rule.

For a long time, the applicability of Berezin's procedure remained confined
essentially to the above two examples, in other words, to Hermitian symmetric
spaces. The~reason was that it is not so easy to prove the formulas
(\ref{tag:Bba})--(\ref{tag:Bbd}) for a general K\"ahler manifold (with the
measures given  by~(\ref{tag:Be})). Doing this is tantamount to obtaining the
asymptotics (as~$h\to0$) of the Berezin transform~(\ref{tag:Bo}), which in turn
depend on the asymptotics of the reproducing kernels~$K_h(x,y)$. For~$\CC^n$
and bounded symmetric domains, these kernels can be computed explicitly, and
turn out to be given~by \begin{equation}  K_\alpha(x,y) = c(\alpha)
e^{\alpha\Phi(x,y)},  \label{tag:Bf}  \end{equation}  where $c(\alpha)$ is a
polynomial in $\alpha$ and $\Phi(x,y)$ is a function analytic in $x,\overline
y$ which coincides with the potential $\Phi(x)$ for~$x=y$. It~follows that
$$ B_\alpha f(x) = c(\alpha) \into f(y) e^{-\alpha S(x,y)} \, dy  $$
where $S(x,y)=\Phi(x,y)+\Phi(y,x)-\Phi(x,x)-\Phi(y,y)$, and one can apply the
standard Laplace (=stationary phase, WJKB) method to get the
asymptotics~(\ref{tag:Bo}).\footnote{The function $S(x,y)$ appeared for the
first time in the paper of Calabi \cite{bib:+Calabi} on imbeddings of K\"ahler
manifolds into~$\CC^n$, under the name of {\sl diastatic function.}}

Thus what we need is an analog of the formula (\ref{tag:Bf}) for a general
K\"ahler manifold. This was first established by Peetre and the second author
for $(\Omg,\omega)$ the annulus in $\CC$ with the Poincar\'e metric and $x=y$
\cite{bib:EP}, and then extended, in turn, to all planar domains with the
Poincar\'e metric \cite{bib:Duke}, to some Reinhardt domains in~$\CC^2$ with a
natural rotation-invariant form~$\omega$ \cite{bib:ET}, and finally to all
smoothly bounded strictly-pseudoconvex domains in $\CC^n$ with K\"ahler form
$\omega$ whose potential $\Phi$ behaves like a power of $\operatorname{dist}
(\cdot,\partial\Omg)$ near the boundary~\cite{bib:FR}~\cite{bib:ESI}.

So~far we have tacitly assumed that the potential $\Phi$ is a globally defined
function on~$\Omg$. We~hasten to remark that almost nothing changes if $\Phi$
exists only locally (which it always does, in view of the K\"ahlerness
of~$\omega$); the only change is that instead of functions one has to consider
sections of a certain holomorphic Hermitian line bundle, whose Hermitian
metric in the fiber is locally given by $e^{-\alpha\Phi(x)}$, and for this
bundle to exist certain cohomology integrality conditions (identical to the
prequantization conditions in the geometric quantization) have to be satisfied.
For~a more detailed discussion of reproducing kernels and of the upper and
lower symbols of operators in the line (or even vector) bundle setting, see
Pasternak-Winiarski \cite{bib:PWin}, Pasternak-Winiarski and Wojcieszynski
\cite{bib:PWinW} and Peetre~\cite{bib:Pee}.

We also remark that Berezin quantization of cotangent bundles (i.e.~$\Omg=T^*
\bfrakQ$ with the standard symplectic form~$\omega$) was announced by
\v{S}ere\v sevskii~\cite{bib:Sher}, who however was able to quantize only
functions polynomial in the moment variables~$p$.

In~the Berezin quantization, the formula (\ref{tag:Da}) is satisfied only in
the following weak sense,
$$ \big\langle K\h_x, \big( [Q\h_\phi,Q\h_\psi] + \tfrac{ih}{2\pi}
Q\h_{\{\phi,\psi\}} \big) K\h_y \big\rangle = O(h^2) \qquad
\forall x,y\in\Omg, \ \forall \phi,\psi\in C^\infty(\Omg).  $$
(We~write $Q\h_\phi$ instead of $Q_\phi$ etc.~in order to make clear the
dependence on~$h$.) A~natural question is whether one can strengthen this to
hold in the operator norm. More specifically, using the lifting $L:f\mapsto T\h
_f$ given by the Toeplitz operators, one would like to replace
(\ref{tag:Bba})~by
\begin{equation}  \Big\| T\h_f T\h_g - \sum_{j=0}^N h^j \,
T\h_{C_j(f,g)} \Big\|  _{\Cal B(\HH_h)} = O(h^{N+1})   \label{tag:Bg}
\end{equation}
for all $N>0$, for some bilinear differential operators $C_j$
satisfying~(\ref{tag:Bbd}). This~is called the {\sl Berezin-Toeplitz\/}
(or~{\sl Wick\/}) quantization. In~the language of the preceding section,
Berezin-Toeplitz quantization (unlike Berezin quantization) is an example of
a strict quantization in the sense of Rieffel. (Here and throughout the rest
of this section, the Toeplitz operators are still taken with respect to the
measures (\ref{tag:Be}) with~$\alpha=1/h$.)

Curiously enough, (\ref{tag:Bg}) was first established not for $\Omg=\CC^n$
with the Euclidean metric, but for the unit disc and the Poincar\'e metric; see
Klimek and Lesniewski~\cite{bib:KLa}. The~same authors subsequently extended
these results to any plane domain using uniformization~\cite{bib:KLb}, and to
bounded symmetric domains with Borthwick and Upmeier~\cite{bib:BLU}.
(Supersymmetric generalizations also exist, see~\cite{bib:BLUR}.) The~case of
$\CC^n$ was treated later by Coburn~\cite{bib:Cob}. For~compact K\"ahler
manifolds (with holomorphic sections of line bundles in place of holomorphic
functions), a very elegant treatment was given by Bordemann, Meinrenken and
Schlichenmaier \cite{bib:BMS} using the theory of generalized Toeplitz
operators of Boutet de Monvel and Guillemin~\cite{bib:BdMG}; see also
Schlichenmaier \cite{bib:Schli} \cite{bib:SchliM}, Karabegov and Schlichenmaier
\cite{bib:KaSchli}, Guillemin \cite{bib:Guill}, Zelditch \cite{bib:Zeld} and
Catlin~\cite{bib:Tian}. The~same approach also works for smoothly bounded
strictly pseudoconvex domains in $\CC^n$ with K\"ahler forms $\omega$ whose
potential behaves nicely at the boundary, see~\cite{bib:ESI}, as well as for
$\Omg=\CC^n$ with the standard (=Euclidean) K\"ahler form~\cite{bib:Borthw}.
For~some generalizations to non-K\"ahler case see Borthwick and
Uribe~\cite{bib:BU}.

We remark that the star products (\ref{tag:Bs}) determined by the $C_j$ in
(\ref{tag:Bg}) and in (\ref{tag:Bba}) are not the same; they are, however,
equivalent, in the following sense. If~one views the Berezin transform
(\ref{tag:Bo}) formally as a power series in $h$ with differential operators on
$\Omg$ as coefficients, then
$$ B_h( f*_{BT} g) = (B_h f) *_B (B_h g) , $$
where $*_B$ and $*_{BT}$ stand for the star products (B=Berezin,
BT=Berezin-Toeplitz) coming from (\ref{tag:Bba}) and~(\ref{tag:Bg}),
respectively. In~the terminology of~\cite{bib:KarA}, the two products are
{\sl duals\/} of each other. See the last page in~\cite{bib:ESI} for the
details. The Berezin-Toeplitz star product $*_{BT}$ is usually called
{\sl Wick,\/} and the Berezin star product $*_B$ {\sl anti-Wick.\/}
(For~$\Omg=\CC^n\simeq\RR^{2n}$, they are further related to the Moyal-Weyl
product $*_{MW}$ from Section~\ref{sec1} by $B_h^{1/2}(f*_{MW}g)=B_h^{1/2}f
*_B B_h^{1/2}g$, or $B_h^{1/2}(f*_{BT}g)=B_h^{1/2}f*_{MW} B_h^{1/2}g$, where
$B_h^{1/2}=e^{h\Delta/2}$ is the square root of $B_h=e^{h\Delta}$.)

Berezin's ideas were initially developed further only in the context of
symmetric (homogeneous) spaces, i.e.~in the presence of a transitive action of
a Lie group. The~coefficients $C_j(\cdot,\cdot)$ are then closely related to
the invariant differential operators on~$\Omg$; see Moreno \cite{bib:Mor},
Moreno and Ortega-Navarro \cite{bib:MorON}, Arnal, Cahen and Gutt
\cite{bib:Arn} and Bordemann et al.~\cite{bib:BorAL} for some interesting
results on star products
in this context. Some connections with Rieffel's $C^*$-algebraic theory can be
found in Radulescu~\cite{bib:Radul}. Formal Berezin and Berezin-Toeplitz star
products on arbitrary K\"ahler manifolds were studied by Karabegov
\cite{bib:KarA}, \cite{bib:KarCf}, Karabegov and Schlichenmaier
\cite{bib:KaSchli} and Reshetikhin and Takhtajan~\cite{bib:ReTa} (cf.~also
Cornalba and Taylor \cite{bib:CornTay} for a formal expansion of the Bergman
kernel);  see~also Hawkins~\cite{bib:Hawk}.

Evidently, a central topic in these developments is the dependence of the
reproducing kernel $K_\mu(x,y)$ of a Bergman space $A^2(\Omg,\mu)$ on the
measure~$\mu$. This dependence is still far from being well understood. For
instance, for $(\Omg,\omega)$ a Hermitian symmetric space (or~$\CC^n$) with the
invariant metric and the corresponding K\"ahler form~$\omega$, $\Phi$ a
potential for $\omega$, and $\nu=\bigwedge^n\omega$ the Liouville (invariant)
measure ($n=\dim_\CC\Omg$), the weight function $w(x)=e^{-\alpha\Phi(x)}$
(with~$\alpha\gg0$) has the property that
$$ K_{w\,d\nu}(x,x) = \frac{\text{const.}}{w(x)} .  $$
The existence of similar weights $w$ on a general K\"ahler manifold is an open
problem. See~Odzi\-je\-wicz \cite{bib:OdzijA}, p.~584, for some remarks and
physical motivation for studying equations of this type. Some results on the
dependence $\mu\mapsto K_\mu$ are in Pasternak-Winiarski~\cite{bib:PWb}.

\section{Prime quantization} \label{sec5}
The most straightforward way of extending (\ref{tag:Schro}) to more general
functions on $\RR^{2n}$ is to specify a {\sl choice of ordering.\/} For
instance, for a polynomial
\begin{equation}  f(p,q)= \sum_{m,k} a_{mk} q^m p^k  \label{tag:PO}
\end{equation}
one can declare that $Q_f=f(Q_p,Q_q)$ with the $Q_q$ ordered to the left of
the~$Q_p$:
\begin{equation}  Q(f) = \sum_{m,k} a_{mk} Q^m_q Q^k_p.  \label{tag:PA}
\end{equation}
(Here $m,k$ are multiindices and we ignore the subtleties concerning the
domains of definition~etc. We~will also sometimes write $Q(f)$ instead of
$Q_f$, for typesetting reasons.) Extending this (formally) from polynomials to
entire functions, in particular to the exponentials $e^{2\pi
i(p\cdot\xi+q\cdot\eta)}$, we~get\footnote{Here we are using the real scalar
product notation $p\cdot\xi=p_1\xi_1+\dots+p_n\xi_n$.}
$$ Q(e^{2\pi i(p\cdot\xi+q\cdot\eta)})
= e^{2\pi i \eta \cdot Q(q)} e^{2\pi i \xi\cdot Q(p)}.  $$
Finally, decomposing an ``arbitrary'' function $f(p,q)$ into exponentials via
the Fourier transform, as in Section~\ref{sec1}, we arrive at a quantization
recipe
\begin{equation}  Q_f \phi(x) = \iint f(p,x) \,e^{2\pi i(x-y)\cdot p /h}
\phi(y) \,d p \,dy. \label{tag:PBB}  \end{equation}
Similarly, using instead of (\ref{tag:PA}) the opposite choice of ordering
\begin{equation}  Q(\sum_{m,k} q^k p^m) = \sum_{m,k} Q_p^k Q_q^m
\label{tag:PC}  \end{equation}
we arrive at
\begin{equation}  Q_f \phi(x) = \iint f(p,y) \,e^{2\pi i(x-y)\cdot p /h}
\phi(y) \,d p \,dy. \label{tag:PD}  \end{equation}
The rules (\ref{tag:PD}) and (\ref{tag:PBB}) are the standard Kohn-Nirenberg
calculi of pseudodifferential operators, see~\cite{bib:KN},
\cite{bib:Foll},~\S23. A~more sophisticated set of ordering rules generalizing
(\ref{tag:PBB}) and (\ref{tag:PD}) can be obtained by fixing a $t\in[0,1]$ and
setting
\begin{equation}  Q_f \phi(x) = \iint f(p,(1-t)x+ty)\,e^{2\pi i(x-y)\cdot p /h}
\phi(y)  \,d p \,dy.  \label{tag:PU}  \end{equation}
The choice $t=\frac12$ gives the Weyl calculus (\ref{tag:WEY}), which can thus
be thought of as corresponding to a ``symmetric'' ordering of $Q_q$ and~$Q_p$.

The drawback of (\ref{tag:PA}) and (\ref{tag:PC}) is that they need not be
self-adjoint operators for real-valued symbols~$f$. This can be remedied by
viewing $\RR^{2n}$ as $\CC^n$ and making the change of coordinates $z=(q+ip)/
\sqrt2$, $\overline z=(q-ip)/\sqrt2$. The operators $Q_z$ and $Q_{\overline
z}=Q_z^*$ are then the {\sl annihilation\/} and {\sl creation\/} operators
$$ Q_z= \frac{Q_q+i Q_p}{\sqrt 2}, \qquad Q_z^*=\frac{Q_q-i Q_p}{\sqrt2}. $$
One can then again assign to a polynomial $f(z,\overline z)=\sum b_{mk} z^m
\overline z^k$ either the operator
$$ Q_f = \sum b_{mk} Q(z)^m Q(z)^{*k}  $$
or the operator
$$ Q_f = \sum b_{mk} Q(z)^{*k} Q(z)^m  $$
which is called the Wick (or normal) and the anti-Wick (anti-normal) ordering,
respectively. The corresponding Wick and anti-Wick calculi are discussed in
\S\ref{sec27} of Folland's book~\cite{bib:Foll}. The anti-Wick calculus turns
out not to be so interesting, but the Wick calculus has an important
reformulation if we replace the underlying Hilbert space $L^2(\RR^n)$, on which
the operators $Q_f$ act, by the {\sl Fock\/} (or~{\sl Segal-Bargmann\/}) space
$A^2(\CC^n,\mu_h)$ of all entire functions on $\CC^n$ square-integrable with
respect to the Gaussian measure $d\mu_h(z):=(\pi h)^{-n} e^{-|z|^2/h} dz$
($dz$~being the Lebesgue measure on~$\CC^n$). Namely, the Bargmann transform
\begin{equation}  \beta: \, L^2(\RR^n)\ni f \longmapsto \beta f(z):= (2\pi
h)^{n/4}  \int_{\RR^n} f(x) e^{2\pi x\cdot z -h \pi^2 x\cdot x -z\cdot z/2h}
\, dx \in A^2(\CC^n,\mu_h)  \label{tag:Barg}  \end{equation}
is a unitary isomorphism and upon passing from $L^2(\RR^n)$ to
$A^2(\CC^n,\mu_h)$ via $\beta$, the operators $Q_f$ become the familiar
Toeplitz operators~(\ref{tag:Bt}):
\begin{equation}  \beta Q_f \beta^{-1} = T_f, \text{ with} \qquad T_f \phi(x)
:= \int_{\CC^n} f(y) \phi(y) K_h(x,y) \, d\mu_h(y),  \label{tag:PE}
\end{equation}
where $K_h(x,y)=e^{x\overline y/h}$ is the reproducing kernel for the
space~$A^2(\CC^n,\mu_h)$. In~this way, we thus recover on $\CC^n$ the
Berezin-Toeplitz quantization discussed in the preceding section.

Another way of writing (\ref{tag:PE})~is
\begin{equation}  T_f = \int_{\CC^n} f(y) \,\Delta_y\,dy,  \label{tag:PF}
\end{equation}
where $\Delta_y=\vert k_y\rangle\langle k_y\vert =\spr{k_y}{\cdot\,}k_y$ is the
rank-one projection operator onto the complex line spanned by the unit vector
\be k_y := \frac{K_h(\,\cdot\,,y)}{\|K_h(\,\cdot\,,y)\|}.  \label{tag:KSKS} \en

This suggests looking, quite generally, for quantization rules of the
form~(\ref{tag:PF}), with a set of ``quantizers'' $\Delta_y$ ($y\in\Omg$)
which may be thought of as reflecting the choice of ordering. This is the basis
of the {\sl prime quantization\/} method introduced in~\cite{bib:AliD}
(see~also \cite{bib:Prugo}), where it is also explained how the choice of the
quantizers (hence also of the ordering) is to be justified on physical grounds.
The~main result of \cite{bib:AliD} is that if the quantizers $\Delta_y$ are
bounded positive operators, $\Delta_y\ge0$, on~some (abstract) Hilbert
space~$\HH$, then there exists a direct integral Hilbert space
$\KK=\into^\oplus \KK_x \,d\nu(x)$ (see~\cite{bib:BratteliR}), where $\KK_x$ is
a family of separable Hilbert spaces indexed by $x\in\Omg$ and $\nu$ is a
measure on~$\Omg$, and an isometry $\iota:\HH\to\KK$ of $\HH$ onto a subspace
of $\KK$ such that
\begin{enumerate}
\item[(i)] $\iota\HH$ is a ``vector-valued'' reproducing kernel Hilbert space,
in the sense that for each $x\in\Omg$ there is a bounded linear operator
$E_x$ from $\iota\HH$ into $\KK_x$ such that for any $f=\into^\oplus f_y\,d\nu
(y) \in\iota\HH$, one has
\begin{equation}  f_x = \into E_x E_y^* f_y \, d\nu(y) \qquad \forall
x\in\Omg. \label{tag:ADa} \end{equation}
\item[(ii)] $\iota\Delta_y\iota^* = E_y^* E_y$.  \end{enumerate}
The operators
$$ T_f = \into f(y) \,\Delta_y \,d\nu(y)  $$
thus satisfy
\begin{equation}  T_f = \into f(y) \,\iota^* E^*_y E_y\iota \,d\nu(y).
\label{tag:ADb}  \end{equation}
If $\KK_x=\CC$ for every $x\in\Omg$, one can identify $\KK$ with
$L^2(\Omg,\nu)$, $\iota$ with an inclusion map of $\HH$ into $\KK$, and $E_x$
with the functional $\spr{K_x}{\cdot\,}$ for some vector $K_x\in\HH$; thus
(\ref{tag:ADa}) becomes
$$ f(x) = \into f(y) \, K(x,y) \,d\nu(y) \qquad\forall f\in\HH,
\text{ where }K(x,y):=\spr{K_x}{K_y},  $$
so $\HH$ is an (ordinary) reproducing kernel Hilbert subspace of
$L^2(\Omg,\nu)$ with reproducing kernel $K(x,y)$, and (\ref{tag:ADb}) reads
$$ T_f = \into f(y) \, \vert K_y\rangle\langle K_y \vert \,d\nu(y),  $$
i.e.~$T_f$ is the Toeplitz type operator
$$ T_f \phi = P(f\phi)   $$
where $P$ is the orthogonal projection of $L^2(\Omg,\nu)$ onto~$\HH$.
In~particular, for $\Omg=\CC^n$ and $\nu$ the Gaussian measure we recover
(\ref{tag:PE}) and~(\ref{tag:PF}).

Note that the Weyl quantization operators~(\ref{tag:WEY}), transferred to
$A^2(\CC^n,\mu_h)$ via the Bargmann transform (\ref{tag:Barg}), can also be
written in  the form~(\ref{tag:PF}), namely (cf.~\cite{bib:Foll}, p.~141)
\begin{equation}  \beta W_f \beta^{-1} = \int_{\CC^n} f(y) \, s_y \, dy ,
\label{tag:PW}  \end{equation}
where $y=q-ip$ ($(p,q)\in\RR^{2n},y\in\CC^n$) and
$$ s_y \phi(z) = \phi(2y-z) e^{2\overline y\cdot(z-y)/h} $$
is the self-adjoint unitary map of $A^2(\CC^n,\mu_h)$ induced by the symmetry
$z\mapsto 2y-z$ of~$\CC^n$. In~contrast to (\ref{tag:PF}), however, this time
the quantizers $s_y$ are not positive operators.

Given $\Delta_y$, one can also consider the ``dequantization'' operator
$T\mapsto\tilde T$,
\begin{equation}  \tilde T(y) := \operatorname{Trace} \, (T\Delta_y),
\label{tag:PG}  \end{equation}
which assigns functions to operators. For the Weyl calculus, it turns out that
$\tilde W_f=f$, a~reflection of the fact that the mapping $f\mapsto W_f$ is a
unitary map from  $L^2(\RR^{2n})$ onto the space of Hilbert-Schmidt operators
(an~observation due to Pool~\cite{bib:Pool}). For the Wick calculus
(\ref{tag:PF}), $\tilde T_f$ is precisely the Berezin transform of~$f$,
discussed above, and the function $\tilde T$ is the lower (covariant, passive)
symbol of the operator~$T$ (and $f$ is the upper (contravariant, active) symbol
of the Toeplitz operator~$T_f$). Using the same ideas as in the previous
section, one can thus try to construct, for a general set of
quantizers~$\Delta_y$, a Berezin-Toeplitz type star product
$$ f *_h g = \sum_{j\ge0} C_j(f,g) \, h^j, \qquad f,g\in C^\infty(\Omg),  $$
by establishing an asymptotic expansion for the product of two operators of the
form~(\ref{tag:PF}),
$$ T_f T_g = \sum_{j\ge0} h^j \, T_{C_j(f,g)} \qquad\text{as }h\to0,  $$
and, similarly, a Berezin-type star product by setting
$$ \tilde T_f *_h \tilde T_g := \widetilde{T_f T_g}.  $$
In~this way we see that the formula (\ref{tag:PF}), which at first glance might
seem more like a mathematical exercise in pseudodifferential operators rather
than a sensible quantization rule, effectively leads to most of the
developments (at least for~$\RR^{2n}$) we did in the previous two sections.

In~the context of $\RR^{2n}$, or, more generally, of a coadjoint orbit of a Lie
group, the ``quantizers'' and ``dequantizers'' above seem to have been first
studied systematically by Gracia-Bondia~\cite{bib:GraBo}; in~a~more general
setting, by Antoine and Ali~\cite{bib:AAli}. Two~recent papers on this topic,
with some intriguing ideas, are Karasev and Osborn~\cite{bib:+KaOsb}. For~some
partial results on the Berezin-Toeplitz star-products for general quantizers,
see Engli\v s~\cite{bib:+Elmp}. The~operators (\ref{tag:PU}) and the
corresponding ``twisted product'' $f\,\sharp\,g$ defined by $Q_{f\sharp g}=
Q_f Q_g$ were investigated by
Unterberger~\cite{bib:Unta} (for~$t=\mfr12$, see H\"ormander~\cite{bib:HormW});
a~relativistic version, with the Weyl calculus replaced by ``Klein-Gordon'' and
``Dirac'' calculi, was developed by Unterberger~\cite{bib:Untr}. The~formula
(\ref{tag:PW}) for the Weyl operator makes sense, in general, on any Hermitian
symmetric space $\Omg$ in the place of~$\CC^n$, with $s_y$ the self-adjoint
unitary isomorphisms of $A^2(\Omg)$ induced by the geodesic symmetry
around~$y$; in~this context, the Weyl calculus on bounded symmetric domains was
studied by Upmeier~\cite{bib:UpmW}, Unterberger and Upmeier \cite{bib:UntU},
and Unterberger \cite{bib:Untw} \cite{bib:UntESI}. Upon rescaling and letting
$h\to0$, one obtains the so-called Fuchs calculus~\cite{bib:UntF}. A~general
study of invariant symbolic calculi (\ref{tag:PF}) on bounded symmetric domains
has recently been undertaken by Arazy and Upmeier~\cite{bib:ArUp}.

An~important interpretation of the above-mentioned equality of the
$L^2(\RR^{2n})$-norm of a function $f$ and the Hilbert-Schmidt norm of the Weyl
operator $W_f$ is the following. Consider once more the~map
$$ \Gamma: f\mapsto T_f  $$
mapping a function $f$ to the corresponding Toeplitz operator (\ref{tag:PE}),
and let $\Gamma^*$ be its adjoint with respect to the $L^2(\RR^{2n})$ inner
product on $f$ and the Hilbert-Schmidt product on~$T_f$. One then checks easily
that $\Gamma^*$ coincides with the dequantization operator~(\ref{tag:PG}). Now
by the abstract Hilbert-space operator theory, $\Gamma$ admits the polar
decomposition
\begin{equation}  \begin{aligned}
& \Gamma = W R,  \qquad\text{with $R:=(\Gamma^*\Gamma)^{1/2}$ and $W$ a partial
isometry with initial} \\
\noalign{\vskip-3pt}
& \text{space $\overline{\operatorname{Ran}\Gamma^*}$ and final space
$\overline{\operatorname{Ran}\Gamma}$.}  \end{aligned}  \label{tag:PH}
\end{equation}
A~simple calculation shows, however, that $\Gamma^*\Gamma$ is precisely the
Berezin transform associated to $A^2(\CC^n,\mu_h)$,
$$ \Gamma^*\Gamma f(y) = \int_{\CC^n} f(x) \, \frac{|K_h(y,x)|^2}{K_h(y,y)}
\, d\mu_h(x) = e^{h\Delta} f(y),  $$
and using the Fourier transform to compute the square root
$(\Gamma^*\Gamma)^{1/2}$ one discovers that $W$ is precisely the Weyl transform
$f\mapsto W_f$. This fact, first realized by Orsted and Zhang~\cite{bib:OeZ}
(see~also Peetre and Zhang~\cite{bib:PeeZh} for a motivation coming from
decompositions of tensor products of holomorphic discrete series
representations), allows us to define an analogue of the Weyl transform by
(\ref{tag:PH}) for any reproducing kernel subspace of any $L^2$ space. For~the
standard scale of weighted Bergman spaces on bounded symmetric domains in
$\CC^n$, this generalization has been studied in Orsted and
Zhang~\cite{bib:OeZ} and Davidson, Olafsson and Zhang~\cite{bib:+Olaf};
the~general case seems to be completely unexplored at present.

From the point of view of group representations, the unit vectors $k_y$
in~(\ref{tag:KSKS}) are the {\sl coherent states\/} in the sense of Glauber
\cite{bib:Glaub}, Perelomov \cite{bib:Perel} and Onofri~\cite{bib:Onof}.
Namely, the group $G$ of all distance-preserving biholomorphic self-maps of
$\CC^n$ (which coincides with the group of orientation-preserving rigid motions
$x\mapsto Ax+b$, $A\in U(n)$, $b\in\CC^n$) acts transitively on $\CC^n$ and
induces a projective unitary representation
$$ U_g: \phi(x) \mapsto \phi(gx) e^{-\spr b{Ax}/h-|b|^2/2h}
\qquad (gx=Ax+b, \  g\in G)  $$
of $G$ in $A^2(\CC^n)$; and the vectors $k_y$ are unit vectors satisfying
\begin{equation}  U_g k_y = \epsilon \, k_{gy}  \label{tag:Coh}  \end{equation}
for some numbers $\epsilon=\epsilon(g,y)$ of unit modulus. Coherent states for
a general group $G$ of transformations acting transitively on a manifold
$\Omg$, with respect to a projective unitary representation $U$ of $G$ in a
Hilbert space~$\HH$, are similarly defined as a family $\{k_y\}_{y\in \Omg}$ of
unit vectors in~$\HH$ indexed by the points of $\Omg$ such that (\ref{tag:Coh})
holds. Choosing a basepoint $0\in \Omg$ and letting $H$ be the subgroup of $G$
which leaves the subspace $\CC k_0$ invariant (i.e.~$g\in H$ iff $U_g
k_0=\epsilon(g) k_0$ for some $\epsilon(g)\in\CC$ of modulus~1), we~can
identify $\Omg$ with the homogeneous space~$G/H$. Suppose that there exists a
biinvariant measure $dg$ on~$G$, and let $dm$ be the corresponding invariant
measure on~$\Omg=G/H$. We~say that the coherent states $\{k_y\}_{y\in \Omg}$
are {\sl square-integrable\/}~if
$$ \into |\spr{k_x}{k_y}|^2 \, dm(y) =:d<\infty   $$
(in view of (\ref{tag:Coh}), the value of the integral does not depend on the
choice of~$x\in \Omg$). If~the representation $U$ is irreducible, it is then
easy to see from the Schur lemma that
$$ \frac1d \into \vert k_y \rangle\langle k_y\vert \, dm(y) = I
\qquad\text{(the identity on $\HH$).}  $$
It~follows that the mapping $\HH\ni f\mapsto f(y):= \spr {k_y}f$ identifies
$\HH$ with a subspace of $L^2(\Omg,dm)$ which is a reproducing kernel space
with kernel $K(x,y)= d^{-1} \spr{k_x}{k_y}$. Thus, in some sense, the
quantizers $\Delta_y$ above and their associated reproducing kernel Hilbert
spaces may be regarded as generalizations of the coherent states to the
situation when there is no group action present. For~more information on
coherent states and their applications in quantization, see for instance
Klauder \cite{bib:Klaud}, Odzijewicz \cite{bib:OdziB}, Unterberger
\cite{bib:UntCS}, Ali and Goldin \cite{bib:AliGo}, Antoine and Ali
\cite{bib:AAli}, Ali~\cite{bib:AliCS}, Bartlett, Rowe and Repka
\cite{bib:+Rowe}, and the survey by Ali, Antoine, Gazeau and
Mueller~\cite{bib:AAGM}, as well as the recent book \cite{bib:AAGbk}, and the
references therein. An~interesting characterization of the cut locus of a
compact homogeneous K\"ahler manifold in terms of orthogonality of coherent
states has recently been given by Berceanu~\cite{bib:Berc}. We~will have more
to say about coherent states in Section~\ref{sec-cohstqua} below.

Another way of arriving at the Toeplitz-type operators (\ref{tag:PE}) is via
geometric quantization. Namely, consider a phase space $(\Omg,\omega)$ which
admits a K\"ahler polarization~$F$, i.e.~one for which $F\cap\overline F=\{0\}$
(hence $F+\overline F=T^*_\CC \Omg$). The~functions constant along $F$ can then
be interpreted as holomorphic functions, the corresponding $L^2$-space becomes
the Bergman space, and the quantum operators (\ref{tag:GQF}) become, as has
already been mentioned above, Toeplitz operators. This link between geometric
and Berezin quantization was discovered by
Tuynman~\cite{bib:TuyGBK}~\cite{bib:TuyCm}, who showed that on a compact
K\"ahler manifold (as well as in some other situations) the operators $Q_f$ of
the geometric quantization coincide with the Toeplitz operators $T_{f+h\Delta
f}$, where $\Delta$ is the Laplace-Beltrami operator. Later on this connection
was examined in detail in a series of papers by Cahen~\cite{bib:Cahen} and
Cahen, Gutt and Rawnsley~\cite{bib:CGR} (parts I and II of \cite{bib:CGR} deal
with compact manifolds, part~III with the unit disc, and part~IV with
homogeneous spaces). See~also Nishioka \cite{bib:Nishi} and
Odzijewicz~\cite{bib:OdziB}.

In~a sense, the choice of polarization in geometric quantization plays a
similar role as the choice of ordering discussed in the paragraphs above, see
Ali and Doebner~\cite{bib:AliD}. Another point of view on the ordering problem
in geometric quantization is addressed in Bao and Zhu~\cite{bib:BaoZ}.

\section{Coherent state quantization}
\label{sec-cohstqua}
The method of coherent state quantization is in some respects a particular
case of the prime quantization from the previous section, exploiting the
prequantization of the projective Hilbert space. Some representative references
are by Odzijewicz~\cite{bib:OdzijA} \cite{bib:OdziB} \cite{bib:OdzijS}
\cite{bib:OdzijD} and Ali~\cite{bib:AliCS} \cite{bib:AAli} \cite{bib:AliM}.
We~begin with a quick review of the symplectic geometry of the projective
Hilbert space.

\subsection{The projective Hilbert space}\label{projhilbsp}
Let $\HH$ be a Hilbert space of dimension~$N$, which could be (countably)
infinite or finite. As~a set, the {\sl projective Hilbert space\/} $\prhs$ will
be identified with the collection of all orthogonal projections onto
one-dimensional subspaces of $\HH$ and  for each non-zero vector $\psi \in \HH$
let  $\Psi = \frac{1}{\parallel \psi \parallel^2}\vert \psi\rangle\langle  \psi
\vert$ denote the corresponding projector. There is a natural  K\"ahler
structure on $\prhs$ as we now demonstrate. An analytic atlas of $\prhs$ is
given by the coordinate charts
\be
  \{ (V_{\Phi}, {\s h}_\phi , \HH_\Phi) \mid \phi \in \HH\setminus\{0\}  \},
\en
where
\be
 V_{\Phi} = \{ \Psi \in \prhs \mid \langle\phi \vert  \psi \rangle \not= 0\}
\en
is an open, dense set in $\prhs$;
\be
 {\HH}_\Phi = (I - \Phi){\HH} = \langle\phi\rangle^\perp
\en
is the subspace of $\HH$ orthogonal to the range of $\Phi$ and ${\s h}_\phi :
V_\Phi \to {\HH}_\Phi$ is the diffeomorphism
\begin{equation}
 {\s h}_\phi (\Psi) = \frac{1}{\langle \hat \phi\vert
 \psi\rangle}(I - \Phi)\psi,
  \quad \hat\phi = \frac{\phi}{\Vert\phi\Vert}.
\end{equation}
Since $V_\Phi$ is dense in $\HH$, it is often enough to consider only one
coordinate chart. Thus, we set $e_0 = \hat \phi$ and  choose an orthonormal
basis $\{e_j\}_{j=1}^{N-1}$ of $\HH_\Phi$ to obtain  a basis of $\HH$ which
will be fixed from now on. We may then identify $\prhs$ with
${\C}{\mathbb P}^N$: For arbitrary $\Psi \in \HH$ we set
\begin{equation}\label{coord}
z_j = \langle e_j |\psi \rangle, \quad j = 1,2, \ldots , N-1,
\end{equation}
and the coordinates of $\Psi \in \prhs$ are the standard homogeneous
coordinates
\begin{equation}
 \langle e_j \vert {\s h}_\phi (\Psi)\rangle =  Z_j = \frac{z_j}{z_0},
   \quad j = 1,2, \ldots , N-1,
 \label{homogcoord}
\end{equation}
of projective geometry.

The projection map $\pi : {\HH}\setminus\{ 0 \}  \to \prhs$  that assigns to
each $\psi \in {\HH}\setminus\{ 0\}$ the corresponding  projector $\Psi \in
\prhs$ is holomorphic in these coordinates. For $\Phi \in \prhs$ we have
$\pi^{-1}(\Psi) = {\C}^\ast \psi$ where $ {\C}^\ast=  {\C} \setminus \{0\}$,
and so  $\pi : {\HH} \setminus \{ 0\} \to \prhs$ is a $GL(1, {\C})$ principal
bundle, sometimes called the {\em canonical line bundle} over  $\prhs$. We will
denote the associated holomorphic line bundle by $\mathbb L(\HH)$ and write
elements in it as $(\Psi,\psi)$, where $\psi \in \Psi(\HH )$. (We~again write
$\pi$ for the canonical projection.) A~local trivialization of $\mathbb L(\HH)$
over $V_\Phi$ is given by the (holomorphic) reference section  $\hat s$ of
$\mathbb L(\HH)$:
\begin{equation}\label{unitsec}
  \hat s(\Psi) =(\Psi,\frac{\psi }{\langle\hat\phi\vert\psi \rangle}),
\end{equation}
and any other section $s: V_\Phi \to \mathbb L(\HH)$ is given by
\begin{equation}\label{kap}
s(\Psi) = (\Psi, \kappa(\Psi))
\end{equation}
where $\kappa : \prhs \to \HH \setminus \{0\}$ is a holomorphic map with
$\frac{\vert \kappa(\Psi) \rangle \langle \kappa(\Psi)\vert}{\Vert\kappa(\Psi)
\Vert^2} = \Psi$. Denote by $s_0$ the zero-section of $\mathbb L(\HH)$.
The~identification map $\imath_\mathbb L : \mathbb L(\HH) \setminus s_0\to \HH$
given~as
\begin{equation}\label{id}
\imath_\mathbb L (\Psi,\psi) = \psi ,
\end{equation}
yields a global coordinatization of $\mathbb L(\HH) \setminus s_0$.
For~any $\psi \in \HH$ let $\langle \psi \vert$ be its dual element.
The restriction of
$\langle \psi \vert$ to the fibre $\pi^{-1} (\Psi^\prime)$ in $\mathbb L(\HH)$,
for arbitrary
$\Psi^\prime \in \prhs$, then
yields a section $s^*_{\Psi^\prime}$ of the  dual bundle
$\mathbb L(\HH)^*$ of  $\mathbb L(\HH)$.
Moreover, the map $\Psi^\prime \mapsto  s^*_{\Psi^\prime}$ is antilinear
between  $\HH$ and
$\Gamma(\mathbb L(\HH)^*)$.  We may hence realize $\HH$ as a space of
holomorphic sections.

The tangent space $T_\Psi\prhs $ to $\prhs$ at the point $\Psi$ has a natural
identification  with $\HH_\Psi$ (obtainable, for example,  by differentiating
curves in $\prhs$ passing  through $\Psi$).
The complex structure of $\HH_\Psi$ then endows the tangent space $T_\Psi\prhs$
with an  integrable complex  structure $J_{\Psi}$, making $\prhs$ into a
K\"ahler manifold. The corresponding  canonical 2-form $\Omega_{FS}$, called
the {\em Fubini-Study 2-form}, is given pointwise by
\begin{equation}
 \Omega_{FS}(X_\Psi, Y_\Psi) =
 \frac{1}{2i} ( \langle \xi \vert \zeta \rangle - \langle
 \zeta\vert \xi \rangle),
\end{equation}
where $  \xi , \zeta  \in \HH_\Psi$  correspond to the tangent vectors $X_\Psi,
Y_\Psi$ respectively. The associated Riemannian metric $g_{FS}$ is given by
\begin{equation}
g_{FS}(X_\Psi, Y_\Psi) = \frac{1}{2} ( \langle \xi \vert \zeta \rangle +
\langle \zeta \vert \xi \rangle) =  \Omega_{FS}(X_\Psi, J_{\Psi}Y_\Psi) .
\end{equation}
In the local coordinates $Z_{j}$, defined in (\ref{homogcoord}), $\Omega_{FS}$
assumes the form
\begin{equation}
\Omega_{FS} = \frac{1}{1+ \Vert{\mathbf Z}\Vert^{2}}\sum_{j,k = 1}^{N-1}
\left[ \delta_{jk} - \frac{Z_{j}\overline{Z}_{k}}
                 {1+ \Vert{\mathbf Z}\Vert^{2}}\right]
d\overline{Z}_{j}\wedge dZ_{k}, \qquad {\mathbf Z}
                  = (Z_{1}, Z_{2}, \ldots , Z_{N-1}).
\end{equation}
Thus, clearly, $d\Omega_{FS} = 0$, implying that $\Omega_{FS}$ is a {\it
closed} 2-form, derivable from the  real K\"ahler potential
\begin{equation}
   \Phi({\mathbf Z}, \overline{\mathbf Z})
   = \log{\lbrack 1+ \Vert{\mathbf Z}\Vert^{2}\rbrack}.
\end{equation}
(That~is, $\Omega_{FS}=\displaystyle\sum_{j,k=1}^{N-1} \frac{\partial^2\Phi}
{\partial \overline Z_j \partial Z_k} \, d\overline Z_j \wedge d Z_k.$)

A~Hermitian metric $H_{FS}$ and a connection $\nabla_{FS}$ on $\mathbb L(\HH)$
can be defined using the inner product of $\HH$: Indeed, since $\pi^{-1}(\Psi)
= \{\Psi\} \times {\C} \psi$, the Hermitian structure $H_{FS}$ is  given
pointwise by
\begin{equation}
H_{FS}((\Psi, \psi),(\Psi, \psi^\prime)) = \langle \psi \vert \psi^\prime
\rangle \end{equation}
for all $(\Psi, \psi),(\Psi, \psi^\prime) \in \pi^{-1}(\Psi)$.  We will use
the identification map $\imath_\mathbb L$ defined in (\ref{id}) to construct a
connection on $\mathbb L(\HH)$. Define the 1-form $\alpha $ on $\HH$ by
\begin{equation}
\alpha(\psi)= \frac{\langle d\psi \vert \psi \rangle}{\|\psi\|^2} .
 \end{equation}
Then the pullback
\begin{equation}
\alpha_{FS} = \imath^*_\mathbb L \alpha
\end{equation}
defines a ${\C}^*$-invariant 1-form on $\mathbb L(\HH)$ whose horizontal space
at $(\Psi, \psi) \in \mathbb L(\HH)$ is $\HH_\Psi$. For~an arbitrary section
$s: V_\Phi \to \mathbb L(\HH) \setminus s_0$ as in (\ref{kap}), the pullback
\begin{equation}\label{pot}
-i\theta_{FS} = s^*\alpha_{FS}
\end{equation}
defines a local 1-form $\theta_{FS}$ on $\prhs$. Pointwise,
\begin{equation}
\theta_{FS}(\Psi) = i\frac{\langle d\kappa(\Psi)\vert
\kappa(\Psi)\rangle}{\Vert \kappa(\Psi) \Vert^2} = i\overline{\partial} \log
\Vert \kappa(\Psi)\Vert^2 ,
\end{equation}
where $\overline{\partial}$ denotes exterior differentiation with respect to
the anti-holomorphic variables. In terms of the coordinatization introduced in
(\ref{coord}), with $f$ as the holomorphic function representing $\kappa$,
we~have
\begin{equation}
\theta_{FS}({\mathbf Z}) = i\frac{d\overline{f({\mathbf Z})}}
  {\overline{f({\mathbf Z})}} + i\frac{\sum_j Z_j\,d\overline{Z}_{j}}
  {1 + \Vert {\mathbf Z}\Vert^2}.
\end{equation}
Furthermore, $\theta_{FS}$ locally defines a compatible connection
$\nabla_{FS}$
\begin{equation}
\nabla_{FS}\,s = -i\theta_{FS} \otimes s,
\end{equation}
and it is easy to verify that
\begin{equation}
\Omega_{FS} = \partial \theta_{FS} =  \mbox{\rm curv}\nabla_{FS},
\end{equation}
where $\partial$ denotes exterior differentiation with respect to the
holomorphic variables and $\mbox{\rm curv}\nabla_{FS}$ is the curvature form
of the line bundle $\mathbb L(\HH)$.

Thus  the Hermitian line bundle $(\mathbb L(\HH), H_{FS}, \nabla_{FS})$ is a
prequantization  of $(\prhs, \Omega_{FS})$ in the sense of geometric
quantization.

\subsection{Summary of coherent state quantization}\label{sec-summcohstquant}

The prequantization of $(\prhs,\Omega_{FS})$ can be exploited to obtain a
prequantization of an arbitrary symplectic manifold $(\bigam ,\Omega)$ whenever
there exists a  {\em symplectomorphism} $\mbox{\rm Coh}$ of $\bigam$
into~$\prhs$. In this case, $\Omega = \mbox{\rm Coh}^* \Omega_{FS}$ and the
line bundle $L:= \mbox{\rm Coh}^*\mathbb L(\HH)$, equipped with the
Hermitian metric $\mbox{\rm Coh}^*H_{FS}$ and (compatible) connection
$\Delta_K:= \mbox{\rm Coh}^*\nabla_{FS}$, is~a prequantization of $\bigam$,
i.e. in particular, $\Omega = \mbox{curv}
(\mbox{\rm Coh}^*\nabla_{FS})$. The expression
\begin{equation}  \theta_K(x):= i(\mbox{\rm Coh}^{*}\theta_{FS})(x),
\end{equation}
defines a 1-form on $L$, for which $\Omega = \,d \theta_K$.
The~Hermitian metric $H_K = \mbox{\rm Coh}^* H_{FS}$ and the compatible
connection $\nabla_K$ are given by
\begin{eqnarray}
H_K((x,\psi),(x,\psi^\prime) ) = \langle \psi, \psi^\prime \rangle\\
\nabla_K\,\mbox{\rm coh} = -i\theta_K \otimes \mbox{\rm coh},
\end{eqnarray}
where $\mbox{\rm coh}$ denotes a smooth section of $L$ and $\mbox{curv}
\nabla_K = \Omega$. More generally, if $\mbox{\rm Coh}: \bigam \to \prhs$ is
only assumed to be a smooth map, not necessarily a symplectomorphism, the above
scheme gives us a prequantization of the symplectic manifold
$(\bigam,\Omega_K)$  where $\Omega_K = \mbox{\rm Coh}^*\Omega_{FS}$. That~is,
one has:

\begin{prop} \label{bundle}
The triple $ (\pi: L  \to  \bigam, H_K, \nabla _K) $,
where $\nabla _K\mbox{\rm coh}  = -i\theta_K\otimes \mbox{\rm coh}$,  is a
Hermitian line bundle with compatible connection, and $\mbox{\rm curv}
\nabla_{K} = \Omega_{K} $.  \end{prop}

To make the connection with coherent states, we note that the elements of
$L$ are pairs $(x, \psi)$ with $\psi \in \HH$ and
$\frac{\vert\psi\rangle\langle\psi\vert}{\Vert\psi\Vert^{2}} = \Psi =
\mbox{\rm Coh}(x)$. Let $U \subset \bigam$ be an open dense set such that the
restriction of $\mathbb L$ to $U$ is trivial. Let $\mbox{\rm coh}: U \to \HH$
be a smooth section of~$L$, that~is, a smooth map satisfying
\begin{equation}
\mbox{\rm Coh}(x) = \frac{\vert \mbox{\rm coh}(x) \rangle \langle \mbox{\rm
coh}(x) \vert }{\Vert \mbox{\rm coh}(x) \Vert^2}
\end{equation}
(such maps can always be found). Let us also write $\eta_x = \mbox{\rm coh}(x),
\quad \forall x \in U$. Assume furthermore that the condition
\begin{equation}
\int_\Gamma \vert \eta_x\rangle \langle \eta_x \vert d\nu(x) = I_{\HH}
\label{tag:KOKO}
\end{equation}
is satisfied, where $I_{\HH}$ is the identity operator on $\HH$ and $\nu$ is
the Liouville measure on $\bigam$, arising from $\Omega$. We call the vectors
$\eta_x$ the {\em coherent states} of the prequantization.

In~terms of the reproducing kernel $K(x,y) = \langle \eta_x \vert \eta_y
\rangle$ and locally on $U$,
$$ \theta_K(x) = d_1 \log K(x_1,x_2) \vert_{x_1 = x_2 = x}   $$
($d_1$ denoting exterior differentiation with respect to $x_1$). Once~we
have~(\ref{tag:KOKO}), we~can define a quantization via the recipe
\begin{equation}
 f \longmapsto Q_f = \int_{\bigam}f(x)
         \vert\eta_x\rangle\langle\eta_x\vert\; d\nu(x) .  \label{tag:RECE}
\end{equation}
Note that this is a particular case of the ``prime quantization'' discussed
in~Section~\ref{sec5}.

As a consequence of Proposition \ref{bundle} we see that $\Omega_{K} $ so
constructed has integral cohomology. Thus the pair $(\bigam, \Omega_{K} )$
satisfies the {\em integrality condition}. We have thus obtained a geometric
prequantization on $(\bigam,\Omega_{K})$ from the natural geometric
prequantization of ($\prhs , \Omega_{FS})$ via the family of coherent states
$\{\eta_x\}$.  While the new two-form $\Omega_{K}$ on $\bigam$  is integral,
this is not necessarily the case for the original form $\Omega$. If~it is, then
there exists a geometric prequantization on $(\bigam,\Omega)$ which we may
compare with the prequantization obtained using the coherent states.
The~original prequantization is said to be {\em projectively induced} if
$\Omega = \Omega_K$; if furthermore, $\bigam$ has a complex structure which is
preserved by $\mbox{\rm Coh}$,  the symplectic manifold $(\bigam, \Omega)$
turns out to be a K\"ahler manifold. For the Berezin quantization, discussed
in Section~\ref{sec4}, the coherent states can be shown to give rise to a
projectively induced prequantization if $\bigam$ is a Hermitian symmetric
space.

It ought to be pointed out that while the map $\mbox{\rm Coh}: \bigam \to
\prhs$ yields a prequantization of $(\bigam, \Omega)$, the method outlined
above does not give an explicit way to determine $\HH$ itself.  However,
starting with the Hilbert space $L^2(\bigam,\nu)$, one can try to obtain
subspaces $\HH_K \subset  L^2(\bigam,\nu)$, for which there are associated
coherent states. Note that (\ref{tag:KOKO}) then means that $\HH_K$ will~be,
in~fact, a~reproducing kernel space (with reproducing kernel $\spr{\eta_x}
{\eta_y}$).

\subsubsection*{Two simple examples}
  Consider a free particle, moving on the configuration space
$\mathbb R^3$. Then,
$\bigam = {\mathbb R}^6$, is the phase space.
This is a symplectic manifold with two-form $\Omega = \sum_{i=1}^3
dp_i \wedge dq_i$. Let $\HH = L^2 (\bigam, d\bp\;d\bq )$ and let us look for
convenient subspaces of it which admit reproducing kernels.
Let~$e:{\mathbb R}^3 \longrightarrow \mathbb C$ be a measurable function,
depending only on the modulus $\Vert \bk \Vert$  and  satisfying
$$ \int_{{\mathbb R}^3} \vert e(\bk )\vert^2 \;d\bk = 1.   $$
For $\ell = 0, 1, 2, \ldots ,$ denote by ${\mathcal P}_{\ell}$ the Legendre
polynomial of order $\ell$,
$$ {\mathcal P}_\ell (x) = \frac 1{2^\ell \ell!} \;\frac {d^\ell}{dx^\ell}
 (x^2 -1 )^\ell . $$
Define
\be
K_{e, \ell}(\bq , \bp; \bq', \bp' ) = \frac {2\ell +1}{(2\pi)^3}\;
\int_{{\mathbb R}^3}      e^{i\bk\cdot (\bq - \bq')}\;{\mathcal P}_\ell
\left( \frac {(\bk - \bp)\cdot (\bk - \bp')}
{\Vert\bk - \bp\Vert\;\Vert \bk - \bp'\Vert}\right)\;\overline{e(\bk - \bp )}
\;e(\bk - \bp )\; d\bk .     \label{quantrepker}   \en
It is then straightforward to verify \cite{bib:AliPr}
that $K_{e, \ell}$ is a reproducing kernel with the usual properties,
\be \begin{aligned}[0pt]
  K_{e, \ell}(\bq , \bp; \bq, \bp ) & >  0, \qquad (\bq, \bp ) \in \bigam ,\\
  K_{e, \ell}(\bq , \bp; \bq', \bp' ) & =  \overline{K_{e, \ell}
(\bq' , \bp'; \bq, \bp )}, \\
  K_{e, \ell}(\bq , \bp; \bq', \bp' ) & =  \int_{{\mathbb R}^6}
  K_{e, \ell}(\bq , \bp; \bq'', \bp'' ) K_{e, \ell}(\bq'' , \bp''; \bq', \bp' )
\; d\bp''\;d\bq'' ,
\end{aligned}   \label{rquantrepker}  \en
and we have the associated family of {\em coherent states},
\be
\mathfrak S =
 \{ \xi_{\bq , \bp} \in \HH \; \vert \; \xi_{\bq , \bp} (\bq' , \bp' )=
 K_{e, \ell}(\bq' , \bp'; \bq, \bp ) , \;\; (\bq, \bp), (\bq', \bp')\in
 \bigam\} \label{quantcohstates}
\en
which span a Hilbert subspace $\HH_{e, \ell} \subset \HH$ and satisfy the
resolution of the identity on~it:
\be
\int_{{\mathbb R}^6}\vert\xi_{\bq , \bp}\rangle\langle\xi_{\bq , \bp}\vert\;
d\bp\;d\bq = I_{e, \ell}.      \label{quantresolid}
\en
Using these coherent states we can do a prime quantization
as~in~(\ref{tag:RECE}),~i.e.,
\be
f \longmapsto Q_f = \int_{{\mathbb R}^6}f(\bq , \bp )
         \vert\xi_{\bq , \bp}\rangle\langle\xi_{\bq , \bp}\vert\; d\bp\;d\bq .
\label{primquant3}
\en
In~particular, we get for the position and momentum observable the operators,
\be
Q_{q_j}\equiv\widehat{q}_j = q_j - i\hbar\frac{\partial}{\partial p_j} , \quad
Q_{p_j}\equiv\widehat{p}_j = - i\hbar\frac{\partial}{\partial q_j}, \qquad
j = 1,2, 3 ,    \label{primquant4}   \en
on~$\HH_{e,\ell}$, so~that
$$ [\widehat{q}_i , \widehat{p}_j ] = i \hbar \delta_{ij}\;I_{e, \ell} . $$
This illustrates how identifying appropriate reproducing kernel Hilbert spaces
can lead to a physically meaningful quantization of the classical system.

Let us next try to bring out the connection between this quantization and
the natural prequantization on $\mathbb C\mathbb P (\HH_{e, \ell})$.
Consider~the~map
\be
\mbox{\rm Coh} : \bigam = {\mathbb R}^6 \longrightarrow \mathbb C\mathbb P
(\HH_{e, \ell}),   \qquad \mbox{\rm Coh}(\bq , \bp ) =   \frac {\vert\xi_{\bq ,
\bp}\rangle\langle\xi_{\bq , \bp}\vert}{\Vert\xi_{\bq , \bp}\Vert^2}.
\label{cohstquant3}
\en
It is straightforward, though tedious, to verify that
\be
   \mbox{\rm Coh}^*\Omega_{FS} = \Omega = \sum_{i=1}^3 dp_i \wedge dq_i .
\label{cohstquant4}
\en
Hence $\Omega$ is projectively induced. The pullback $L =
\mbox{\rm Coh}^* \mathbb L(\HH_{e, \ell})$ of the canonical line bundle
$\mathbb L (\HH_{e,\ell})$ (over $\mathbb C\mathbb P (\HH_{e, \ell})$) under
$\mbox{\rm Coh}$ gives us a line bundle over $\bigam = {\mathbb R}^6$.

Take a reference section $\widehat{s}(\bq , \bp ) = \xi_{\bq , \bp}$
in $L$. Square-integrable sections of this bundle form a Hilbert space
$\HH_{L}$, with scalar product
$$ \langle s_1 \vert s_2 \rangle = \int_{{\mathbb R}^6}
 \overline{\Psi_1 (\bq , \bp )}\; \Psi_2 (\bq , \bp )
 K_{e, \ell}(\bq, \bp; \bq ,\bp )\;d\bq\;d\bp ,
 \qquad s_i (\bq , \bp ) = \widehat{s}\Psi_i , \;\; i=1,2, $$
and again, $\HH_{L}$ is naturally (unitarily) isomorphic to $L^2
(\bigam , d\bq\;d\bp )$. We~take the symplectic potential
$$ \theta = \sum_{i=1}^3 p_i\; dq_i ,$$
so that $\Omega = d\theta$, and thus we obtain a prequantization,
as~in~\S\ref{sec21}, yielding the position and momentum operators
$$ \widehat{q}_j = -i\hbar \frac {\partial}{\partial p_j} + q_j , \qquad
   \widehat{p}_j = -i\hbar \frac {\partial}{\partial q_j} ,  $$
which are the same as in (\ref{primquant4}), but now act on the (larger) space
$L^2(\bigam , d\bq\;d\bp )$.

\medskip

Our second example, following \cite{bib:Gaz} and \cite{bib:GazP}, is somewhat
unorthodox and makes use of a construction of coherent states associated to the
principal series representation of $SO_0(1,2)$. The quantization is performed
using (\ref{tag:RECE}). The coherent states in question are defined on the
space $S^1 \times \mathbb R = \{ x \equiv (\beta,J) \, | \, 0 \leq \beta < 2\pi
, \, J \in \R \}$, which is the phase space of a particle moving on the unit
circle. The $J$ and $\beta$ are canonically conjugate variables and define the
symplectic form $dJ\wedge d\beta$. Let $\HH$ be an abstract Hilbert space and
let $\{\psi_n\}_{n=0}^\infty$ be an orthonormal basis of~it. Consider next the
set of functions,
\begin{equation}
  \phi_n (x) = e^{(-\epsilon n^2/2)} \,e^{n(\epsilon J +  i \beta)},
  \qquad n = 0,1,2, \ldots \;,
\label{tag:CS-fcns}
\end{equation}
defined on $S^1 \times \mathbb R$, where $\epsilon > 0$ is a parameter which
can be arbitrarily small. These functions are orthonormal with respect to the
measure,
$$
d\mu(x) = \sqrt{\frac{\epsilon}{\pi}}\,\frac{1}{2\pi}
e^{-\epsilon J^2} \, dJ\, d\beta \; .
$$
Define the normalization factor,
\begin{equation}
 \mathcal{N}(J) = \sum_{n=0}^\infty \vert\phi_n (x)\vert^2 =
\sum_{n=0}^\infty e^{(-\epsilon n^2)} \,e^{2n\epsilon J } < \infty\;
\label{tag:norci}
\end{equation}
(which is proportional to an elliptic Theta function), and use it to construct
the coherent states
\begin{equation}
\eta_x := \eta_{J, \beta} = \frac{1}{\sqrt{{\mathcal N} (J)}} \sum_{n=0}^\infty
    \overline{\phi_n (x)}\psi_n
 = \frac{1}{\sqrt{{\mathcal N} (J)}} \sum_{n=0}^\infty
    e^{(-\epsilon n^2/2)} \,e^{n(\epsilon J - i \beta)}\; \psi_n\; .
\label{tag:ccs}
\end{equation}
These are easily seen to satisfy $\Vert \eta_{J, \beta}\Vert = 1$ and the
resolution of the identity
\begin{equation}
 \int_{S^1\times \mathbb R}\vert \eta_{J,\beta} \rangle
 \langle \eta_{J, \beta} \vert\;
 \mathcal{N}(J)\;d\mu (x) = I_\HH \; ,
\label{tag:cylresolid}
\end{equation}
so that the map
$$ W:\HH \longrightarrow L^2 (S^1 \times \mathbb R , \; \mathcal N(J)\;d\mu ),
\quad \text{where} \quad
 (W\phi )(J, \beta ) = \langle \eta_{J,\beta} \mid \phi\rangle  \; , $$
is a linear isometry onto a subspace of $L^2 (S^1 \times \mathbb R , \;
\mathcal N (J) \;d\mu )$. Denoting this subspace by
$\HH_\text{hol}$, we see that it consists of functions of the type,
$$ (W\phi )(J, \beta ) = \frac1{\sqrt{\Cal N(J)}} \sum_{n=0}^\infty c_n z^n
:= \frac{F(z)}{\sqrt{\Cal N(J)}} , $$
where we have introduced the complex variable $z = e^{\epsilon J + i \beta}$
and $c_n = e^{-\epsilon n^2 /2}\langle \psi_n \mid \phi \rangle$. The function
$F(z)$ is entire analytic and the choice of the subspace $\HH_\text{hol}
\subset L^2 (S^1 \times \mathbb R , \; \mathcal N (J) \;d\mu )$ --- that is,
of the coherent states (\ref{tag:ccs}) --- is then akin to choosing a
polarization.

In view of (\ref{tag:KOKO}) and (\ref{tag:RECE}), the quantization rule for
functions $f$ on the phase space $S^1\times \mathbb R$ becomes
\begin{equation}
Q_f := \int_{S^1\times \mathbb R} f(J, \beta )\; | \eta_{J, \beta}\rangle
\langle \eta_{J, \beta}| \, \mathcal{N}(J)\; d\mu(x) .  \label{tag: oper}
\end{equation}
For $f(J, \beta ) = J$,
\begin{equation}
  Q_J = \int_{S^1\times \mathbb R} J\, | \eta_{J,\beta} \rangle
   \langle \eta_{J,\beta} |\;\mathcal{N}(J)\;d\mu (x)
  = \sum_{n=0}^\infty n\, | \psi_n\rangle \langle \psi_n|\; .
\label{tag:Jsym}
\end{equation}
This is just the angular momentum operator, which as an operator on
$\HH_\text{hol}$ is seen to assume the form
$Q_J = -i\frac{\partial}{\partial \beta}$.
For an arbitrary function of $\beta$, we get similarly
\begin{equation}
 Q_{f(\beta)} = \int_{S^1\times \mathbb R}  f(\beta) \,  |\eta_{J,\beta}\rangle
  \langle \eta_{J,\beta} |\; \mathcal{N}(J)\; d\mu(x)  = \sum_{n,n'}
  e^{-\frac{\epsilon}{4}\,(n-n')^2} \,c_{n-n'}(f)| \psi_n\rangle
 \langle \psi_{n'} |\; ,
\label{tag:f(beta)}
\end{equation}
where $c_{n}(f)$ is the $n$th Fourier coefficient of $f$. In particular, we
have for the ``angle'' operator:
\begin{equation}
\label{tag:opangle}
 Q_{\beta} = \pi I_{\HH} +  \sum_{n\neq n'} i
 \frac{e^{-\frac{\epsilon}{4}(n-n')^2}}{n-n'}\, | \psi_n \rangle
 \langle \psi_{n'} |,
\end{equation}
and for the ``fundamental Fourier harmonic'' operator
\begin{equation}
\label{tag:opfourier}
Q_{e^{i\beta}} =  e^{-\frac{\epsilon}{4}}\, \sum_{n=0}^\infty
 | \psi_{n + 1}\rangle \langle \psi_n |\; ,
\end{equation}
which, on $\HH_\text{hol}$, is the operator of multiplication by  $e^{i\beta}$
up to the factor $e^{-\frac{\epsilon}{4}}$ (which can be made arbitrarily close
to unity). Interestingly, the commutation relation
\begin{equation}
 \lbrack Q_J, \;Q_{e^{i\beta}} \rbrack = Q_{e^{i\beta}}\;,
\label{tag:cylccr}
\end{equation}
is ``canonical'' in that it is in exact correspondence with the classical
Poisson bracket
$$\{ J, e^{i\beta}\} = i e^{i\beta}\; . $$

\section{Some other quantization methods} \label{sec6}
Apart from geometric and deformation quantization, other quantization methods
exist; though it is beyond our expertise to discuss them all here, we~at least
briefly indicate some references.

For quantization by {\sl Feynman path integrals,\/} a standard reference is
Feynman and Hibbs~\cite{bib:FeyH} or Glimm and Jaffe~\cite{bib:Jaffe}; a~recent
survey is Grosche and Steiner~\cite{bib:Grosche}. Path integrals
are discussed also in Berezin's book~\cite{bib:BerBk}, and a local deformation
quantization formula resembling the Feynman expansion in a 2d quantum field
theory lies also at the core of Kontsevich's construction~\cite{bib:Kon} of
star product on any Poisson manifold. (More precisely, Kontsevich's formula is
an expansion of a certain Feynman integral at a saddle point, see Cattaneo and
Felder~\cite{bib:CattFeld}.) Connections between Feynman path integrals,
coherent states, and the Berezin quantization are discussed in Kochetov and
Yarunin \cite{bib:KoYar}, Odzijewicz \cite{bib:OdziB}, Horowski, Kryszen and
Odzijewicz  \cite{bib:OdziC}, Klauder~\cite{bib:Klaud}, Chapter~V in Berezin
and Shubin \cite{bib:+BerShu}, Marinov~\cite{bib:MarinB}, Charles
\cite{bib:+Charl}, and Bodmann~\cite{bib:Bodm}. For a discussion of
Feynman path integrals in the context of geometric quantization, see Gawedzki
\cite{bib:GaweF}, Wiegmann \cite{bib:Weigh}, and Chapter~9 in the book of
Woodhouse~\cite{bib:Woodh}.

Another method is the {\sl asymptotic quantization\/} of Karasev and Maslov
\cite{bib:KMab}. It~can be applied on any symplectic manifold, even when no
polarization exists and the geometric quantization is thus inapplicable. It~is
based on patching together local Weyl quantizations in Darboux coordinate
neighbourhoods, the result being a quantization rule assigning to any $f\in
C^\infty(\Omg)$ a Fourier integral operator on a sheaf of function spaces
over $\Omg$ such that the condition (\ref{tag:Da}) is satisfied. The main
technical point is the use of the Maslov canonical operator (see
e.g.~Mishchenko, Sternin and Shatalov~\cite{bib:Mis}). The main disadvantage of
this procedure is its asymptotic character: the operators gluing together the
local patches into the sheaf are defined only modulo~$O(h)$, and so essentially
everything holds just modulo~$O(h)$ (or, in an improved version, module
$O(h^\infty)$ or modulo the smoothing operators). The~ideas of Karasev and
Maslov were further developed in their book \cite{bib:KMbk} (see also
Karasev~\cite{bib:Karax}), in Albeverio and Daletskii \cite{bib:AlbD}, and
Maslov and Shvedov~\cite{bib:MaslSh}. A~good reference is Patissier and Dazord
\cite{bib:PatDa}, where some obscure points from the original exposition
\cite{bib:KMab} are also clarified. For comparison of this method with the
geometric and deformation quantizations, see Patissier~\cite{bib:Patis}.

We remark that this asymptotic quantization should not be confused with the
``asymptotic quantization'' which is sometimes alluded to in the theory of
Fourier integral operators and of generalized Toeplitz operators (in~the sense
of Boutet de Monvel and Guillemin), see e.g.~Boutet de Monvel \cite{bib:BdM} or
Bony and Lerner~\cite{bib:BLe} (though the two are not totally unrelated).
Another two asymptotic quantizations exist in coding theory (see
e.g.~Neuhoff~\cite{bib:Neuhof}, Gray and Neuhoff~\cite{bib:+GrNeu}) and in
quantum gravity (Ashtekar~\cite{bib:Ashte}).

{\sl Stochastic quantization\/} is based, roughly speaking, on viewing the
quantum indeterminacy as a stochastic process, and applying the methods of
probability theory and stochastic analysis. They are actually two of the kind,
the geometro-stochastic quantization of Prugove\v cki~\cite{bib:Prug} and the
stochastic quantization of Parisi and Wu~\cite{bib:PaWu}. The~former arose,
loosely speaking, from Mackey's systems of imprimitivity $(U,E)$
(Mackey~\cite{bib:Mack} --- see the discussion of Borel quantization in
\S\ref{sec-synborquant} above), with $U$ a unitary representation of a symmetry
group and $E$ a projection-valued measure satisfying $U_g E(m) U^*_g =E(gm)$
for any Borel set~$m$, by demanding that $E$ be not necessarily projection but
only positive-operator valued (POV) measure; this leads to appearance of
reproducing kernel Hilbert spaces and eventually makes contact with the prime
quantization discussed in the preceding section. See~Ali and Prugove\v
cki~\cite{bib:AliPr}; a~comparison with Berezin quantization is available in
Ktorides and Papaloucas~ \cite{bib:KtoP}. The~stochastic quantization of Parisi
and Wu originates in the analysis of perturbations of the equilibrium solution
of a certain parabolic stochastic differential equation (the~Langevin
equation), and we won't say anything more about it but refer the interested
reader to Chaturvedi, Kapoor and Srinivasan~\cite{bib:ChKS}, Damgaard and
H\"uffel \cite{bib:Damg}, Namsrai \cite{bib:Namsr}, Mitter \cite{bib:Mitt},
or Namiki~\cite{bib:Namik}. A~comparison with geometric quantization appears in
Hajra and Bandyopadhyay \cite{bib:HajBan} and Bandyopadhyay~\cite{bib:+Bandy}.
Again, the term ``stochastic quantization'' is sometimes also used as a synonym
for the {\sl stochastic mechanics\/} of Nelson~\cite{bib:NelsM}.

Finally, we mention briefly the method of {\sl quantum states\/} of
Souriau~\cite{bib:SouQS}. It~builds on the notions of diffeological space and
diffeological group, introduced in~\cite{bib:SouDfg}, which are too technical
to describe here, and uses a combination of methods of harmonic and convex
analysis. See~the expository article \cite{bib:SouExp} for a summary of later
developments. Currently, the connections of this method with the other
approaches to quantization seem unclear (cf.~Blattner~\cite{bib:Blatt}).

   The subject of quantization is vast and it is not the ambition,
nor within the competence, of~the present authors to write a
comprehensive overview, so we better stop our exposition at this
point, with an apology to the reader for those topics that were
omitted, and to all authors whose work went unmentioned. We have
not, for instance, at all touched the important and fairly complex
problem of quantization with constraints, including BFV and BRST
quantizations (see~Sniatycki \cite{bib:SniaCo}, Tuynman
\cite{bib:TuyBRST}, Ibort \cite{bib:Ibort}, Batalin and Tyutin
\cite{bib:BataTyu}, Batalin, Fradkin and Fradkina \cite{bib:BFF},
Kostant and Sternberg \cite{bib:+KoStern}, Grigoriev and
Lyakhovich~\cite{bib:GriLya}) and the relationship between quantization and
reduction (Sjamaar~\cite{bib:+Sjam}, Tian and Zhang \cite{bib:+TiZha}, Jorjadze
\cite{bib:+Jorj}, Bordemann, Herbig and Waldmann \cite{bib:+BoHeWa}, Mladenov
\cite{bib:+Mlad}, Huebschmann \cite{bib:+Hueb}, Vergne~\cite{bib:+Verg});
or~quantum field theory and field quantization (Greiner and
Reinhardt~\cite{bib:GreiRei}, Borcherds and Barnard~\cite{bib:BaBo}),~etc. Some
useful surveys concerning the topics we have  covered, as well as some of those
that we have not, are Sternheimer \cite{bib:Stw}, Weinstein \cite{bib:WeinAst},
Fernandes \cite{bib:Fern}, Echeverria-Enriquez et
al.~\cite{bib:EEMLRRVM}, Sniatycki \cite{bib:SniaA}, Ali
\cite{bib:AliSurv}, Blattner \cite{bib:Blatt}, Tuynman \cite{bib:TuyWis},
Borthwick \cite{bib:+Bwsrv}, and the books of Fedosov \cite{bib:FedosBk},
Landsman \cite{bib:LandsBk}, Bates and Weinstein \cite{bib:BaWei}, Souriau
\cite{bib:SouSD}, Perelomov \cite{bib:Perel}, Bandyopadhyay \cite{bib:+Bandy},
Greiner and Reinhardt \cite{bib:GreiRei} and Woodhouse \cite{bib:Woodh}
mentioned above.

\bigskip

\noindent\textsc{Acknowledgements.} This survey is based on an appendix to the
habilitation thesis of the second author~\cite{bib:EDrSc} and on lecture notes
from a course on quantization techniques given at Cotonou, Benin, by the first
author \cite{bib:AliBen}.
The authors would like to thank G.~Tuynman for many helpful conversations
on geometric quantization and record their gratitude to J.-P Antoine,
J.-P. Gazeau and G.A. Goldin, for constructive feedback on the manuscript.

\end{document}

\bye